\documentclass[traditabstract]{aa}
\usepackage{graphicx}
\usepackage{txfonts}
\usepackage{amssymb}
\usepackage[below]{placeins}
\usepackage{natbib}
\usepackage{longtable,lscape}
\bibpunct{(}{)}{;}{a}{}{,}

\usepackage{grffile}
\usepackage{epstopdf}
\graphicspath{{}}

\usepackage[table]{xcolor}
\definecolor{tableShade}{HTML}{E4E4E4}

\def\mpe               {1}
\def\lmu               {2}
\def\annarbor          {3}
\def\excellence        {4}
\def\maryland          {5}
\def\uilli             {6}
\def\NCSA              {7}
\def\UNDakota          {8}
\def\IAPFrance         {9}
\def\Fermilab          {10}
\def\UCOLick           {11}
\def\UChicago          {12}
\def\CUTaiwan          {13}
\def\saclay            {14}

\begin{document}


\title{The XMM-BCS galaxy cluster survey}
\subtitle{I. The X-ray selected cluster catalog from the initial 6 deg$^2$}
\authorrunning{R. \v{S}uhada et al.}

\author{R.~\v{S}uhada\inst{\mpe,\lmu}\thanks{email: rsuhada@usm.lmu.de},
J.~Song \inst{\annarbor},
H.~B\"ohringer\inst{\mpe},
J.~J.~Mohr\inst{\mpe,\lmu,\excellence},
G.~Chon\inst{\mpe},
A.~Finoguenov \inst{\mpe,\maryland},
R.~Fassbender\inst{\mpe,\maryland},
S.~Desai\inst{\lmu,\uilli},
R.~Armstrong\inst{\NCSA},
A.~Zenteno\inst{\lmu,\excellence},
W.~A.~Barkhouse\inst{\UNDakota},
E.~Bertin\inst{\IAPFrance},
E.~J.~Buckley-Geer\inst{\Fermilab},
S.~M.~Hansen\inst{\UCOLick},
F.~W.~High\inst{\UChicago},
H.~Lin\inst{\Fermilab},
M.~M\"uhlegger\inst{\mpe},
C.~C.~Ngeow\inst{\CUTaiwan},
D.~Pierini\thanks{Visiting astronomer at MPE.},
G.~W.~Pratt\inst{\saclay},
M.~Verdugo\inst{\mpe}, and
D.~L.~Tucker\inst{\Fermilab}
}

\institute{
\inst{\mpe} Max-Planck-Institut f\"ur extraterrestrische Physik,
Giessenbachstr. 1, 85748 Garching, Germany \\
\inst{\lmu} Department of Physics, Ludwig-Maximilians-Universit\"{a}t,
Scheinerstr. 1, 81679 Munich, Germany \\
\inst{\annarbor} University of Michigan, Physics Department, 450
Church Street, Ann Arbor, MI 48109-1040, USA \\
\inst{\excellence} Excellence Cluster Universe, Boltzmannstr. 2, 85748
Garching, Germany \\
\inst{\maryland} University of Maryland, Baltimore County, 1000
Hilltop Circle, Baltimore, MD 21250, USA \\
\inst{\uilli} Department of Astronomy, 1002 W. Green Street, Urbana,
IL 61801, USA \\
\inst{\NCSA} National Center for Supercomputing Applications,
University of Illinois, 1205 West Clark Street, Urbana, IL 61801, USA \\
\inst{\UNDakota} Department of Physics \& Astrophysics, University of
North Dakota, Grand Forks, ND 58202 \\
\inst{\IAPFrance} Institut d'Astrophysique de Paris, UMR 7095 CNRS,
Universit\'e Pierre et Marie Curie, 98 bis boulevard Arago, F-75014
Paris, France \\
\inst{\Fermilab} Fermi National Accelerator Laboratory, P.O. Box 500,
Batavia, IL 60510 \\
\inst{\UCOLick} University of California Observatories \& Department
of Astronomy, University of California, Santa Cruz, CA 95064 \\
\inst{\UChicago} University of Chicago, 5640 South Ellis Avenue,
Chicago, IL 6063 \\
\inst{\CUTaiwan} Graduate Institute of Astronomy, National Central
University, No. 300 Jonghda Rd, Jhongli City 32001, Taiwan \\
\inst{\saclay} Laboratoire AIM, IRFU/Service d'Astrophysique - CEA/DSM
- CNRS - Universit\'{e} Paris Diderot, B\^{a}t. 709, CEA-Saclay,
F-91191 Gif-sur-Yvette Cedex, France \\
}

\date{Received/accepted}



 \abstract
{
The XMM-\emph{Newton} - Blanco Cosmology Survey project (XMM-BCS) is a
coordinated X-ray, optical and mid-infrared cluster survey in a field
also covered by Sunyaev-Zel'dovich effect (SZE) surveys by the South
Pole Telescope and the Atacama Cosmology Telescope.  The aim of the
project is to study the cluster population in a 14~deg$^2$ field
(center: $\alpha\approx$ 23:29:18.4, $\delta\approx$ -54:40:33.6).
The uniform multi-wavelength coverage will also allow us for the first
time to comprehensively compare the selection function of the
different cluster detection approaches in a single test field and
perform a cross-calibration of cluster scaling relations.

In this work, we present a catalog of 46 X-ray selected clusters from
the initial 6~deg$^2$ survey core. We describe the XMM-BCS source
detection pipeline and derive physical properties of the clusters.  We
provide photometric redshift estimates derived from the BCS imaging
data and spectroscopic redshift measurements for a low redshift subset
of the clusters. The photometric redshift estimates are found to be
unbiased and in good agreement with the spectroscopic values.

Our multi-wavelength approach gives us a comprehensive look at the
cluster and group population up to redshifts $z \approx 1$. The median
redshift of the sample is 0.47 and the median mass
M$_{500}\approx1\times10^{14}$ M$_{\odot}$ ($\sim2$~keV). From the
sample, we derive the cluster $\log N - \log S$ using an approximation
to the survey selection function and find it in good agreement with
previous studies.

We compare optical mass estimates from the Southern Cosmology Survey
available for part of our cluster sample with our estimates derived
from the X-ray luminosity. Weak lensing masses available for a subset
of the cluster sample are in agreement with our estimates.  Optical
masses based on cluster richness and total optical luminosity are
found to be significantly higher than the X-ray values.

The present results illustrate the excellent potential of medium-deep,
X-ray surveys to deliver cluster samples for cosmological
modelling. In combination with available multi-wavelength data in
optical, near-infrared and SZE, this will allow us to probe the
dependence of the selection functions on relevant cluster observables
and provide thus an important input for upcoming large-area
ðmulti-wavelength cluster surveys.

}



   \maketitle
%

\section{Introduction}
\label{sec:intro}
The formation of the cold dark matter (CDM) dominated large-scale
structure of the Universe is hierarchical with smallest objects
collapsing first. With passing time more and more massive structures
are able to decouple from the Hubble flow and enter the non-linear
regime, collapse and eventually virialize. The statistical properties
of the matter density field (e.g. its power spectrum) as well as the
growth of the structures are strongly dependent on the background
cosmology and can be thus used to put constraints on cosmological
models.

From this point of view, clusters occupy a very important place in the
structure formation scenario, by being the most recent (i.e. redshifts
$z \lesssim 2$ - coincident with the onset of the dark energy
dominance) and thus also the most massive structures ($10^{13} -
10^{15}$ M$_{\odot}$) to virialize. The cluster abundance is therefore
exponentially sensitive to the growth of the large scale-structure and
to the underlying cosmological parameters
\citep{haiman01,majumdar03,haiman05}.

The key parameter in cosmological tests of this type - the total mass
of clusters (identified with dark matter halos) - is itself not
directly observable. Fortunately, in first approximation, clusters are
virialized and their growth is gravitationally driven and therefore
self-similar. This allows us to link their mass to some suitable
observable quantity originating from the baryonic components of a
cluster - its galaxy population and the intra-cluster medium
(ICM). The ICM is directly observable in X-rays or through the
distortion of the Cosmic Microwave Background (CMB) imprinted by the
ICM thermal electron population via inverse Compton scattering
\citep[the so-called Sunyaev-Zel'dovich effect (SZE),][]{sunyaev72}.

Since the ICM closely traces the DM potential, it offers better
(i.e. lower scatter) mass-proxies than those available from optical
observations of the cluster's galaxy population
\citep[e.g.][]{reyes08}.  In X-rays, the simplest and observationally
least expensive mass-proxy is the X-ray luminosity $L_{\mathrm{X}}$
\citep{reiprich02, pratt09, mantz10}.

For the SZE experiments the most direct way to estimate the cluster
mass is from the source signal-to-noise ratio
\citep[e.g.][]{williamson11, vanderlinde10} and more importantly,
through the integrated Compton parameter $Y_{\mathrm{SZ}}$. Numerical
simulations suggest that $Y_{\mathrm{SZ}}$ is an excellent proxy of
cluster mass \citep{dasilva04,motl05,nagai06}. First cross-comparisons
with X-ray and SZE studies are generally finding good agreement
between the mass estimates and no significant deviation from the
self-similar predictions \citep{planck11c, planck11b, planck11,
  melin11, andersson10, marrone09, bonamente08}.

If deeper X-ray observations are available, we can use the
spectroscopic temperature $T_{\mathrm{X}}$, gas mass $M_{\mathrm{g}}$
and their combination $Y_{\mathrm{X}}=T_{\mathrm{X}} M_{\mathrm{g}}$
\citep[the X-ray analogue to the $Y_{\mathrm{SZ}}$
  parameter,][]{kravtsov06, vikhlinin09, arnaud10} as good mass
proxies. Using the $Y_{\mathrm{X}}$ parameter \citet{vikhlinin09} put
a strong constraint on the cosmological parameters including the dark
energy equation of state. From a methodological point of view, this is
interesting for two reasons: \textbf{1)} it shows that useful
cosmological constrains can be obtained already from relatively small
samples of clusters of galaxies, demonstrating the exceptional
potential of this type of cosmological tests; and \textbf{2)} already
this modest sample is practically systematics-limited, especially due
to uncertainties in the mass estimation.

There are many factors that affect the scaling relations and the
intrinsic scatter of the cluster populations around these relations:
the presence of cool cores \citep{markevitch98, ohara06, motl05,
  pratt09}, substructures and the cluster's dynamical state
\citep{boehringer10b, jeltema08} and additional non-gravitational
physics \citep{nagai06}, etc. In addition, one has to account for the
Malmquist and Eddington bias when determining the scaling relations
from an X-ray selected sample of clusters by proper treatment of the
selection and mass functions \citep[especially for
  $L_{\mathrm{X}}$,][]{pacaud07, vikhlinin09, pratt09, mantz10,
  mantz10a}.  As our cluster samples cover broader redshift ranges
potential deviations from the self-similar evolution of the scaling
relations also become an important question.

In summary, in order to be able to well constrain cosmological models
with cluster samples we need: \textbf{1)} large cluster samples
covering redshifts beyond unity; \textbf{2)} good knowledge of the
cluster selection function's dependence on relevant observables and
the distributions of these observables in the cluster population;
\textbf{3)} a reliable, low scatter mass-proxy with a known evolution
in the redshift range of interest.

Surveying for clusters in SZE has a large potential with regards to
all three requirements, having an almost redshift independent
selection very close to a selection function with a fixed mass limit
at all redshifts and a robust mass-proxy in the $Y_{\mathrm{SZ}}$
parameter.  Two ground-based large-area cluster surveys are currently
underway: one by the South Pole Telescope (SPT) and one by the Atacama
Cosmology Telescope (ACT). Both have already provided their first
SZE-selected cluster samples
\citep[][]{williamson11,vanderlinde10,marriage10,staniszewski09} as
well as observations of already known clusters \citep{plagge10,
  hincks10}. Also the \emph{Planck} space mission has delivered its
first cluster catalog \citep{planck11}.

While the SZE surveying approach is a very interesting new channel to
perform cluster cosmology, there is still much work to be done at
these early stages to understand the systematics like the influence of
radio/sub-mm sources and primary CMB fluctuations on the selection,
the mass calibration and sensitivity to cluster outskirts.

A multi-wavelength follow-up program of SZE selected clusters is
essential, but selection function studies require also comparison of
blind surveys. To this end we are conducting the XMM-BCS cluster
survey.  The survey field covers a 14 deg$^2$ area in the overlap
region of the SPT and ACT surveys.  The field has full coverage with
the 4m CTIO telescope at Cerro Tololo, Chile, in the framework of the
Blanco Cosmology Survey (BCS) in \emph{griz} bands and \emph{Spitzer}
observations in the mid-infrared (mid-IR). With this optical to mid-IR
coverage we are able to provide robust photometric redshift estimates
out to redshifts $\approx0.8$ ($\approx 1$ once also the
\emph{Spitzer} data is included).  The X-ray coverage consists of
XMM-\emph{Newton} observations split into two distinct parts.  The 6
deg$^2$ core of the X-ray survey field was observed with 42
individual, standard pointings (with $\sim10$~ks effective exposure
time).  In this work, we present an initial cluster catalog based on
these observations.

After SPT commenced its operations, it was soon found that the mass
threshold of contemporary SZE surveys is higher than expected.  In
order to offer a larger overlap between the SZE and X-ray selected
cluster samples, we carried out an extension of the X-ray survey by
covering an additional 8 deg$^2$ in three large-area fields utilizing
the new mosaic mode type of observations. These observations allowed
us to cover a significantly larger area in a very time-efficient
way. First results as well as details on the analysis of this type of
XMM-\emph{Newton} observations are described in \citet{suhada10}.  We
demonstrate there the feasibility of blindly detecting clusters found
with current generation SZE experiments in only $\sim3$~ks long
XMM-\emph{Newton} observations (including tentative spectroscopic
temperature measurements) in the case of two SPT detected clusters.
The final 14 deg$^2$ X-ray cluster catalog is expected to roughly
double the number of clusters in the present sample and this sample
will then be interesting also for its cosmology-constraining power.

The paper is organized as follows: in Sect.~\ref{sec:pipe} and
\ref{sec:detectpipe} we describe the analysis of the X-ray
observations and cluster detection pipeline. The optical data,
photometric redshift estimation and spectroscopic campaign are
detailed in Sect.~\ref{sec:photoz}.  In Sect.~\ref{sec:results} we
provide our cluster sample, the physical parameters of the detected
clusters and determine the survey's preliminary statistical
properties. We also cross-correlate our cluster catalog with known
sources and carry out a detailed comparison with the optically
selected sample of \citet{menanteau09} and \citet{menanteau10} (M09
and M10 hereafter).  Sect.~\ref{sec:discuss} discusses the X-ray error
budget and gives an outlook on the upcoming work in the context of the
XMM-BCS survey. We give our conclusions in Sect.~\ref{sec:results}. In
the appendices we provide ancillary information for the individual
clusters, a preliminary comparison of our simplified sensitivity
function calculations with realistic simulations and a
cross-comparison with the XMM-LSS cluster survey.

We adopt a $\Lambda$CDM cosmology with $(\Omega_{\Lambda}, \Omega_{M}, w, H_0) = (0.7, 0.3,
-1, 70$ km s$^{-1}$~Mpc$^{-1})$.  Estimated physical parameters are
given in apertures corresponding to overdensities by factors 200 and
500 with respect to the \emph{critical} density of the Universe at the
redshift of a given cluster. Throughout the article we refer to
objects in our sample as "clusters" regardless of their mass. The term
"group" will be used to refer to systems with masses $M_{200}
\lesssim10^{14}$~M$_{\odot}$. We will refer to individual objects by
their identification number (ID). Proper object names are listed in
Table~\ref{tab:flags}.

\section{XMM-\emph{Newton} data reduction}
\label{sec:pipe}
\begin{figure*}[ht!]
\begin{center}
\includegraphics[width=\textwidth]{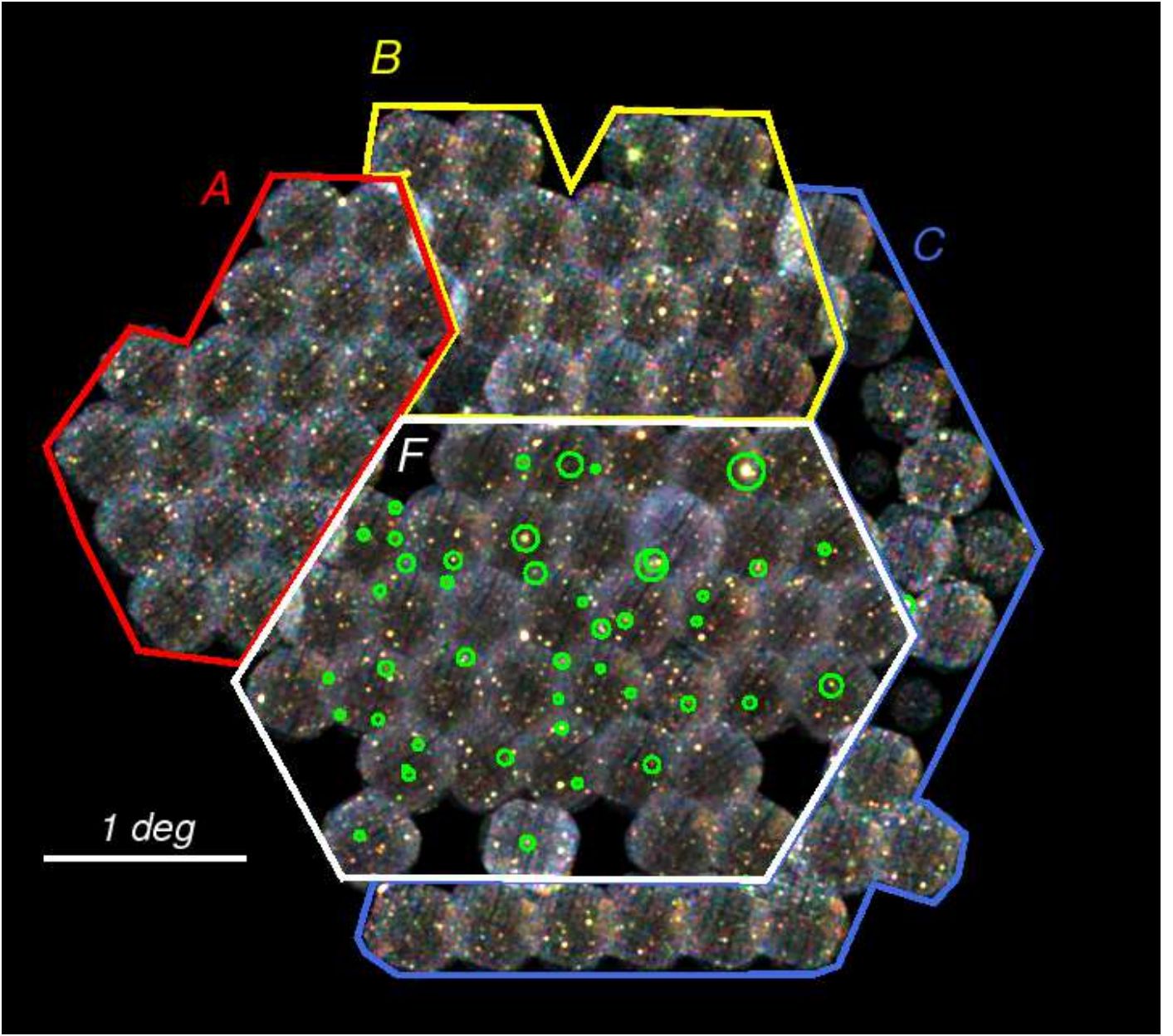}
\end{center}
\caption[The XMM-BCS mosaic]{Mosaic X-ray image of the 14~deg$^2$
  XMM-\emph{Newton} sky survey. The false color image was constructed
  from the surface brightness in the $0.3-0.5$, $0.5-2.0$ and
  $2.0-4.5$~keV bands.  The white region (F) marks the 6~deg$^2$ core
  of the survey presented in this work.  Regions A, B and C constitute
  the extension of the survey by mosaic mode observations.  The
  missing fields have significant losses due to soft proton
  flares. Bluer fields are affected by enhanced background.  Green
  circles mark the positions of the present cluster sample and have a
  radius equal to $r_{500}$.}
\label{fig:mosaic}
\end{figure*}

\begin{figure*}[ht!]
\begin{center}
\includegraphics[width=\textwidth]{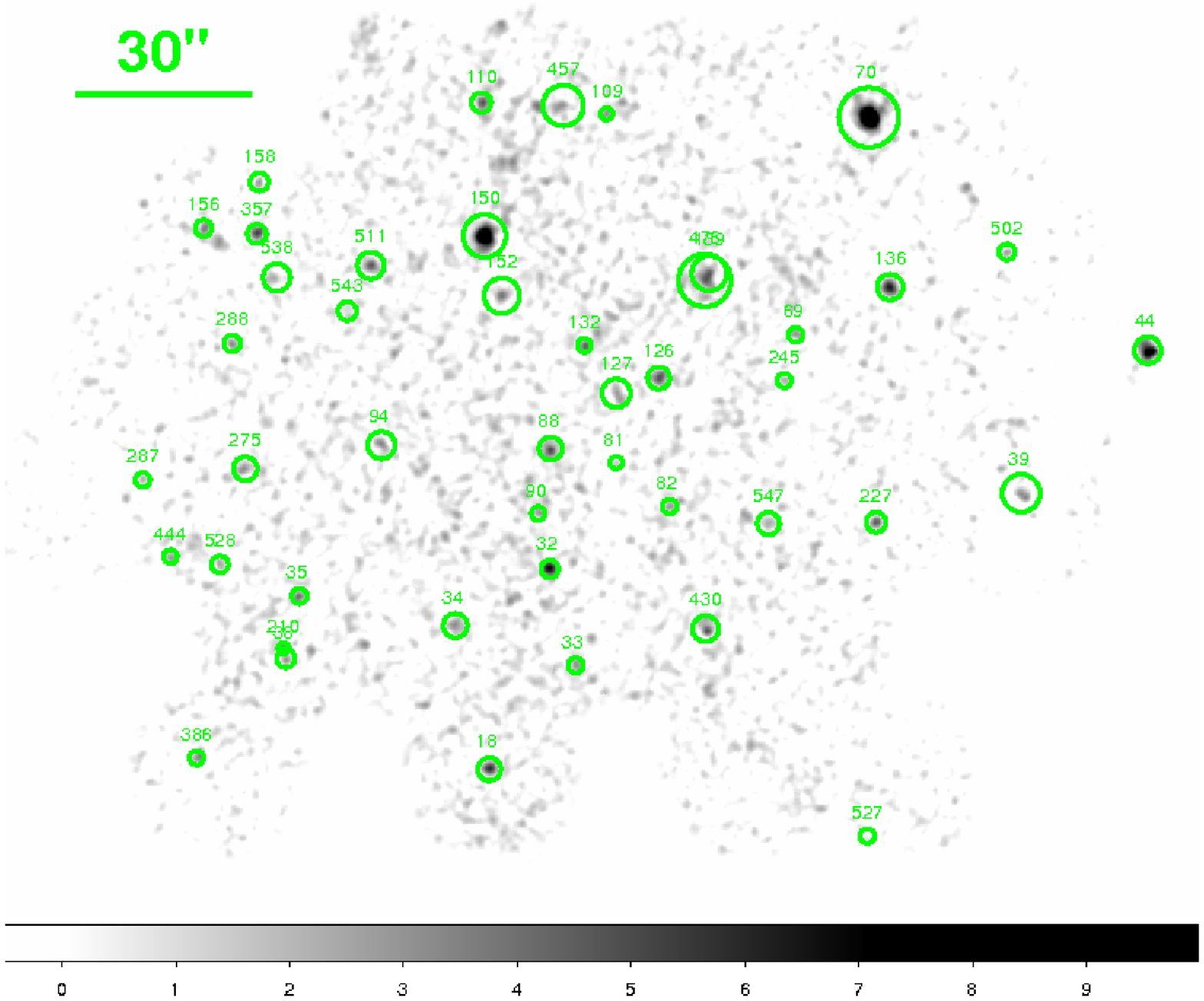}
\end{center}
\caption[The XMM-BCS signal-to-noise map]{The $0.5-2$~keV band
  signal-to-noise ratio map of the XMM-BCS core region (region F in
  Fig.~\ref{fig:mosaic}) smoothed with a Gaussian of
  $32^{\prime\prime}$ width. Circles indicate the $r_{500}$ radii of
  the detected clusters and are labeled with the cluster ID number in
  the catalog. Point sources have been subtracted using the method of
  \citet{finoguenov09}.}
\label{fig:s2nmap}
\end{figure*}

The XMM-\emph{Newton} coverage of the XMM-BCS survey core consists of
42 partially overlapping pointings with offsets of 22.8 arcmin
covering a total area of about 6 deg$^2$ (see Fig.~\ref{fig:mosaic}).
The observing time was allocated in the frame of an
\emph{XMM-\emph{Newton} Large Program} during AO6. Four additional
observations were carried out in AO7 to replace fields with large
losses due to soft-proton flaring. The observation of field F09
(Table~\ref{tab:obslog}) was carried out in two parts.  The total
observing time amounts to $\sim580$ ks, with an average total nominal
time per pointing of $\sim15$~ks (including instrument setup time and
high background periods).  Table~\ref{tab:obslog} displays the basic
information about the individual pointings. The THIN filter was used
in all observations. The EPIC PN camera was operated in full frame
mode.

The full XMM-BCS X-ray field is displayed in
Fig.~\ref{fig:mosaic}. The core region presented in this work is
inside the white boundaries (region F). Regions A, B and C mark the
three mosaic extensions of the survey. The five missing fields in
region F have been completely lost due to flaring (F03, F05, F42) or
had large time losses due to flaring and have a very high residual
quiescent soft proton contamination (F07 and F13). The point source
subtracted signal-to-noise map of the core region is displayed in
Fig.~\ref{fig:s2nmap} \citep[see e.g.][]{finoguenov09, bielby10}.

\begin{table*}
 \centering
 \caption[XMM-\emph{Newton} observation log]{The individual
   XMM-\emph{Newton} pointings. Quoted exposures are effective
   exposures with high background periods filtered out.}
\vspace{0.1cm}
 \label{tab:obslog}
\begin{footnotesize}
\begin{tabular}{l l l l  l l l  l l l}
\hline
\hline
\multicolumn{2}{c}{Field ID} & RA (J2000) & Dec (J2000) &
\multicolumn{3}{c}{Exposure times (ks)} \\
OBSID & Internal & & & PN &MOS1 & MOS2 \\
\hline
0505380101 & F01  & 23:21:38.4 & -56:07:34.4  & \multicolumn{3}{l}{observation
lost due to flaring} \\
0554561001 & F01b & 23:22:00.1 & -56:09:03.3  & 7.8   & 10.4  & 10.4 \\
0505380201 & F02  & 23:24:23.5 & -56:07:13.2  & \multicolumn{3}{l}{observation
lost due to flaring} \\
0554560201 & F02b & 23:24:43.8 & -56:09:03.0  & 10.2  & 13.2  & 13.2 \\
0505380301 & F03  & 23:27:07.0 & -56:07:16.3  & \multicolumn{3}{l}{observation
lost due to flaring} \\
0505380401 & F04  & 23:29:50.6 & -56:07:16.0  & 5.2   & 6.9   & 6.9  \\
0554560901 & F04b & 23:30:11.7 & -56:09:01.2  & 3.1   & 12.7  & 12.7 \\
0505380501 & F05  & 23:32:34.6 & -56:07:12.8  & \multicolumn{3}{l}{observation
lost due to flaring} \\
0505380601 & F06  & 23:35:39.3 & -56:08:18.7  & 5.6   & 10.6  & 10.6 \\
0505380701 & F07  & 23:20:49.3 & -55:45:35.1  & 4.4   & 9.6   & 9.6  \\
0505380801 & F08  & 23:23:31.4 & -55:45:39.2  & 9.3   & 11.0  & 9.8  \\
0505380901 & F09  & 23:26:12.7 & -55:46:10.2  & 2.3   & 6.0   & 6.0  \\
0505384801 & F09b & 23:26:11.6 & -55:46:30.2  & 7.3   & 9.8   & 9.8  \\
0505381001 & F10  & 23:28:55.3 & -55:45:39.2  & 9.7   & 12.6  & 12.6 \\
0505381101 & F11  & 23:31:37.8 & -55:45:39.7  & 7.2   & 9.7   & 9.7  \\
0505381201 & F12  & 23:34:19.5 & -55:45:42.6  & 10.8  & 13.5  & 13.5 \\
0505381301 & F13  & 23:37:01.4 & -55:45:39.2  & 2.3   & 10.6  & 10.6 \\
0505381401 & F14  & 23:19:29.9 & -55:23:01.1  & 10.8  & 13.9  & 13.9 \\
0505381501 & F15  & 23:22:09.7 & -55:23:23.1  & 7.4   & 9.9   & 9.9  \\
0505381601 & F16  & 23:24:50.3 & -55:23:26.3  & 3.2   & 11.7  & 11.7 \\
0505381701 & F17  & 23:27:29.7 & -55:23:45.9  & 7.3   & 10.0  & 10.0 \\
0505381801 & F18  & 23:30:10.5 & -55:23:41.1  & 11.3  & 15.2  & 15.2 \\
0505381901 & F19  & 23:32:51.0 & -55:23:38.5  & 7.4   & 8.9   & 8.9  \\
0505382001 & F20  & 23:35:31.3 & -55:23:44.6  & 10.5  & 13.9  & 13.9 \\
0505382101 & F21  & 23:38:12.0 & -55:23:43.7  & 4.8   & 8.2   & 8.2  \\
0505382201 & F22  & 23:18:20.7 & -55:00:13.1  & 11.8  & 14.3  & 14.3 \\
0505382301 & F23  & 23:20:58.9 & -55:00:36.3  & 7.3   & 10.0  & 10.0 \\
0505382401 & F24  & 23:23:37.8 & -55:00:35.5  & 7.5   & 10.0  & 10.0 \\
0505382501 & F25  & 23:26:16.6 & -55:00:42.1  & 15.2  & 20.6  & 20.6 \\
0505382601 & F26  & 23:28:55.2 & -55:00:49.1  & 9.4   & 12.1  & 12.1 \\
0505382701 & F27  & 23:31:34.3 & -55:00:51.0  & 6.1   & 11.9  & 11.9 \\
0505382801 & F28  & 23:34:12.9 & -55:00:55.7  & 7.1   & 9.8   & 9.8  \\
0505382901 & F29  & 23:36:51.9 & -55:00:54.2  & 7.3   & 9.9   & 9.9  \\
0505383001 & F30  & 23:19:41.6 & -54:37:27.7  & 12.7  & 16.4  & 16.4 \\
0505383101 & F31  & 23:22:18.6 & -54:37:53.3  & 7.4   & 10.0  & 10.0 \\
0505383201 & F32  & 23:24:56.1 & -54:37:52.3  & 11.6  & 13.6  & 13.6 \\
0505383301 & F33  & 23:27:32.7 & -54:38:04.7  & 13.0  & 15.9  & 15.9 \\
0505383401 & F34  & 23:30:10.6 & -54:38:00.9  & 9.1   & 11.9  & 11.9 \\
0505383501 & F35  & 23:32:29.0 & -54:36:00.3  & \multicolumn{3}{l}{observation
lost due to flaring} \\
0554560601 & F35b & 23:32:47.7 & -54:38:05.8 & 7.7 &  11.4 & 11.1  \\
0505383601 & F36  & 23:35:25.6 & -54:37:57.3  & 8.6   & 11.8  & 11.8 \\
0505383701 & F37  & 23:21:08.8 & -54:15:02.4  & 7.5   & 9.9   & 9.9  \\
0505383801 & F38  & 23:23:44.6 & -54:15:01.5  & 8.7   & 11.5  & 11.5 \\
0505384901 & F39  & 23:25:58.1 & -54:14:20.2  & 5.5   & 6.8   & 6.8  \\
0505384001 & F40  & 23:28:56.6 & -54:15:15.2  & 9.4   & 12.2  & 12.2 \\
0505384101 & F41  & 23:31:32.4 & -54:15:13.7  & 9.9   & 12.5  & 12.5 \\
0505384201 & F42  & 23:33:49.9 & -54:13:13.3  & \multicolumn{3}{l}{observation
lost due to flaring} \\
\hline
\end{tabular}
\end{footnotesize}
\end{table*}

The EPIC data was processed with the XMM-\emph{Newton} Standard
Analysis System (SAS) version 7.1.0\footnote{We provide a test
    using the current SAS 11.0.0 in Sect.~\ref{sec:systematics}.}. We
reduced and calibrated the raw observational data files with the SAS
tasks \texttt{epchain} for the EPIC PN detector and \texttt{emchain}
for both MOS detectors. Events in bad pixels, bad columns and close to
the chip gaps are excluded from further analysis.

\begin{figure*}[t!]

\begin{center}

\includegraphics[width=0.41\textwidth]{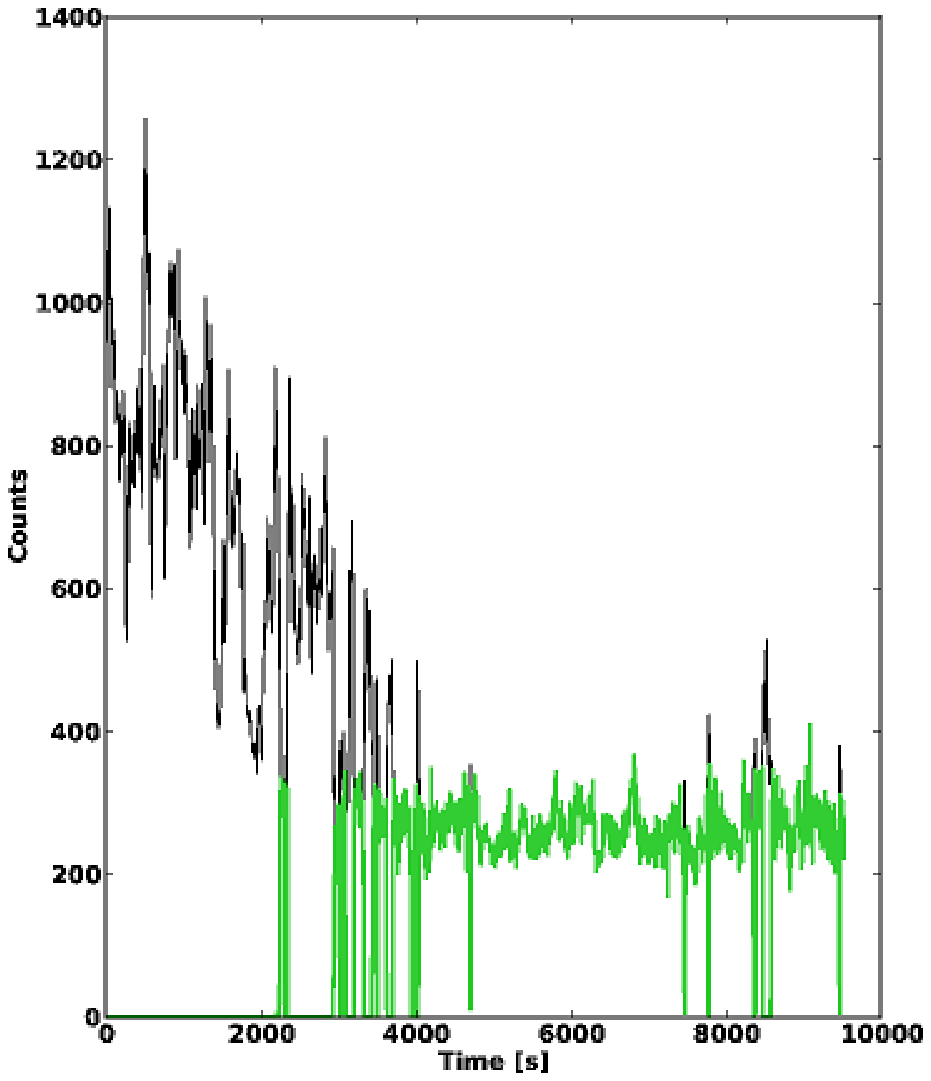}
\includegraphics[width=0.58\textwidth]{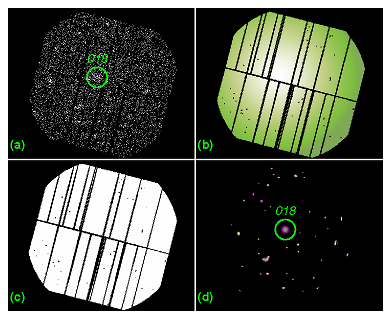}

\end{center}

\caption[Pipeline products]{\textbf{Left:} The black line shows the 10
  second-binned PN light curve in the $0.3-10$~keV band for the field
  F04. The beginning of the observation was affected by flaring.  The
  green curve shows the light curve after the two-step cleaning (see
  Sect.~\ref{sec:pipe}), which safely removed all contaminated time
  intervals. \textbf{Right:} Examples of the detection pipeline
  products for field F04 in the $0.5-2$~keV band of the PN detector:
  \textbf{a)} counts image, \textbf{b)} double-component background
  model, \textbf{c)} binary detection mask, \textbf{d)} reconstruction
  of all the detected sources.  The green circle (2 arcmin radius)
  marks the cluster ID 018.}
\label{fig:pipeprod}
\end{figure*}

The event lists were screened for high background periods caused by
soft proton flares with a two-step sigma clipping algorithm
\citep{pratt03}. We reject time intervals with background count rates
above the $3 \sigma$ limit from the mean level in the $12 -14$ keV
band for PN and $10 - 12$~keV band for MOS1 and MOS2. The mean
background count rate is determined by fitting a Gaussian model to the
distribution of counts in the light curve binned in 100 second
intervals. After this first cleaning step, we apply the same $3
\sigma$ clipping procedure in the $0.3 - 10$~keV band on 10 second
binned light curves to conservatively remove time intervals affected
by low energy flares. An example of a two-step cleaned light curve is
displayed in Fig.~\ref{fig:pipeprod}.

Time lost due to flaring in our observations amounts typically to
$\sim20\%$ of the full effective observing time. Six observations of
the initial fields from AO6 were too heavily affected by the flaring
even after the two step cleaning. Three of these fields have been
replaced by observations in AO7 (F01b, F02b, F35b); the partially lost
field F04 was reobserved as well.

Detection and analysis of faint diffuse sources like clusters of
galaxies in shallow surveys can be additionally affected by low energy
soft protons with a roughly constant flux. This so-called quiescent
soft proton background can not be detected based on light curve
screening due to its small temporal variations, especially not in
observations with a short duration. In order to characterize possible
contamination from this part of the non-X-ray background, we applied
the diagnostics developed by \citet{deluca04}, based on flux ratios
inside and outside the field of view of each detector.
The vast majority of fields is not contaminated by the quiescent soft
proton background at all in any of the detectors. Four fields (F04,
F06, F16, F25) have a slight\footnote{Slight contamination means
  $<15\%$ increase in the background with respect to normal levels
  \citep[for details see][]{deluca04}. This enhancement is modeled in
  first approximation by our composite background model (see text
  further and Sect.~\ref{sec:detect}).}  contamination with negligible
effect on data analysis and derived results. Fields F07, F13 have
significant time losses due to flaring periods (particularly in PN)
and in addition are now found to have strong residual quiescent
contamination ($>30\%$). There is no cluster in the present sample
found in these fields. The PN camera in field F32 is also
significantly affected ($\sim39\%$ background enhancement). We
identified two clusters (ID~476 and 139) in this observation. Since
the results from the PN and MOS cameras for these sources are in
agreement within the error bars we conclude that our background model
was able to account for the enhancement. For more details on these
sources see Sect.~\ref{sec:objectnotes}.

The double component background model (see Sect.~\ref{sec:detect})
used for the source detection and characterization can in principle
account to first order for such an enhanced background by increasing
the normalization of the background model. The vignetting function of
such particle background has a different shape than the vignetting of
the X-ray photons, but it is known only tentatively. We expect the
errors from such first order approximation to be small compared to
other sources of uncertainty (including the shot noise).  We
thus decide to include into our analysis also fields with a strong
residual quiescent contamination, but parameters derived for sources
in these fields should be handled with caution.

We treat out-of-time-events (OOTE) for the PN detector in a standard
way. For each observation, we generate an OOTE eventlist with the
\texttt{epchain} and remove time periods identified in the two step
cleaning process of the main PN eventlist. Whenever an image is
extracted from the PN eventlist, we extract also an image with the
same selection criteria from the OOTE eventlist, scale this image with
a factor of 0.063 (full frame readout mode) and subtract it from the
PN image.

\section{Source detection}
\label{sec:detectpipe}
As the main source detection algorithm we utilize the sliding box
technique and a maximum likelihood source fitting in their improved
implementation in the SAS tasks \texttt{eboxdetect} and
\texttt{emldetect}.  A detailed description of the work flow and
configuration of our detection pipeline, originally developed for the
XMM-\emph{Newton} Distant Cluster Project (XDCP), can be found in
\citet{fassbender-thesis} and \citet[][submitted]{fassbender11c}; here
we only summarize the main steps.

Source detection is carried out in three different schemes:
\newline
\textbf{(i)} the standard three band scheme: provides continuous,
non-overlapping coverage in three energy bands: $0.3 - 0.5$~keV, $0.5
- 2.0$~keV and $2.0 - 4.5$~keV.
\newline
\textbf{(ii)} the optimized single band scheme: covers the $0.35 -
2.4$~keV band and was chosen to maximize the signal-to-noise-ratio for
clusters of galaxies with a large range of redshifts and temperatures
\citep[see also][]{scharf02}. This bandpass is expected to maximize
the signal-to-noise-ratio especially for high redshift systems
($z\gtrsim 1$)
\newline
\textbf{(iii)} the five band spectral matched scheme: uses five
partially overlapping bands ($0.3-0.5$, $0.35-2.4$, $0.5-2.0$,
$2.0-4.5$ and $0.5-7.5$ keV). This scheme is equivalent to a single
band detection in the full $0.3-7.5$ keV range, where the energy
intervals in the overlaps have higher weighting. The shape of the
weighting function roughly mimics the expected continuum spectrum
shape of a hot cluster \citep{fassbender-thesis}. This setup was used
only to confirm detections from the first two schemes and we do not
use any results derived from it in the current work.

\subsection{Source list generation}
\label{sec:detect}
In order to obtain the raw source lists, we extract images from the
cleaned eventlist for each detector and each band required in the
given detection scheme (e.g. in the three band scheme three images for
each detector, in total nine images per field). We run the sliding box
detection algorithm (\texttt{eboxdetect} in the so-called \emph{local}
mode) on these images. The background for each potential source is
estimated only locally in a detection cell of $5 \times 5$ pixels in 4
successive runs with the number of pixels per cell doubled in each
iteration.  Sources detected by this procedure are then excised,
creating an image usable for proper background estimation.

We model the background of each detector and band individually with a
\emph{double component background model}. This background model is a
linear combination of two templates based on vignetted and unvignetted
exposure maps, taking into account the sky X-ray background (vignetted
component) and the particle and instrumental background (unvignetted
in the first approximation).

The final sliding box detection is then run utilizing the fitted
background model instead of a locally estimated background. For all
sources above the detection threshold we carry out a maximum
likelihood fitting (with the \texttt{emldetect} task). A beta profile
\citep{cavaliere76} with a fixed beta value of $\beta=2/3$ convolved
with the two dimensional point-spread function (PSF) is fitted to each
source. The fit is carried out for all three detectors and all the
bands in the given detection scheme \emph{simultaneously}.  The free
parameters of the fit are the source position, normalization of the
model (for each detector and band) and the core radius,
$\theta_{\mathrm{c}}$, characterizing the source extent. If the extent
of the source is not statistically significant, the source is refitted
as a point source with extent fixed to zero.

The \emph{detection} likelihood of a source is given by the
\texttt{det\_ml} parameter in the \texttt{eboxdetect} and
\texttt{emldetect} tasks, defined as $\mathrm{det\_ml} = -\ln
P_{\mathrm{rand}}$, where $P_{\mathrm{rand}}$ is the probability of
observed counts arising from pure random Poissonian fluctuations. In
each step of the detection process, the minimum detection likelihood
is set to 6, roughly equivalent to a $\gtrsim 3 \sigma$ detection in
terms of signal-to-noise ratio.

The \emph{extent} likelihood \texttt{ext\_ml}, defined analogously to
characterize the probability of the source being extended, is required
to be $\geq 3$ in the three-band scheme and $\geq 5$ in the single
band scheme (corresponding approximately to minimum extent
significances of $\sim2 \sigma$ and $\sim3 \sigma$, respectively).

For a more detailed discussion and justification of the chosen
detection schemes and thresh\-olds we refer to
\citet{fassbender-thesis}, who also demonstrates the performance of
the described source detection methods on over 450 archival
XMM-\emph{Newton} observations in the framework of the XDCP project.
A description of the used SAS tasks can be found in the SAS 7.1.0
reference manual.\footnote{\texttt{xmm.esac.esa.int/sas/7.1.0/}}

In the current work, we aim for the best possible survey completeness
including the high redshift end of the cluster distribution and
reliable source classification especially close to the detection
thresholds.  This is also helped by combining different detection
schemes and setting relatively low extent thresholds. The increasing
source contamination close to the detection threshold is treated with
careful screening using the optical data and ancillary X-ray
information (e.g. quality flags described in
Appendix~\ref{sec:flags}).

The detected sources create a raw master list of extended source
\emph{candidates}.  Each of these candidates is then screened visually
with optical imaging data (4 band BCS imaging) and accepted to the
final cluster catalog only if a significant overdensity of galaxies in
the photometric redshift space is found (Sect.~\ref{sec:photoz}). The
available \emph{Spitzer} imaging for the whole field will be used in
the future to confirm $z>$1 systems, where the depth of the BCS
imaging is not sufficient anymore.

The purely X-ray based selection function will be developed in
subsequent work based on simulations, where completeness and
contamination of different detection schemes will be studied. Guided
by extensive simulations of X-ray observations \citep{muehlegger10},
we will get a high precision description of the survey selection
function. A statistically well defined cluster sample will be drawn
from the current catalog (plus its 8~deg$^2$ extension) and used to
study the evolution of the cluster X-ray luminosity function and
perform cosmological tests.

\subsubsection{Treatment of MOS CCDs in anomalous state}
\label{sec:hc}
\begin{figure*}[ht!]
\includegraphics[width=0.58\textwidth]{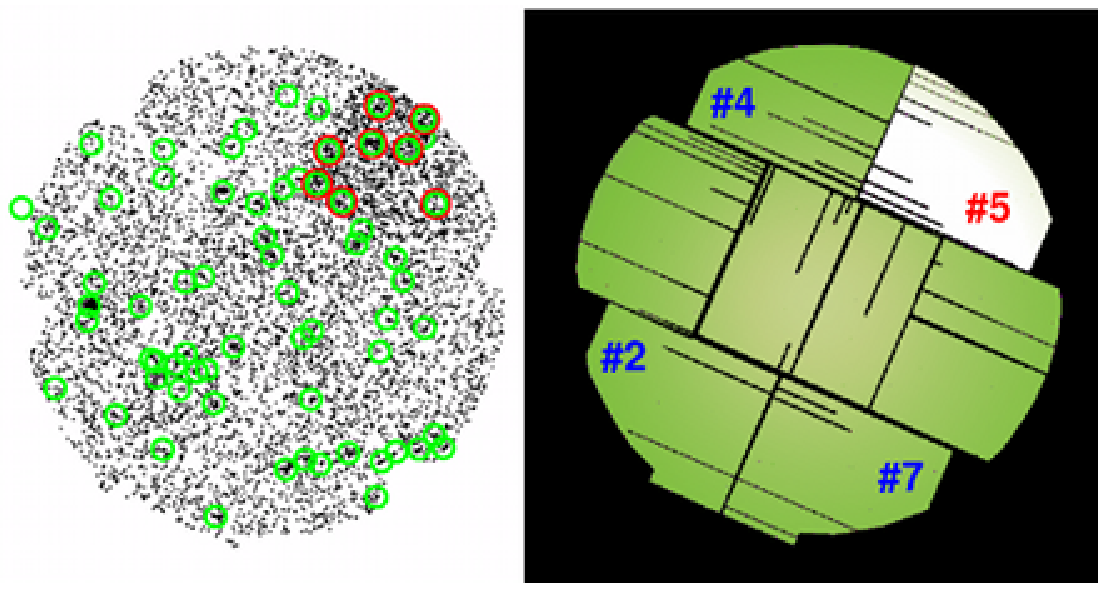}
\includegraphics[width=0.40\textwidth]{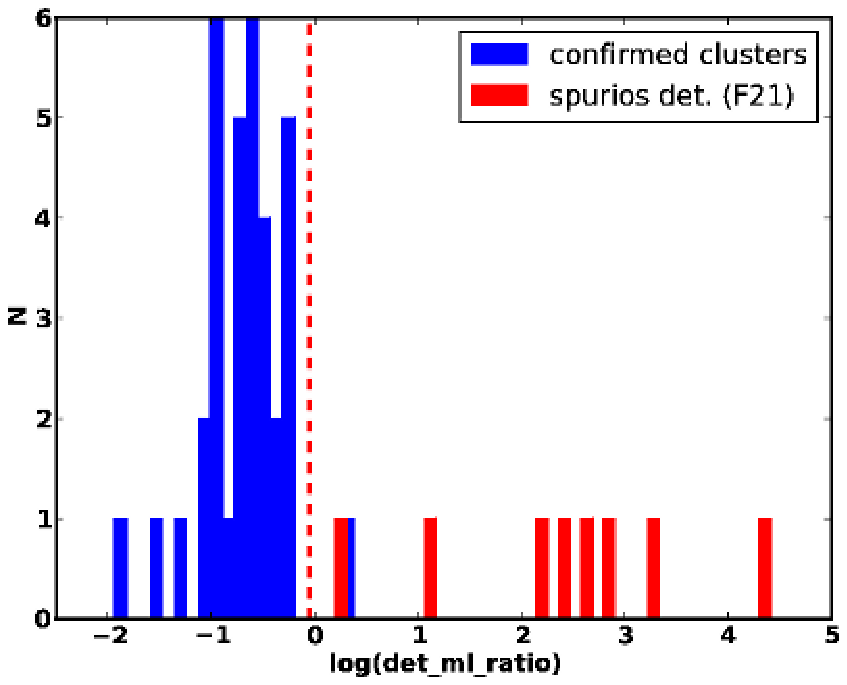}
\caption[Hot chip treatment]{\textbf{Left:} Image of field F21 taken
  by the MOS2 camera in the $0.5-2.0$~keV band.  The MOS2 CCD\#5 is
  visibly in an anomalously high ("hot") state with an enhanced
  background. Sources detected in this field are marked by green
  circles. Sources with red circles were automatically flagged as
  possibly spurious detections caused by the presence of the hot chip.
  \textbf{Middle:} A composite background model for the same detector
  and band created by fitting the double component model independently
  to the CCD\#5 and the rest of the chips.  The three blue-marked
  chips are the reference chips used to identify hot chips in the
  observations.  \textbf{Right:} The ratio of the total detection
  likelihood (log scale) from the MOS2 CCD\#5 in the $0.3-0.5$ and
  $0.5-2.0$ keV bands to the total detection likelihoods from all
  other detectors and bands (log scale). Blue bars show the confirmed
  clusters from our sample, the red bars the 8 flagged sources from
  field F21 (from the left panel). The vertical line marks where the
  soft band MOS2 detection constitutes 90\% of the total detection
  likelihoods in all detectors. The flagged sources were confirmed as
  spurious by the optical data.  A single confirmed cluster (ID 275)
  appears above the threshold, but is \emph{not} flagged as spurious
  since it would have been above the detection likelihood even without
  the MOS2 detection (i.e. not meeting all the required criteria
  described in Sect.~\ref{sec:hc}).
}
\label{fig:hcfig}
\end{figure*}
A special note is required concerning the anomalous states of CCD\#4
of the MOS1 and CCD\#5 of the MOS2 detectors and their effect on
extended source detections. Half of our fields have the MOS2 CCD\#5 in
the anomalous state and $\sim20\%$ have an anomalous MOS1 CCD\#4 (some
observations are affected by both). These anomalous ("hot") states are
characterized by high overall background count rates with atypical
hardness ratios. The most affected are the soft bands \citep[in our
 case $0.3-0.5$~keV and $0.5-2$~keV bands, see
 also][]{kuntzsnowden08}.

We check for the presence of a hot chip in an observation by comparing
count rates extracted from the suspected chip and the mean count rate
of three other chips in symmetrical positions around the central chip
(i.e. the mean count rate of CCD\#2, CCD\#4, CCD\#7 of MOS2 and
CCD\#3, CCD\#5, CCD\#7 of MOS1).  These reference chips were selected,
because they best match the area, shape and position of the affected
chips (see middle panel of Fig.~\ref{fig:hcfig}).  The count rates
calculated in the $0.3-2.0$~keV band from the three reference chips
are then averaged to reduce shot noise and a chip is flagged hot, if
its count rate is more than $10\%$ higher than the mean count rate
from the reference chips. This threshold is chosen to be very
conservative and was found to perform excellently, since chips in
anomalous states have typically count rates $50 - 100\%$ higher than
the reference rate. The algorithm also automatically flags
observations where a bright source lies on a reference chip. In these
cases we manually excise the source and repeat the calculation to
obtain an unaffected count rate ratio.

The exceptionally high background of the hot chips leads to many
spurious extended source detections, when left untreated (see
Fig.~\ref{fig:hcfig}).  We flag sources as possibly spurious
detections caused by the presence of a hot chip if at the same time:
1) they lie on a chip that was flagged hot, 2) are extended, 3) the
detection likelihood from the given hot MOS detector in the soft bands
(sum of the $0.3-0.5$ and $0.5-2.0$~keV bands) accounts for more than
$90\%$ of the total detection likelihood and 4) the source would be
under our detection threshold without the detection on the affected
chip.  We still visually checked every flagged source also in the
optical images and confirmed the classification of these sources as
spurious.

An example of this procedure can be seen in Fig.~\ref{fig:hcfig}. The
observation of field F21 has a hot MOS2 CCD\#5, clearly visible as an
enhanced background (in the raw image in the left panel and in the
model background in the middle).  The 8 extended sources detected on
this chip were flagged as potentially spurious based on the described
criteria. The detection likelihood ratio (the MOS 2 detection
likelihood in the soft bands over the total detection likelihood) of
these 8 sources are displayed on the left panel of
Fig.~\ref{fig:hcfig} (red) as compared to the sample of confirmed
clusters in our sample (blue).

A similar criterion can be applied in principle also to spurious point
source detections.  An additional improvement can be achieved by
weighting the input detection likelihoods by the number of pixels in
the detection aperture in order to avoid a possible bias, if a source
has a low detection likelihood in one of the reference detectors only
because it falls on a chip gap or is (partially) out of the
field-of-view.

We also make an attempt to model the high background of the hot chips
by fitting the double component model to a hot chip (in first
approximation) and another double component model to the remaining
chips. The two parts of the background model are then combined to
create a composite background map for the full detector area (middle
panel of Fig.~\ref{fig:hcfig}). All the extended sources on hot chips
flagged as spurious with the described detection likelihood test, are
not detected when the composite background maps are utilized,
confirming the reliability of our classification. The effect of using
a composite background instead of a standard background on detections
coming from the remaining, non-anomalous chips is minor, since the two
background models in these areas differ typically by less than $5\%$,
and only the softest bands of each detection scheme are affected. For
the source characterization in observations affected by hot chips we
use exclusively composite background maps.

\subsection{Growth curve analysis}
\label{sec:gca}

\begin{figure}[t!]
\begin{center}
\includegraphics[width=0.5\textwidth]{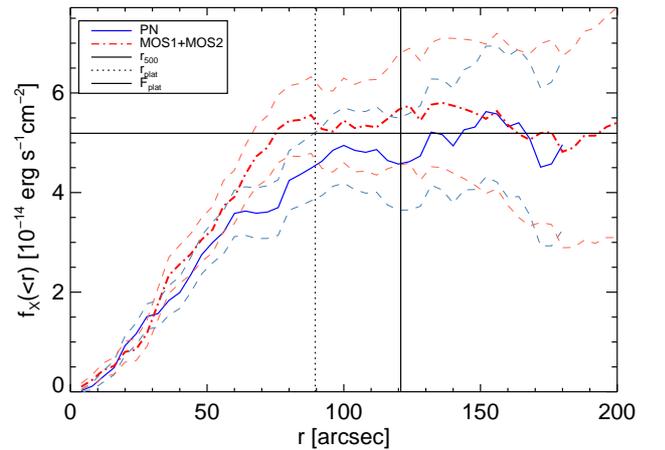}
\end{center}
\caption[Growth curve of cluster ID 018]{Example of the growth curve
  analysis of source ID 018 (photo-z=0.39). The cluster's redshift and
  luminosity are close to the median values of the entire sample. The
  curves show the encircled cumulative flux as a function of radius
  (PN: blue curve, combined MOS: red, dot-dashed).  The PN and MOS
  curves are in good agreement. Dashed lines mark the flux measurement
  error bars which include the Poisson noise and an additional 5\%
  systematic error from the background estimation. The estimated
  plateau flux is F$_{plat} = 5.19 \times 10^{-14}$~erg s$^{-1}$
  cm$^{-2}$ (horizontal line), reached at
  $r_{\mathrm{plat}}\sim90$~arcsec. The vertical line signifies the
  estimated $r_{500}$ radius of the source, $r_{500} = 0.6$~Mpc ($\sim
  117$~arcsec). In this case, the plateau radius is slightly smaller
  than $r_{500}$ and the flux and luminosity for $r_{500}$ had to be
  extrapolated from their plateau values. The required extrapolation
  is only $\sim2\%$ in this case. See Sect.~\ref{sec:physpar} for
  details.}
\label{fig:gca}
\end{figure}

The X-ray count rate is the most direct cluster observable. With an
estimate of the energy conversion factor at hand (see also
Sect.~\ref{sec:physpar}) we can calculate from it the X-ray flux
$F_{\mathrm{X}}$, which in turn can be converted to X-ray luminosity
$L_{\mathrm{X}}$. The luminosity is a key parameters since it allows
us to calculate other important physical parameters (particularly the
mass of the system) from scaling relations.

A typical cluster of galaxies in relatively shallow observations like
ours appears as a faint diffuse source with typically $\gtrsim100$
source photons registered (total from all three standard bands and
detectors).  Thus in order to get a reliable measurement of the flux
and trace the emission of the cluster as far out as possible, we have
to employ a robust method. In this work we utilize the \emph{growth
  curve method} developed for the REFLEX and NORAS cluster samples
derived from the \emph{ROSAT} all sky survey by
\citet[][]{boehringer00}.  Here we summarize the procedure.

For each source, we extract images, exposure maps and background maps
in the $0.5 - 2.0$~keV band, excluding all point sources detected by
the pipeline. MOS1 and MOS2 products are then directly co-added, since
the difference in their response matrices is small. We run the growth
curve program on the PN and co-added MOS images independently.

In this analysis we use the X-ray center coordinates obtained from the
beta model fitting procedure in the source detection step. We also
explored the possibility of recentering by minimizing the dipole
moment of the count distribution \citep[see
  e.g.][]{boehringer10b}. This procedure usually yielded centers very
close to the best-fit beta model coordinates, but for faint sources
often completely diverged.  The best-fit coordinates were always found
to be a good description of the detected X-ray emission centroid.

Counts are extracted from the image in concentric rings starting from
the center and scaled by the exposure time. In this way we obtain the
total (source + background) count rate profile.  The expected
background count rate is estimated from the model background map and
subtracted for each ring from the total count rate, obtaining the
source count rate profile. The \emph{growth curve} is the cumulative
background subtracted source count rate profile. An example of
  a growth curve is displayed in Fig.~\ref{fig:gca} (shown here in
  flux units using the energy conversion factor calculated as
  described in Sect.~\ref{sec:physpar}).

We term the radius of the full aperture inside which a stable growth
curve can be obtained, the \emph{extraction radius} $r_{\mathrm{ext}}$
(typically $150-200$~arcsec).  It is adjusted for each source
individually (increased for brightest, most extended sources or
trimmed for sources close to the edge of FOV or to a partially blended
systems) and includes the source itself as well as enough sky region
to check the reliability of the double component background
subtraction.

If the background model describes the local background accurately, the
growth curve levels off to a flat plateau at the outer edge of the
source. To estimate the total detected cluster emission, we first
calculate the \emph{significance radius} $r_{\mathrm{sig}}$, defined
as the radius outside which the source signal increases less than the
$1\sigma$ uncertainty in the count rate. The significance radius thus
gives the outermost radius where the potential increase of the growth
curve becomes less than $1\sigma$ significant.

To alleviate the effect of shot noise, $r_{\mathrm{sig}}$ is
determined by smoothing the growth curve in 20 and 48 arcsec windows
(5 and 12 pixels respectively). For most clusters the two estimates
are in agreement. In the remaining cases, the local background usually
exhibits irregular features not captured by the double component model
and we select the more appropriate $r_{\mathrm{sig}}$ and plateau
after visual inspection.

In addition, a single multiplicative correction factor to the
background model can be set, if the plateau exhibits a significant
residual slope. This additional factor corrects the overall
normalization of the double component model locally inside
$r_{\mathrm{ext}}$. The average background correction factors are
$-2\%$ (i.e. a $2\%$ decrease compared to the default double component
background) for PN and $0\%$ for MOS (with standard deviations $7\%$
and $8\%$, respectively).  More than $3/4$ of the present sample have
correction factors smaller than $10\%$.  \citet{reiprich01} and
\citet{reiprich02} used a similar correction procedure utilizing a
second order polynomial to obtain stable plateaus.  In our case, a
simple correction factor turned out to be sufficient and not leading
to background over-fitting.

After setting the background correction, the total source count rate
is estimated as the count rate of the plateau. The flat plateau of the
growth curve outside $r_{\mathrm{sig}}$ is then fitted with a line. If
the slope of the line is less than $0.8\%$ per radial bin, the plateau
fit is accepted and the plateau count rate $CTR_{plat}$ is estimated
as the mean of the fitted line. If the slope is still not negligible,
an additional attempt is made to find a stable plateau by iteratively
removing the outermost and innermost (still outside
$r_{\mathrm{sig}}$) bins.  We note that in $\sim80\%$ of cases the
first simple fit is fully acceptable and no further iterations are
necessary. For more detailed description of the iterative process and
quality flags of the plateau fit see Sect.~\ref{sec:flags} in the
appendix. The \emph{plateau radius} $r_{\mathrm{plat}}$, is defined
simply as the radial distance where the growth curve first reaches
$CTR_{plat}$.

We provide a performance test of our X-ray photometry method on the
example of the XMM-LSS cluster catalog \citep{pacaud07} in
Appendix~\ref{sec:lsscomp_photometry}.  The main advantages of the
growth curve method thus are: \textbf{(i)} Excellent sensitivity
allowing us to trace cluster emission to the outermost faint
outskirts. \textbf{(ii)} It makes no assumptions about the source
profile unlike methods based on beta model fitting, which is fully
degenerate in the regime with $<~400 - 500$ counts and is known not to
be an appropriate description of cluster emission for irregular and
cool core clusters. \textbf{(iii)} The method allows to check and
correct the background modelling which is done for the whole field of
view, by adjusting several parameters to the conditions local to each
analyzed source.  \textbf{(iv)} The PN and combined MOS growth curves
are treated completely independently. Their comparison provides us
with an important consistency check and allows us to treat instrument
specific features in the background separately.

\subsection{Physical parameter estimation}
\label{sec:physpar}

With a stable PN and MOS growth curve at hand we determine all the
relevant physical parameters of the clusters (see
Table~\ref{tab:physpar}) in an iterative way:

\textbf{1)} The physical parameters are set to their initial values
$(r_{500}, T_{500}) = (0.5$~Mpc$, 2.5$~keV).

\textbf{2)} The physical aperture radius is converted to arcseconds
using the assumed cosmological parameters. The total count rate inside
this radius is estimated from the PN and MOS growth curves in the
$0.5-2$~keV band.

\textbf{3)} The count rates are converted to flux with an energy
conversion factor (ECF) calculated assuming a MEKAL spectral model
\citep{mewe85,kaastra92,liedahl95} with abundance set to 0.3 times
solar abundance, temperature equal to the trial $T_{500}$ value and
redshift set to the photo-z value (or spectroscopic redshift where
available).

To account for the spatial variation of the spectral response of the
detectors we calculate a response matrix for each source individually
in a 150 arcsec aperture centered on the source for the THIN
filter. The MOS2 response matrix is used to calculate the ECF of the
co-added MOS count rates.

This procedure is used to estimate the $0.5-2$~keV and bolometric
fluxes for PN and MOS and the same model is also used to convert the
fluxes to luminosities.

\textbf{4)} In some cases the estimated value of $r_{500}$ in the
given iteration is larger than $r_{\mathrm{plat}}$ (i.e. larger than
the region with directly measurable emission), and therefore an
extrapolation factor has to be applied to the flux and luminosity
estimates. We correct for the missing flux between the plateau radius
and current iteration estimate of $r_{500}$ by extrapolating the
source emission assuming a beta model.  The $\beta$ and
$r_{\mathrm{core}}$ parameters of the beta model are calculated using
the scaling relations of \citet{reiprich01} \citep[see
  also][]{reiprich02, finoguenov07}:

\begin{equation}
\label{eq:beta}
r_{core} = 0.07\times r_{500}\left( \frac{T}{1\,\mathrm{keV}}\right)^{0.63}\,\,\,\mathrm{and}\,\,\,\,\,
 \beta=0.4\left( \frac{T}{1\,\mathrm{keV}}\right)^{1/3}.
\end{equation}

\clearpage
\begin{landscape}
\begin{table}[ht!]
\rowcolors{1}{tableShade}{white}
 \centering
\caption[Physical parameters of the clusters sample]{\small{Physical
    parameters of the clusters sample. Fluxes and luminosities are
    given for the $0.5-2$~keV band. See Appendix~\ref{sec:objectnotes}
    for notes on individual sources.  Notes: $^{\dag}$~the cluster was
    detected in observations strongly affected by flaring;
    $^{\diamond}$~the cluster is heavily affected by blending with a
    nearby source.  Notes for redshifts: $^a$ spectroscopic redshift;
    $^b$ a high redshift system for which a secure photometric
    redshift estimate is not possible from the current photometric
    catalog (the listed parameters are tentative estimates).}}
\vspace{0.1cm}
 \label{tab:physpar}
   \begin{scriptsize}
\begin{tabular}{lcccccccccccc}
\hline
\hline
   \multicolumn{1}{c}{ID} &
   \multicolumn{1}{c}{R.A. (J2000)} &
   \multicolumn{1}{c}{Dec (J2000)} &
   \multicolumn{1}{c}{z} &
   \multicolumn{1}{c}{$r_{\mathrm{plat}}$} &
   \multicolumn{1}{c}{$F_{plat}$} &
   \multicolumn{1}{c}{$r_{500}$} &
   \multicolumn{1}{c}{$F_{500}$} &
   \multicolumn{1}{c}{$L_{500}$} &
   \multicolumn{1}{c}{$T_{500}$} &
   \multicolumn{1}{c}{$M_{500}$} &
   \multicolumn{1}{c}{$Y_{500}$} &
   \multicolumn{1}{c}{M$_{200}$}  \\
&(deg) &(deg) & (photo.) &(arcmin/$r_{500}^{-1}$) &($10^{-14}$~erg s$^{-1}$ cm$^{-2}$)
 & (kpc) & ($10^{-14}$~erg s$^{-1}$cm$^{-2}$)  & ($10^{43}$~erg s$^{-1})$
 & (keV) & $(10^{13}$ M$_{\odot}$) & ($10^{13}$ M$_{\odot}$~keV) & $(10^{13}$ M$_{\odot}$) \\
\hline
     011$^{\dag}$ &$ 351.8070 $&$ -56.0615 $&$ 0.97 \pm 0.10 $&$ 0.7/0.6 $&$ 2.80 \pm 0.42 $&$ 577 \pm 54 $&$ 2.85 \pm 0.43 $&$ 11.8 \pm 1.8 $&$ 3.4 \pm 1.0 $&$ 16.4 \pm 4.6 $&$ 4.8 \pm 2.9 $&$ 24.3 \pm 6.8 $ \\
 018 &$ 352.4828 $&$ -56.1360 $&$ 0.39 \pm 0.04 $&$ 1.5/0.7 $&$ 5.14 \pm 0.50 $&$ 633 \pm 58 $&$ 5.21 \pm 0.51 $&$ 2.6 \pm 0.3 $&$ 2.3 \pm 0.7 $&$ 10.9 \pm 3.0 $&$ 1.8 \pm 1.1 $&$ 15.2 \pm 4.2 $ \\
 032 &$ 352.1778 $&$ -55.5662 $&$ 0.83 \pm 0.07 $&$ 1.6/1.0 $&$ 8.06 \pm 0.68 $&$ 702 \pm 64 $&$ 8.00 \pm 0.66 $&$ 21.5 \pm 1.8 $&$ 4.2 \pm 1.2 $&$ 25.0 \pm 6.9 $&$ 10.8 \pm 6.4 $&$ 37.0 \pm 10.2 $ \\
 033 &$ 352.0448 $&$ -55.8400 $&$ 0.79 \pm 0.05 $&$ 1.1/0.9 $&$ 2.94 \pm 0.25 $&$ 593 \pm 54 $&$ 2.94 \pm 0.25 $&$ 7.6 \pm 0.6 $&$ 3.0 \pm 0.9 $&$ 14.4 \pm 4.0 $&$ 3.5 \pm 2.1 $&$ 21.0 \pm 5.8 $ \\
 034 &$ 352.6538 $&$ -55.7270 $&$ 0.28 \pm 0.02 $&$ 1.7/0.8 $&$ 2.84 \pm 0.45 $&$ 536 \pm 50 $&$ 2.89 \pm 0.46 $&$ 0.7 \pm 0.1 $&$ 1.5 \pm 0.4 $&$ 5.8 \pm 1.6 $&$ 0.5 \pm 0.3 $&$ 7.9 \pm 2.2 $ \\
 035 &$ 353.4388 $&$ -55.6387 $&$ 0.67 \pm 0.05 $&$ 1.2/0.9 $&$ 2.02 \pm 0.32 $&$ 560 \pm 53 $&$ 1.95 \pm 0.31 $&$ 3.6 \pm 0.6 $&$ 2.4 \pm 0.7 $&$ 10.5 \pm 3.0 $&$ 1.8 \pm 1.1 $&$ 15.0 \pm 4.2 $ \\
 038$^{\diamond}$ &$ 353.5130 $&$ -55.8156 $&$ 0.39 \pm 0.05 $&$ 1.1/0.7 $&$ 1.30 \pm 0.20 $&$ 503 \pm 47 $&$ 1.33 \pm 0.20 $&$ 0.7 \pm 0.1 $&$ 1.5 \pm 0.4 $&$ 5.4 \pm 1.5 $&$ 0.4 \pm 0.3 $&$ 7.5 \pm 2.1 $ \\
 039 &$ 349.8214 $&$ -55.3244 $&$ 0.18 \pm 0.04 $&$ 2.3/0.7 $&$ 9.12 \pm 0.56 $&$ 586 \pm 53 $&$ 9.33 \pm 0.57 $&$ 0.8 \pm 0.1 $&$ 1.6 \pm 0.5 $&$ 6.8 \pm 1.9 $&$ 0.7 \pm 0.4 $&$ 9.2 \pm 2.5 $ \\
 044 &$ 349.2212 $&$ -54.9036 $&$ 0.44 \pm 0.02 $&$ 2.4/1.1 $&$ 17.14 \pm 1.15 $&$ 787 \pm 72 $&$ 16.75 \pm 1.10 $&$ 10.5 \pm 0.7 $&$ 3.6 \pm 1.0 $&$ 22.2 \pm 6.1 $&$ 7.5 \pm 4.4 $&$ 31.6 \pm 8.6 $ \\
 069 &$ 350.9631 $&$ -54.8923 $&$ 0.75 \pm 0.07 $&$ 1.7/1.3 $&$ 2.22 \pm 0.43 $&$ 560 \pm 53 $&$ 2.03 \pm 0.35 $&$ 4.8 \pm 0.8 $&$ 2.6 \pm 0.7 $&$ 11.6 \pm 3.3 $&$ 2.2 \pm 1.4 $&$ 16.7 \pm 4.7 $ \\
 070 &$ 350.6286 $&$ -54.2691 $&$ 0.152^{a}     $&$ 2.9/0.6 $&$ 80.78 \pm 1.64 $&$ 815 \pm 74 $&$ 83.43 \pm 1.70 $&$ 5.0 \pm 0.1 $&$ 2.9 \pm 0.8 $&$ 17.9 \pm 4.9 $&$ 4.5 \pm 2.6 $&$ 24.6 \pm 6.7 $ \\
 081 &$ 351.8470 $&$ -55.2624 $&$ 0.85 \pm 0.12 $&$ 0.4/0.4 $&$ 1.01 \pm 0.17 $&$ 497 \pm 47 $&$ 1.05 \pm 0.18 $&$ 3.5 \pm 0.6 $&$ 2.3 \pm 0.7 $&$ 9.1 \pm 2.6 $&$ 1.4 \pm 0.9 $&$ 13.2 \pm 3.7 $ \\
 082 &$ 351.5779 $&$ -55.3859 $&$ 0.63 \pm 0.05 $&$ 0.9/0.7 $&$ 1.32 \pm 0.23 $&$ 526 \pm 50 $&$ 1.34 \pm 0.23 $&$ 2.2 \pm 0.4 $&$ 2.0 \pm 0.6 $&$ 8.3 \pm 2.4 $&$ 1.1 \pm 0.7 $&$ 11.8 \pm 3.3 $ \\
 088 &$ 352.1748 $&$ -55.2234 $&$ 0.43 \pm 0.04 $&$ 2.5/1.3 $&$ 8.50 \pm 1.27 $&$ 679 \pm 63 $&$ 7.43 \pm 1.01 $&$ 4.6 \pm 0.6 $&$ 2.7 \pm 0.8 $&$ 14.1 \pm 3.9 $&$ 3.0 \pm 1.8 $&$ 19.8 \pm 5.5 $ \\
 090 &$ 352.2366 $&$ -55.4081 $&$ 0.58 \pm 0.02 $&$ 0.8/0.7 $&$ 0.76 \pm 0.18 $&$ 481 \pm 47 $&$ 0.78 \pm 0.18 $&$ 1.1 \pm 0.2 $&$ 1.6 \pm 0.5 $&$ 6.0 \pm 1.8 $&$ 0.6 \pm 0.4 $&$ 8.3 \pm 2.4 $ \\
 094 &$ 353.0185 $&$ -55.2120 $&$ 0.269^{a}     $&$ 1.3/0.6 $&$ 4.83 \pm 0.53 $&$ 583 \pm 54 $&$ 5.02 \pm 0.55 $&$ 1.1 \pm 0.1 $&$ 1.7 \pm 0.5 $&$ 7.4 \pm 2.1 $&$ 0.8 \pm 0.5 $&$ 10.1 \pm 2.8 $ \\
 109 &$ 351.9058 $&$ -54.2705 $&$ 1.02^{b}      $&$ 0.7/0.7 $&$ 1.45 \pm 0.21 $&$ 510 \pm 48 $&$ 1.47 \pm 0.21 $&$ 7.3 \pm 1.1 $&$ 2.8 \pm 0.8 $&$ 12.0 \pm 3.4 $&$ 2.6 \pm 1.6 $&$ 17.7 \pm 5.0 $ \\
 110 &$ 352.5161 $&$ -54.2388 $&$ 0.47 \pm 0.06 $&$ 1.4/0.8 $&$ 3.77 \pm 0.49 $&$ 605 \pm 56 $&$ 3.47 \pm 0.47 $&$ 2.7 \pm 0.4 $&$ 2.3 \pm 0.7 $&$ 10.4 \pm 2.9 $&$ 1.7 \pm 1.0 $&$ 14.6 \pm 4.1 $ \\
 126 &$ 351.6393 $&$ -55.0206 $&$ 0.42 \pm 0.02 $&$ 1.9/1.0 $&$ 5.48 \pm 0.41 $&$ 643 \pm 59 $&$ 5.38 \pm 0.40 $&$ 3.2 \pm 0.2 $&$ 2.4 \pm 0.7 $&$ 11.8 \pm 3.2 $&$ 2.1 \pm 1.3 $&$ 16.5 \pm 4.5 $ \\
 127 &$ 351.8492 $&$ -55.0648 $&$ 0.207^{a}     $&$ 0.8/0.3 $&$ 2.58 \pm 0.29 $&$ 499 \pm 46 $&$ 2.91 \pm 0.33 $&$ 0.4 \pm 0.1 $&$ 1.2 \pm 0.4 $&$ 4.4 \pm 1.2 $&$ 0.3 \pm 0.2 $&$ 5.9 \pm 1.6 $ \\
 132 &$ 352.0084 $&$ -54.9292 $&$ 0.96 \pm 0.17 $&$ 1.3/1.1 $&$ 2.91 \pm 0.34 $&$ 571 \pm 53 $&$ 2.74 \pm 0.31 $&$ 11.1 \pm 1.2 $&$ 3.3 \pm 0.9 $&$ 15.7 \pm 4.4 $&$ 4.4 \pm 2.7 $&$ 23.3 \pm 6.5 $ \\
 136 &$ 350.5036 $&$ -54.7500 $&$ 0.36 \pm 0.02 $&$ 2.4/1.1 $&$ 8.29 \pm 0.76 $&$ 673 \pm 62 $&$ 8.15 \pm 0.68 $&$ 3.4 \pm 0.3 $&$ 2.5 \pm 0.7 $&$ 12.6 \pm 3.5 $&$ 2.4 \pm 1.4 $&$ 17.6 \pm 4.8 $ \\
 139$^{\dag\diamond}$  &$ 351.3953 $&$ -54.7212 $&$ 0.169^{a} $&$ 1.9/0.6 $&$ 4.83 \pm 0.65 $&$ 517 \pm 48 $&$ 5.01 \pm 0.67 $&$ 0.4 \pm 0.1 $&$ 1.3 \pm 0.4 $&$ 4.6 \pm 1.3 $&$ 0.3 \pm 0.2 $&$ 6.2 \pm 1.7 $ \\
 150 &$ 352.5015 $&$ -54.6184 $&$ 0.176^{a}     $&$ 2.9/0.8 $&$ 18.25 \pm 0.89 $&$ 654 \pm 59 $&$ 18.48 \pm 0.91 $&$ 1.6 \pm 0.1 $&$ 2.0 \pm 0.6 $&$ 9.4 \pm 2.6 $&$ 1.3 \pm 0.7 $&$ 12.9 \pm 3.5 $ \\
 152 &$ 352.4168 $&$ -54.7886 $&$ 0.139^{a}     $&$ 0.9/0.3 $&$ 2.50 \pm 0.41 $&$ 448 \pm 42 $&$ 2.97 \pm 0.49 $&$ 0.2 \pm 0.1 $&$ 1.0 \pm 0.3 $&$ 2.9 \pm 0.8 $&$ 0.1 \pm 0.1 $&$ 3.9 \pm 1.1 $ \\
 156 &$ 353.8815 $&$ -54.5865 $&$ 0.67 \pm 0.06 $&$ 0.8/0.6 $&$ 3.35 \pm 0.24 $&$ 614 \pm 56 $&$ 3.43 \pm 0.24 $&$ 6.0 \pm 0.4 $&$ 2.8 \pm 0.8 $&$ 13.9 \pm 3.8 $&$ 3.2 \pm 1.9 $&$ 19.9 \pm 5.5 $ \\
 158 &$ 353.6032 $&$ -54.4586 $&$ 0.55 \pm 0.03 $&$ 1.4/0.9 $&$ 3.57 \pm 0.54 $&$ 617 \pm 58 $&$ 3.58 \pm 0.54 $&$ 4.0 \pm 0.6 $&$ 2.5 \pm 0.7 $&$ 12.1 \pm 3.4 $&$ 2.3 \pm 1.4 $&$ 17.2 \pm 4.8 $ \\
 210 &$ 353.5240 $&$ -55.7859 $&$ 0.83 \pm 0.09 $&$ 0.7/0.7 $&$ 0.55 \pm 0.11 $&$ 451 \pm 43 $&$ 0.56 \pm 0.11 $&$ 1.9 \pm 0.4 $&$ 1.9 \pm 0.5 $&$ 6.6 \pm 1.9 $&$ 0.8 \pm 0.5 $&$ 9.5 \pm 2.7 $ \\
 227 &$ 350.5425 $&$ -55.4199 $&$ 0.346^{a}     $&$ 1.1/0.6 $&$ 1.51 \pm 0.19 $&$ 506 \pm 47 $&$ 1.56 \pm 0.20 $&$ 0.6 \pm 0.1 $&$ 1.4 \pm 0.4 $&$ 5.3 \pm 1.5 $&$ 0.4 \pm 0.3 $&$ 7.2 \pm 2.0 $ \\
 245 &$ 351.0160 $&$ -55.0225 $&$ 0.62 \pm 0.03 $&$ 0.8/0.7 $&$ 0.97 \pm 0.18 $&$ 500 \pm 48 $&$ 0.99 \pm 0.18 $&$ 1.6 \pm 0.3 $&$ 1.8 \pm 0.5 $&$ 7.0 \pm 2.0 $&$ 0.8 \pm 0.5 $&$ 9.9 \pm 2.8 $ \\
 275 &$ 353.6991 $&$ -55.2736 $&$ 0.29 \pm 0.03 $&$ 1.7/0.8 $&$ 2.93 \pm 0.45 $&$ 541 \pm 51 $&$ 2.95 \pm 0.45 $&$ 0.8 \pm 0.1 $&$ 1.5 \pm 0.4 $&$ 6.0 \pm 1.7 $&$ 0.5 \pm 0.3 $&$ 8.2 \pm 2.3 $ \\
 287 &$ 354.2119 $&$ -55.2988 $&$ 0.57 \pm 0.04 $&$ 0.9/0.7 $&$ 0.98 \pm 0.35 $&$ 501 \pm 54 $&$ 1.00 \pm 0.36 $&$ 1.3 \pm 0.5 $&$ 1.7 \pm 0.5 $&$ 6.7 \pm 2.1 $&$ 0.7 \pm 0.5 $&$ 9.3 \pm 3.0 $ \\
 288 &$ 353.7523 $&$ -54.9164 $&$ 0.60 \pm 0.04 $&$ 1.8/1.2 $&$ 2.75 \pm 0.63 $&$ 580 \pm 56 $&$ 2.44 \pm 0.48 $&$ 3.4 \pm 0.7 $&$ 2.4 \pm 0.7 $&$ 10.7 \pm 3.1 $&$ 1.9 \pm 1.2 $&$ 15.3 \pm 4.4 $ \\
 357 &$ 353.6200 $&$ -54.6066 $&$ 0.48 \pm 0.06 $&$ 2.0/1.2 $&$ 3.56 \pm 0.47 $&$ 596 \pm 55 $&$ 3.16 \pm 0.40 $&$ 2.6 \pm 0.3 $&$ 2.2 \pm 0.6 $&$ 10.1 \pm 2.8 $&$ 1.6 \pm 0.9 $&$ 14.2 \pm 3.9 $ \\
 386 &$ 353.9763 $&$ -56.0928 $&$ 0.53 \pm 0.05 $&$ 0.7/0.6 $&$ 0.66 \pm 0.17 $&$ 468 \pm 47 $&$ 0.67 \pm 0.18 $&$ 0.8 \pm 0.2 $&$ 1.5 \pm 0.4 $&$ 5.2 \pm 1.6 $&$ 0.4 \pm 0.3 $&$ 7.2 \pm 2.2 $ \\
 430 &$ 351.3891 $&$ -55.7327 $&$ 0.206^{a}     $&$ 1.3/0.6 $&$ 1.59 \pm 0.33 $&$ 455 \pm 44 $&$ 1.69 \pm 0.36 $&$ 0.2 \pm 0.1 $&$ 1.0 \pm 0.3 $&$ 3.3 \pm 1.0 $&$ 0.2 \pm 0.1 $&$ 4.4 \pm 1.3 $ \\
 444 &$ 354.0839 $&$ -55.5189 $&$ 0.71 \pm 0.05 $&$ 0.9/0.7 $&$ 1.27 \pm 0.25 $&$ 521 \pm 50 $&$ 1.29 \pm 0.26 $&$ 2.8 \pm 0.6 $&$ 2.2 \pm 0.6 $&$ 8.8 \pm 2.5 $&$ 1.3 \pm 0.8 $&$ 12.6 \pm 3.6 $ \\
 457 &$ 352.1177 $&$ -54.2472 $&$ 0.1^{a}       $&$ 1.6/0.5 $&$ 1.98 \pm 0.52 $&$ 387 \pm 39 $&$ 2.29 \pm 0.60 $&$ 0.1 \pm 0.1 $&$ 0.7 \pm 0.2 $&$ 1.8 \pm 0.5 $&$ <0.1 $&$ 2.4 \pm 0.7 $ \\
 476$^{\dag\diamond}$  &$ 351.4166 $&$ -54.7412 $&$ 0.102^{a} $&$ 2.3/0.5 $&$ 10.27 \pm 1.18 $&$ 508 \pm 47 $&$ 11.04 \pm 1.27 $&$ 0.3 \pm 0.1 $&$ 1.2 \pm 0.3 $&$ 4.1 \pm 1.1 $&$ 0.2 \pm 0.1 $&$ 5.5 \pm 1.5 $ \\
 502 &$ 349.9334 $&$ -54.6400 $&$ 0.55 \pm 0.05 $&$ 0.7/0.5 $&$ 1.55 \pm 0.14 $&$ 540 \pm 50 $&$ 1.61 \pm 0.14 $&$ 1.9 \pm 0.2 $&$ 2.0 \pm 0.6 $&$ 8.1 \pm 2.2 $&$ 1.1 \pm 0.6 $&$ 11.4 \pm 3.2 $ \\
 511 &$ 353.0628 $&$ -54.7006 $&$ 0.269^{a}     $&$ 1.5/0.6 $&$ 4.41 \pm 0.71 $&$ 574 \pm 54 $&$ 4.53 \pm 0.73 $&$ 1.0 \pm 0.2 $&$ 1.7 \pm 0.5 $&$ 7.1 \pm 2.0 $&$ 0.7 \pm 0.4 $&$ 9.7 \pm 2.7 $ \\
 527 &$ 350.5456 $&$ -56.3127 $&$ 0.79 \pm 0.06 $&$ 1.0/0.8 $&$ 2.26 \pm 0.37 $&$ 568 \pm 54 $&$ 2.27 \pm 0.37 $&$ 6.0 \pm 1.0 $&$ 2.7 \pm 0.8 $&$ 12.6 \pm 3.6 $&$ 2.7 \pm 1.7 $&$ 18.3 \pm 5.2 $ \\
 528 &$ 353.8357 $&$ -55.5442 $&$ 0.35 \pm 0.02 $&$ 0.8/0.5 $&$ 0.63 \pm 0.21 $&$ 441 \pm 46 $&$ 0.68 \pm 0.22 $&$ 0.3 \pm 0.1 $&$ 1.1 \pm 0.3 $&$ 3.5 \pm 1.1 $&$ 0.2 \pm 0.1 $&$ 4.8 \pm 1.5 $ \\
 538 &$ 353.5258 $&$ -54.7310 $&$ 0.20 \pm 0.02 $&$ 1.6/0.7 $&$ 1.99 \pm 0.81 $&$ 469 \pm 52 $&$ 2.07 \pm 0.85 $&$ 0.2 \pm 0.1 $&$ 1.1 \pm 0.3 $&$ 3.6 \pm 1.2 $&$ 0.2 \pm 0.1 $&$ 4.8 \pm 1.6 $ \\
 543 &$ 353.1806 $&$ -54.8297 $&$ 0.57 \pm 0.03 $&$ 1.7/1.0 $&$ 4.15 \pm 1.13 $&$ 629 \pm 61 $&$ 4.07 \pm 0.90 $&$ 4.9 \pm 1.1 $&$ 2.7 \pm 0.8 $&$ 13.2 \pm 3.8 $&$ 2.8 \pm 1.7 $&$ 18.7 \pm 5.5 $ \\
 547 &$ 351.0815 $&$ -55.4305 $&$ 0.241^{a}     $&$ 1.1/0.5 $&$ 1.05 \pm 0.33 $&$ 443 \pm 46 $&$ 1.12 \pm 0.35 $&$ 0.2 \pm 0.1 $&$ 1.0 \pm 0.3 $&$ 3.1 \pm 1.0 $&$ 0.1 \pm 0.1 $&$ 4.2 \pm 1.3 $ \\

\hline
\end{tabular}
\end{scriptsize}

\end{table}
\end{landscape}

The good sensitivity of the growth curve method allows us to trace
cluster emission out to large radii, therefore the required
extrapolation is typically minor. The mean correction is
$\sim5\%/\sim6\%$ for PN/MOS ($\sim46\%$ at maximum).

In cases when $r_{500} \le r_{\mathrm{plat}}$ no extrapolation is
needed and the flux and luminosity estimates are independent of any
assumption about the spatial distribution of the source emission.

\textbf{5)} The source flux and luminosity are then obtained by
averaging the PN and MOS fluxes (luminosities) weighted by their
inverse squared errors. Sources for which the PN and MOS estimates do
not agree or one of the estimates is missing (e.g. source outside of
the FOV of a given detector) are flagged (Table~\ref{tab:flags} in the
appendix). An X-ray photometric quality flag is also assigned to each
source based on the quality of the plateau fit, portion of pixels
outside the detection mask, presence of anomalous features in the
X-ray background and visual screening.

\textbf{6)} We then use this total (i.e. camera averaged) bolometric
luminosity value to calculate the temperature and mass of the cluster
for the next iteration by utilizing the $L-T$ and $L-M$ scaling
relations of \citet{pratt09}:

\begin{equation}
  M = 2\times 10^{14}\, \mathrm{M_{\odot}} \left(\frac{h(z)^{-7/3}L}{1.38\times10^{44}\, \mathrm{erg\, s}^{-1}}\right)^{1/2.08}
\end{equation}

\begin{equation}
 T = 5\, \mathrm{keV} \left(\frac{h(z)^{-1}L}{7.13\times10^{44}\, \mathrm{erg\, s}^{-1}}\right)^{1/3.35}
\end{equation}

These relations were obtained by the BCES orthogonal fits
\citep{bces}, which do not treat $T_{500}$ (or $M_{500}$) as the
independent variable, since our independent variable is in fact
$L_{\mathrm{X}} (<r_{500})$.  At this stage it is impossible to safely
detect and remove emission from possible cooling cores because of the
limited resolution of XMM-\emph{Newton}. Therefore we opt not to do so
and use the relations that include also the core regions.

The $Y_{500}$ parameter is calculated from the \citet{pratt09} relation:
\begin{equation}
 Y_X = 2\times 10^{14}\, \mathrm{M_{\odot}\,keV} \left(\frac{h(z)^{-9/5}L}{5.35\times10^{44}\, \mathrm{erg\, s}^{-1}}\right)^{1/1.04}.
\end{equation}

\textbf{7)} The $M_{500}$ estimate is then used to obtain a new
$r_{500}$ radius from $r_{500}=\sqrt[3]{3M_{500}/4\pi500\rho_C(z)}$,
where $\rho_C(z)$ is the critical density of the Universe at redshift
$z$. This new $r_{500}$ aperture is used along the updated $T_{500}$
value to recalculate the fluxes and luminosities by repeating steps
$1-7$. The process is repeated until convergence to a final solution.

The final parameters are listed in Table~\ref{tab:physpar}. The table
also includes mass estimates in the $r_{200}$ aperture. An NFW profile
\citep{navarro97} was assumed in order to extrapolate the mass from
the $r_{500}$ to $r_{200}$ aperture. The extrapolation factor was
calculated using the approximation derived by \citet{hukravtsov03},
where the parameters of the NFW profile are iteratively estimated from
the final (i.e. converged) $M_{500}$ mass by utilizing the
\citet{bullock01} relation for the concentration parameter
calculation.

We provide a discussion on the scaling relations utilized here and the
error budget of our iterative method in Sect.\ref{sec:systematics}.

\section{Photometric redshift estimation}
\label{sec:photoz}

In order to measure the photometric redshifts (photo-zs) of the X-ray
selected systems in our sample, we applied the red-sequence redshift
estimator to the Blanco Cosmology Survey (BCS) imaging data which
covers two $50$~deg$^2$ patches of the southern sky and includes the
full area of the present XMM-\emph{Newton} survey.  The BCS is a 60
nights NOAO survey program carried out from 2005-2008 on the Blanco 4m
telescope at Cerro Tololo Interamerican Observatory.  This $griz$
survey was tuned to the required depths to follow the passively
evolving cluster red sequence population to $L^*+1$ at 5$\sigma$
significance out to $z=1$.  The data were acquired using two offset
layers of imaging in $g$ and $r$ band, and three offset layers of
imaging in $i$ and $z$ band.

The BCS data have been processed and calibrated using a development
version of the Dark Energy Survey data management system \citep[DESDM
  v4,][]{mohr08}.  The core processing includes flat fielding, bias
subtraction, illumination and fringe corrections.  Astrometric
refinement uses the USNO-B2 catalog and the AstrOmatic tool SCAMP
\citep{bertin06}.  Reduced single epoch images are combined into deep,
coadded images using the AstrOmatic tool SWarp together with DESDM
code that homogenized the PSF to the median seeing within each
tile/band combination.  Model fitting photometry using bulge+disk
decomposition was carried out using the extended version of SExtractor
\citep{bertin96} developed within DESDM to enable PSF corrected model
fitting to all detected objects on an image. Photometric calibration
was carried out using PSF model magnitudes calibrated using the
stellar locus in the color space defined by $grizH$ where the $H$ band
data from 2MASS \citep{skrutskie00} provides the overall photometric
scale \citep{armstrong10}.  A similar approach has been used for the
processing and calibration of all SPT cluster followup and redshift
estimation using Blanco 4m data, and the results have enabled detailed
studies of cluster galaxy properties \citep{zenteno11} as well as
precise photometric redshift estimation \citep{high10,williamson11}.

The full description of our photometric redshift method is provided in
\citet{song11}; here we give its brief summary.  The red-sequence
redshift estimator utilizes all available filters, ($g$-, $r$-, $i$-,
and $z$) to search for redshift peaks in the density distribution of
galaxies within a radius of 0.8~Mpc centered on the X-ray detection.
To define the red-sequence at each redshift slice, we assume a single
stellar population (SSP) model by \citet{bruzual03} with a single
burst of star formation at $z_{f}=3$ and passive evolution of red
galaxies thereafter. The SSP models are run with six distinct
metallicities in order to be able to model a tilted red sequence. The
models are calibrated to reproduce the tilt of the color-magnitude
relation for the Coma cluster \citep[e.g.][]{bower92}.

A single stellar population (SSP) model assuming a single burst of
star formation at $z_{f}=3$ and passive evolution of red galaxies
thereafter, by \citet{bruzual03} is used to define the red-sequence at
each redshift slice, which is calibrated to Coma cluster.

\begin{figure*}[t!]
\begin{center}
\includegraphics[width=0.49\textwidth]{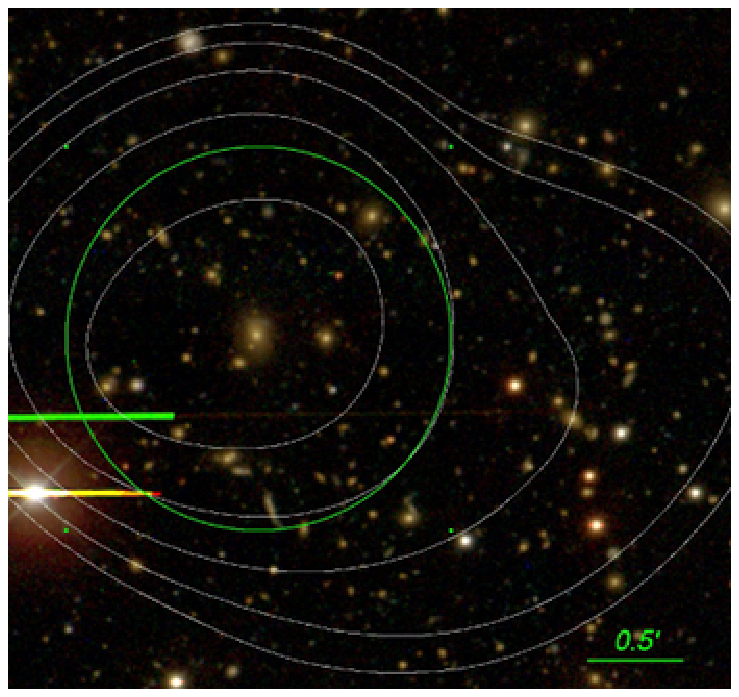}
\includegraphics[width=0.49\textwidth]{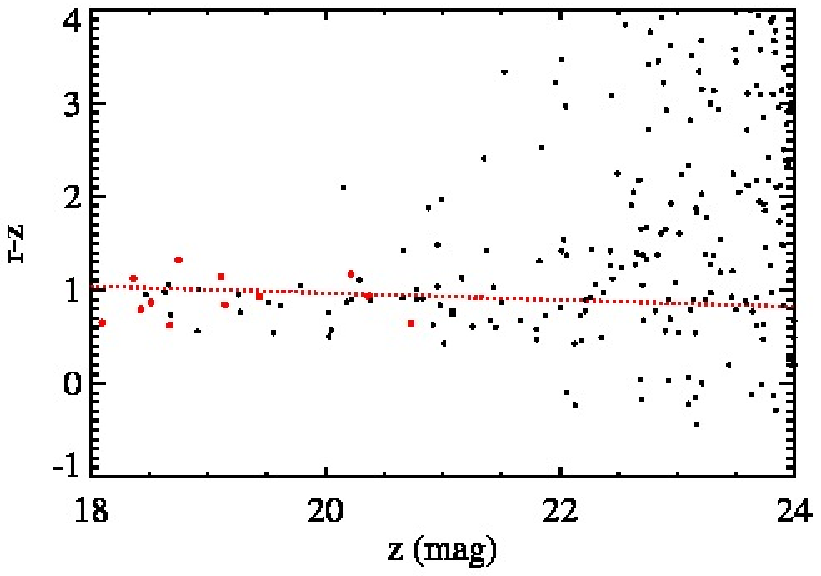}
\end{center}
\caption[Optical image of cluster ID 018]{\textbf{Left:} Pseudo-color
  image of source ID 018 from the Blanco Cosmology Survey in the $gri$
  bands. X-ray contours are overlaid in white. The green circle shows
  the estimated $0.5 \times r_{500}$.  \textbf{Right:} Color-magnitude
  diagram ($r-z$ vs. $z$) for the cluster ID 018. The red points show
  the member galaxies used for the photo-z estimate and the red dotted
  line indicates the single stellar population model at cluster
  redshift $z=0.39$ (see text for details).}
\label{fig:018opt}
\end{figure*}

The contribution of background galaxies is estimated from a
surrounding $36^{\prime} \times 36^{\prime}$ sky patch and
statistically subtracted. For each X-ray cluster candidate the whole
redshift range from $z=0$ to $z=1.05$ is scanned through using
simultaneously two colors that bracket the $4000~\AA$ break for a
given redshift. This suppresses false overdensity peaks at
transitional redshifts where the $4000~\AA$ break moves between two
adjacent bands (e.g. the transition between the $g$ and $r$ band
around $z\approx0.35$).  Once a peak in redshift space is identified,
we refine the redshift estimate by fitting a Gaussian function to the
redshift density distribution. We then select cluster members in a
stripe (0.05 width in color) around the estimated red-sequence. The
final cluster redshift value is calculated as the inverse color error
weighted mean redshift of the selected member galaxies. This assures
that the reliability of the photo-z values for the whole system is
always better than for any individual galaxy.  In a few cases two or
more solutions were found by our algorithm. For these systems we
visually check the obtained redshift distributions and select the more
likely solution given the positions of galaxies with respect to the
X-ray emission.  An example of a final color-magnitude diagram is
shown in Fig.~\ref{fig:018opt} for cluster ID 018 (its redshift is
close to the median redshift of the cluster sample).

\begin{figure*}[t!]
\begin{center}
\includegraphics[width=0.49\textwidth]{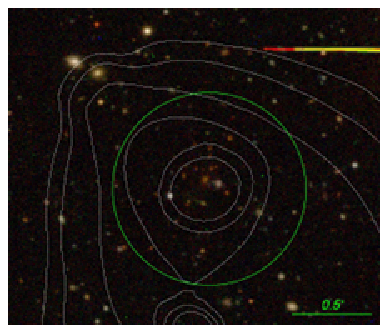}
\includegraphics[width=0.49\textwidth]{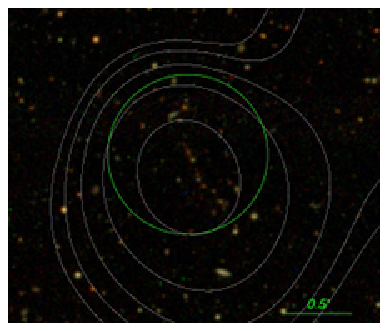}
\end{center}
\caption[BCS color images of two high-z systems]{Pseudo-color images
  in the \emph{gri} bands of the two X-ray detected (white contours)
  systems with secure photo-z values above $z>0.9$.  Green circles
  mark the $0.5 \times r_{500}$ radius. Both clusters have a BCG
  coincident with the center of the X-ray detection.}
\label{fig:highz}
\end{figure*}

The described photo-z estimation method allows us to measure the
cluster redshift with good precision up to $z\approx0.8$ even for low
richness systems. The overall uncertainty of the photo-zs is on the
$\sim10\%$ level.  While care was taken to obtain reliable results
also for $z\gtrsim0.8$ systems (see Fig.~\ref{fig:highz} for two
examples), here the already obtained \emph{Spitzer} mid-infrared
observations will provide an important improvement in subsequent work.
The final photometric redshifts are presented in
Table~\ref{tab:physpar}. A more detailed analysis of optical
counterparts for our systems including optical luminosity and richness
estimates will be presented in a companion paper (Song et al., in
prep.).

\begin{table}
\centering
\caption[Spectroscopic redshifts]{Spectroscopic redshift for 12
  clusters in the redshift range $z=0-0.4$.  The redshifts were
  obtained from long-slit spectroscopic observations at the NTT
  telescope. The redshifts of the BCG galaxies are in the
  $z_{\mathrm{spec}}^A$ column. Four clusters have a redshift for one
  additional member galaxy ($z_{\mathrm{spec}}^B$). Photometric
  redshifts $z_{\mathrm{photo}}$ are taken from
  Table~\ref{tab:physpar}.  For five systems we also provide the
  photometric redshifts from the SCS survey (M09, M10).}
\vspace{0.1cm}
\label{tab:specz}
\begin{tabular}{rrrrrr}
\hline
\hline
  \multicolumn{1}{c}{ID} &
  \multicolumn{1}{c}{$z_{\mathrm{spec}}^A$} &
  \multicolumn{1}{c}{$z_{\mathrm{spec}}^B$} &
  \multicolumn{1}{c}{$z_{\mathrm{photo}}$} &
  \multicolumn{1}{c}{$z_{\mathrm{photo}}^{\mathrm{SCS}}$} \\
\hline
  70 & 0.152 & 0.152  & $0.17 \pm 0.03$ & 0.12 \\
  94 & 0.269 &        & $0.29 \pm 0.04$ &      \\
  127 & 0.207 & 0.209 & $0.22 \pm 0.02$ &      \\
  139 & 0.169 &       & $0.18 \pm 0.01$ &      \\
  150 & 0.176 & 0.173 & $0.20 \pm 0.02$ & 0.14 \\
  152 & 0.139 &       & $0.17 \pm 0.02$ &      \\
  227 & 0.346 &       & $0.35 \pm 0.04$ &      \\
  430 & 0.206 & 0.205 & $0.18 \pm 0.01$ &      \\
  457 & 0.1 &         & $0.10 \pm 0.01$ &      \\
  476 & 0.102 &         & $0.10 \pm 0.01$ & 0.1  \\
  511 & 0.269 &       & $0.26 \pm 0.02$ & 0.2  \\
  547 & 0.241 &       & $0.22 \pm 0.02$ & 0.18 \\
\hline
\end{tabular}
\end{table}

\subsection{Spectroscopic redshifts}
\label{sec:specz}

Spectroscopic redshifts are required to identify the clusters as
compact objects, to derive precise physical parameters and later for
cosmological modeling. In order to make a first step towards these
goals we have carried out spectroscopic observations of a subsample of
our clusters in the redshift range $z=0-0.4$.

The observations made use of the EFOSC2 instrument at the 3.6~m New
Technology Telescope (NTT) in La Silla, Chile. The observations were
carried out in September 2010, with typical exposure times of 840
seconds (two spectra per cluster, 420 seconds each). Our long-slit
observations have been obtained using Grism \#4 (wavelength range
$4085-7520\AA$).  The slits ($1.5^{\prime\prime}$ width) were placed
on the BCG and an additional suitable cluster member candidate by
rotating the slit. The BCG galaxies in these systems could be easily
identified as the brightest red-sequence galaxies always coincident
with the X-ray centroid, allowing us to safely anchor the cluster
redshifts.

A standard reduction process was applied to the data using IRAF
tasks.\footnote{\texttt{iraf.noao.edu}} The observations were bias
subtracted, cleaned from cosmic rays, and flat fielded.  For each
galaxy we have obtained two spectra which were sky subtracted and
combined to increase the signal-to-noise ratio.  The wavelength
calibration was carried out by comparison with exposures of He and Ar
lamps.

The final spectra were then correlated with a database of galaxy
templates. In almost all cases the H and K lines and the
$4000\AA$~break were visible and used for visual check of the template
correction results.  Spectroscopic redshifts have been secured for a
total of 12 BCG galaxies.  Due to relatively short exposures used only
four systems a second member galaxy in the slit had good
signal-to-noise ratio in order to safely measure its redshift. In all
four cases the galaxies were found to have concordant redshifts with
the BCG value.  The spectroscopic redshifts of the galaxies are
summarized in Table~\ref{tab:specz} along with our photo-z
estimates. We compare the two redshift sets in
Sect.~\ref{sec:speczcompare}.

\subsection{Comparison of spectroscopic and photometric redshifts}
\label{sec:speczcompare}

\begin{figure*}[t]
\begin{center}
\includegraphics[width=0.48\textwidth]{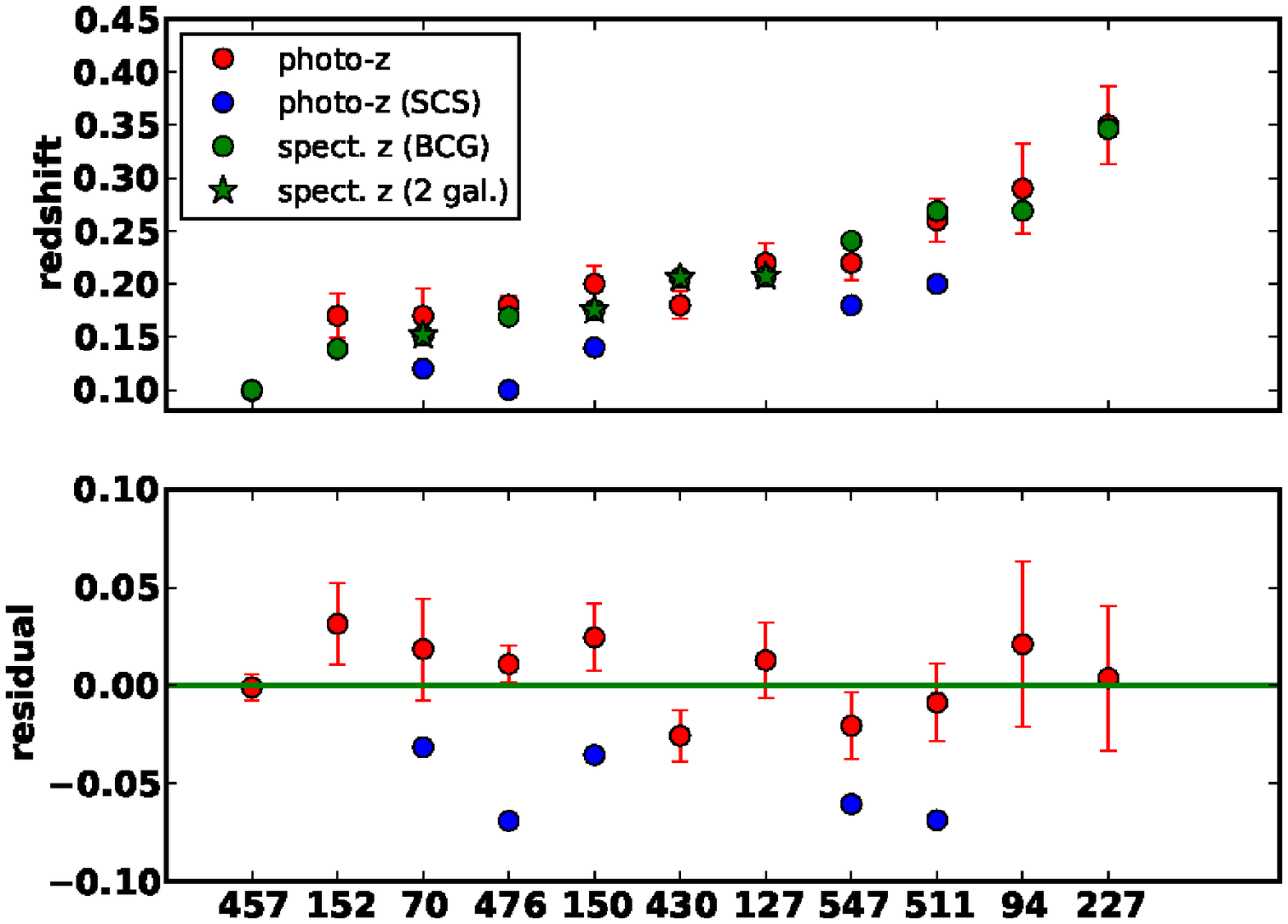}
\includegraphics[width=0.48\textwidth]{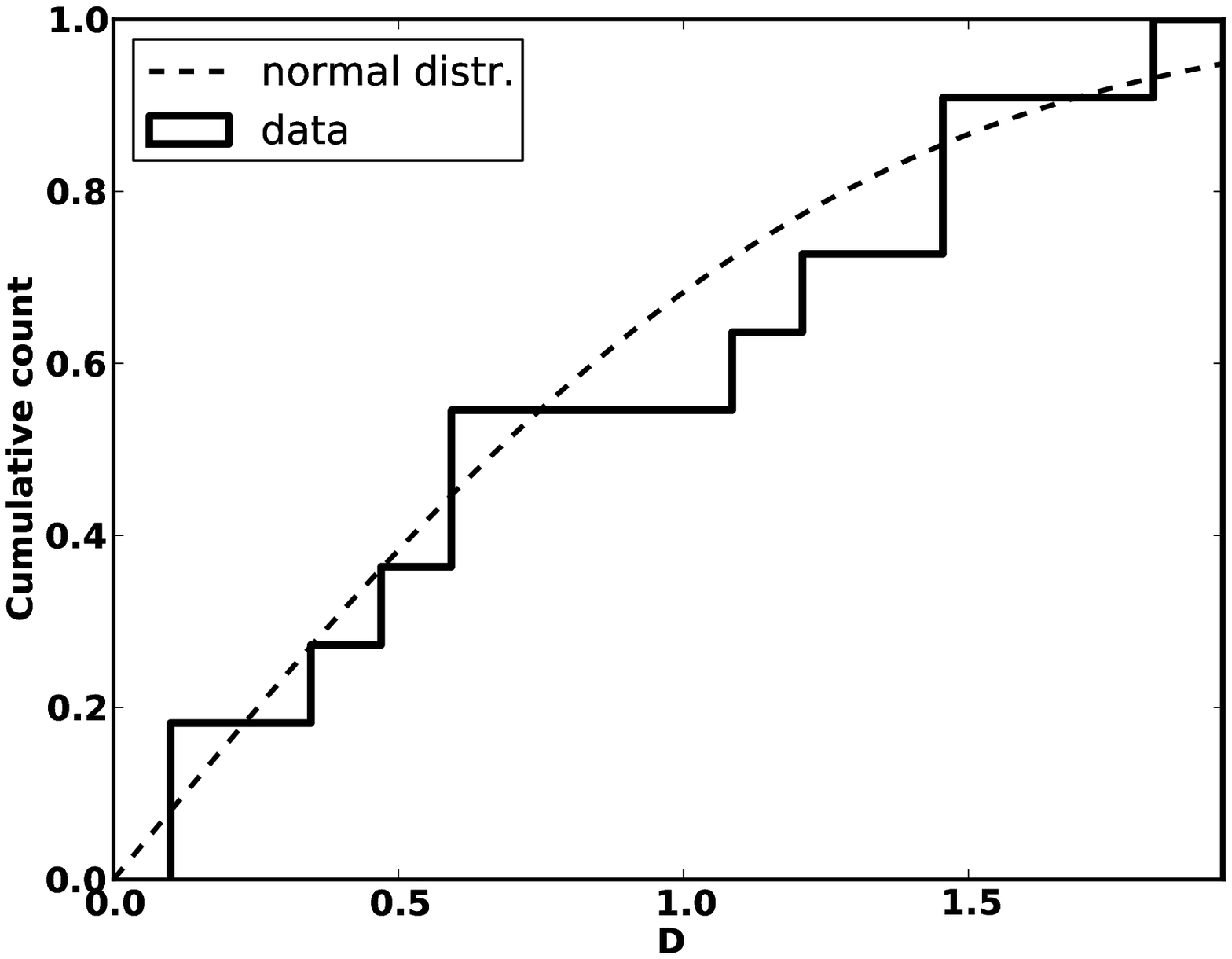}
\end{center}
\caption[Comparison of spectrometric and photometric
  redshifts]{Consistency test of our photometric redshift estimates
  with spectroscopic measurements in the redshift range $z=0-0.4$.
  \textbf{Left:} Comparison of our photometric redshift estimates
  (red, $1\sigma$ error bars) with spectroscopic values (green). Green
  stars mark clusters for which we have two concordant galaxy
  redshifts, while green circles indicate clusters for which only the
  BCG has a spectroscopic redshift. The photometric redshifts obtained
  by the SCS survey (M09, M10) are shown in blue.  The x-axis displays
  the cluster ID number. The objects are sorted in increasing redshift
  order. The bottom panel shows the residuals of the photo-z values
  with respect to the spectroscopic measurement.  \textbf{Right:}
  Cumulative histogram of the difference between the photometric and
  spectroscopic redshift normalized by the $1\sigma$ uncertainty of
  the photo-z values, i.e.  $D =
  |z_{\mathrm{photo}}-z_{\mathrm{spec}}|/\sigma_{\mathrm{photo}}$.
  The dashed line shows the expectation for the Gaussian distribution.
  Both curves are in good agreement, with a Kolmogorov-Smirnov test
  confirming that the distribution of the $D$ values is Gaussian at
  the $96\%$ confidence level.}
\label{fig:speczcalib}
\end{figure*}

For a subsample of 12 clusters ($z<0.4$) we have obtained
spectroscopic redshifts of their BCG galaxy and in four systems also
for one additional member galaxy (Sect.~\ref{sec:specz}). We compare
the spectroscopic redshifts with our photo-z values in
Fig.~\ref{fig:speczcalib} (left). Our photometric values (red points)
agree well within the error bars with the spectroscopic redshifts of
the BCG (green points, brighter green points mark the clusters with
two concordant redshifts).  Blue points mark the photo-zs for five of
the systems obtained by the SCS survey (M09, M10).  These values
exhibit a systematic bias toward lower redshifts, with a mean relative
difference of $19\%$. A similar trend is also visible in
Fig.~\ref{fig:scsphotoz} (top left) where we compare our photo-z
values with the SCS measurements for all clusters common to both
samples.

The right panel of Fig.~\ref{fig:speczcalib}, displays the comparison
of the absolute difference of our photometric and spectroscopic
estimates in units of photo-z error, $D =
|z_{\mathrm{photo}}-z_{\mathrm{spec}}|/\sigma_{\mathrm{photo}}$. A
comparison with a Gaussian expectation shows an agreement at the
$96\%$ confidence level, confirming both the good precision of our
photo-z estimates and the realistic description of their errors.

The present spectroscopic sample covers only part of the redshift
range and does not allow us to check the photometric redshift
calibration at higher redshift. However, the good agreement at low z
supports the photo-z method used. We also note, that same photo-z
method has been applied to a large number of SPT selected clusters
extending to beyond $z=1$, and the agreement between spectroscopic and
photometric redshifts has been excellent \citep{high10, williamson11}.

\section{Results}
\label{sec:results}
\begin{figure*}[t!]
\begin{center}
\includegraphics[width=0.49\textwidth]{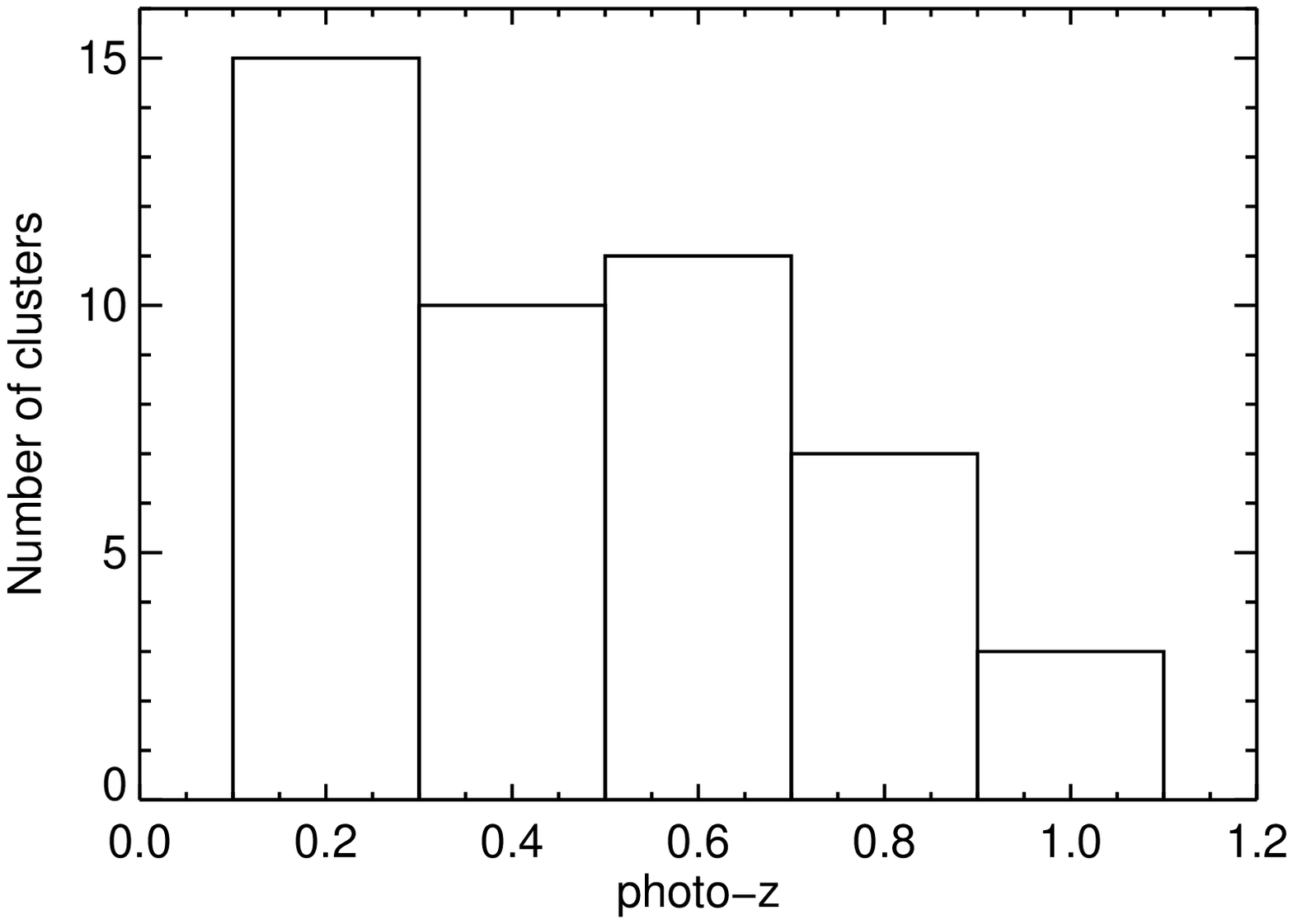}
\includegraphics[width=0.49\textwidth]{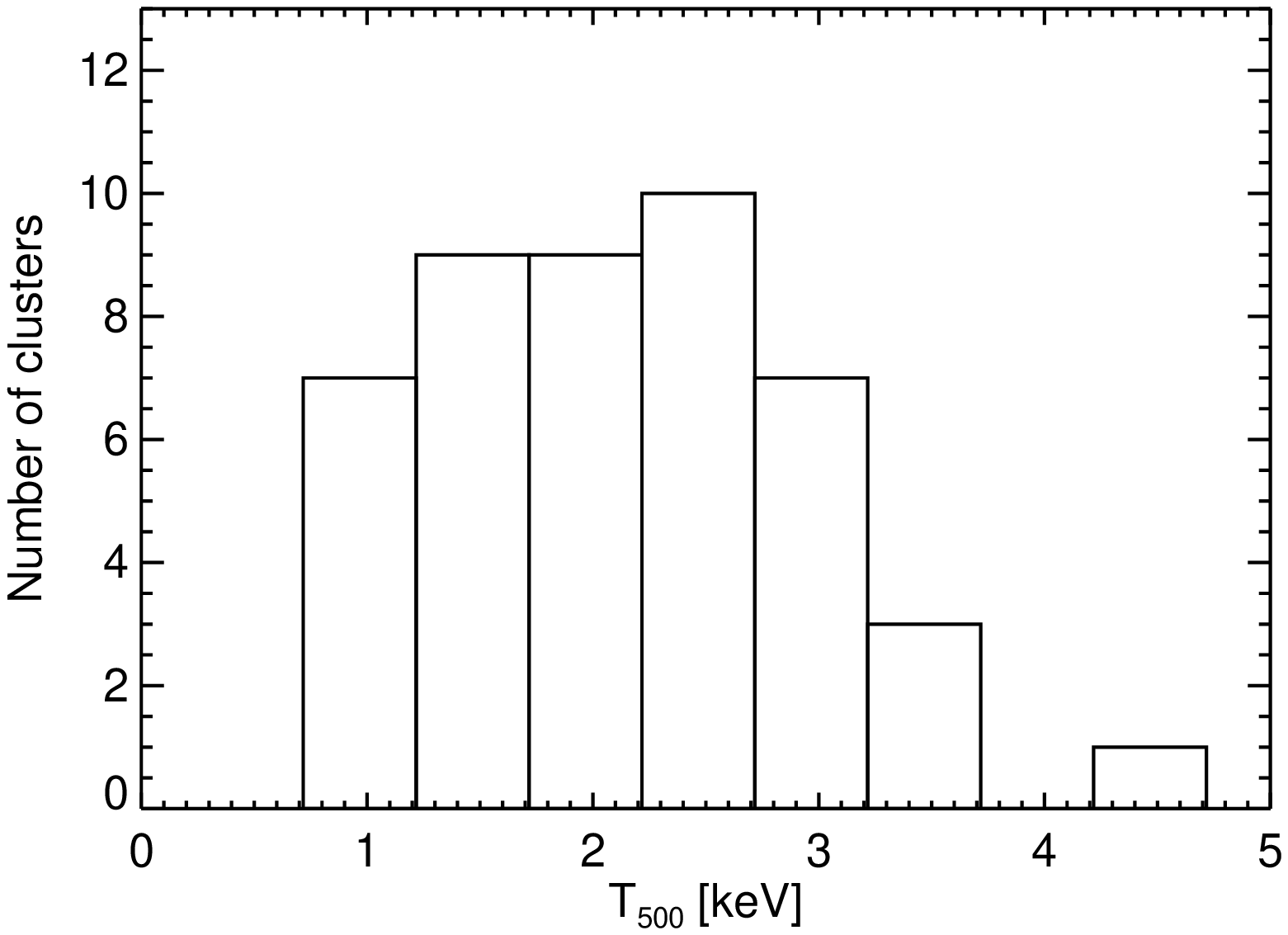}
\includegraphics[width=0.49\textwidth]{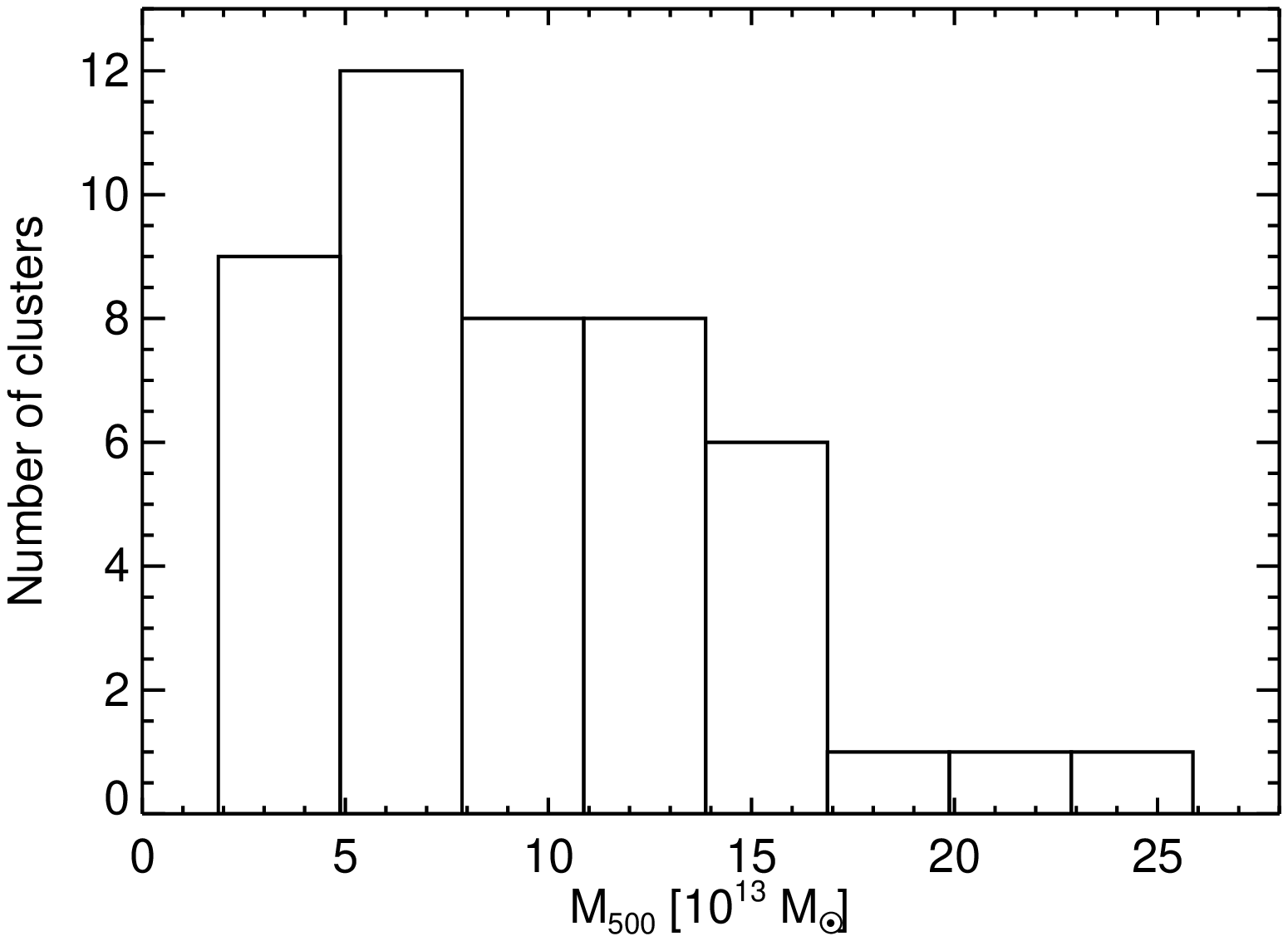}
\end{center}
\caption[Redshift, mass and temperature distributions of the cluster
  sample]{\textbf{Top left:} The redshift distribution of the 46
  cluster sample based on the photometric redshifts obtained with the
  red sequence fitting method. \textbf{Top right:} The X-ray
  temperature distribution estimated from the L-T scaling relation of
  \citet{pratt09}.  \textbf{Bottom:} Distribution of the cluster
  masses in the $r_{500}$ aperture calculated from the luminosities
  using the L$-$M scaling relation from \citet{pratt09} (see
  Sect.~\ref{sec:physpar}).}
\label{fig:histoz}
\end{figure*}

\subsection{Galaxy cluster sample}
Table~\ref{tab:physpar} provides the physical properties determined
for the 46 clusters in the present sample.  The measured X-ray
luminosity of the systems (Sect.~\ref{sec:physpar}) and the
photometric and spectroscopic redshifts (Sect.~\ref{sec:photoz}
and~\ref{sec:specz}) are used as inputs for the cluster scaling
relations to estimate further physical parameters. Ancillary X-ray
information on the individual clusters can be found in
Table~\ref{tab:flags}.

The redshift, temperature and mass distributions are shown in
Fig.~\ref{fig:histoz}. We display the X-ray luminosity of our systems
as a function of redshift in Fig.~\ref{fig:lxz}.  The median redshift
of the cluster sample is $z=0.47$. Six of the systems have photometric
redshifts $z>0.8$. Three of these have redshifts consistent with
$z=1$, although the photo-z uncertainty in this regime is large. The
median temperature of the clusters is $\sim 2$~keV and the median
M$_{500}$ mass $9 \times 10^{13}$~M$_{\odot}$ (based on luminosity
scaling relations). We are thus able to probe the cluster/group
transition regime practically at all redshifts out to $z\approx1$.

\subsection{Survey sky-coverage}
\label{sec:skycov}
\begin{figure*}[t!]
\begin{center}
\includegraphics[width=0.5\textwidth]{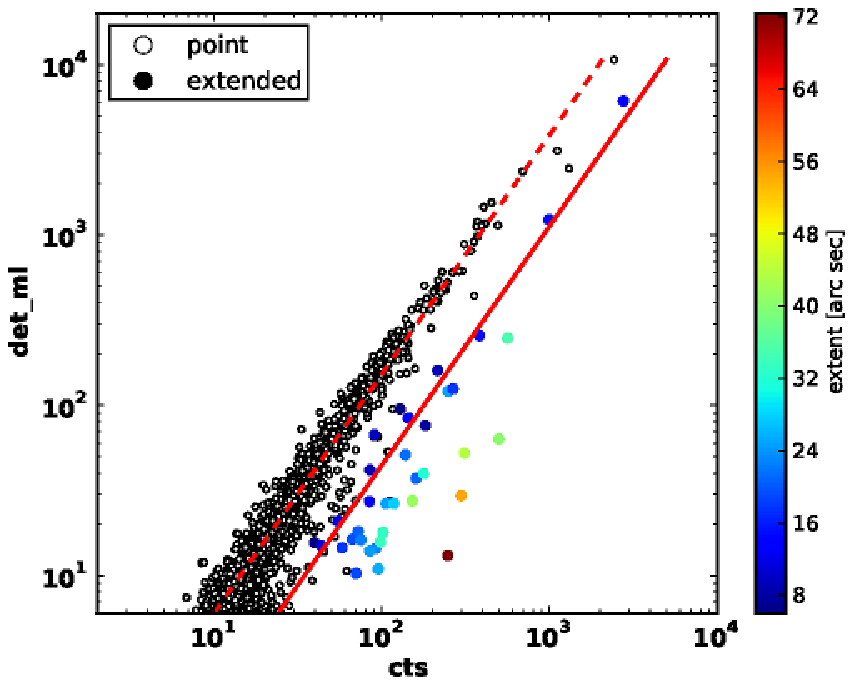}
\includegraphics[width=0.49\textwidth]{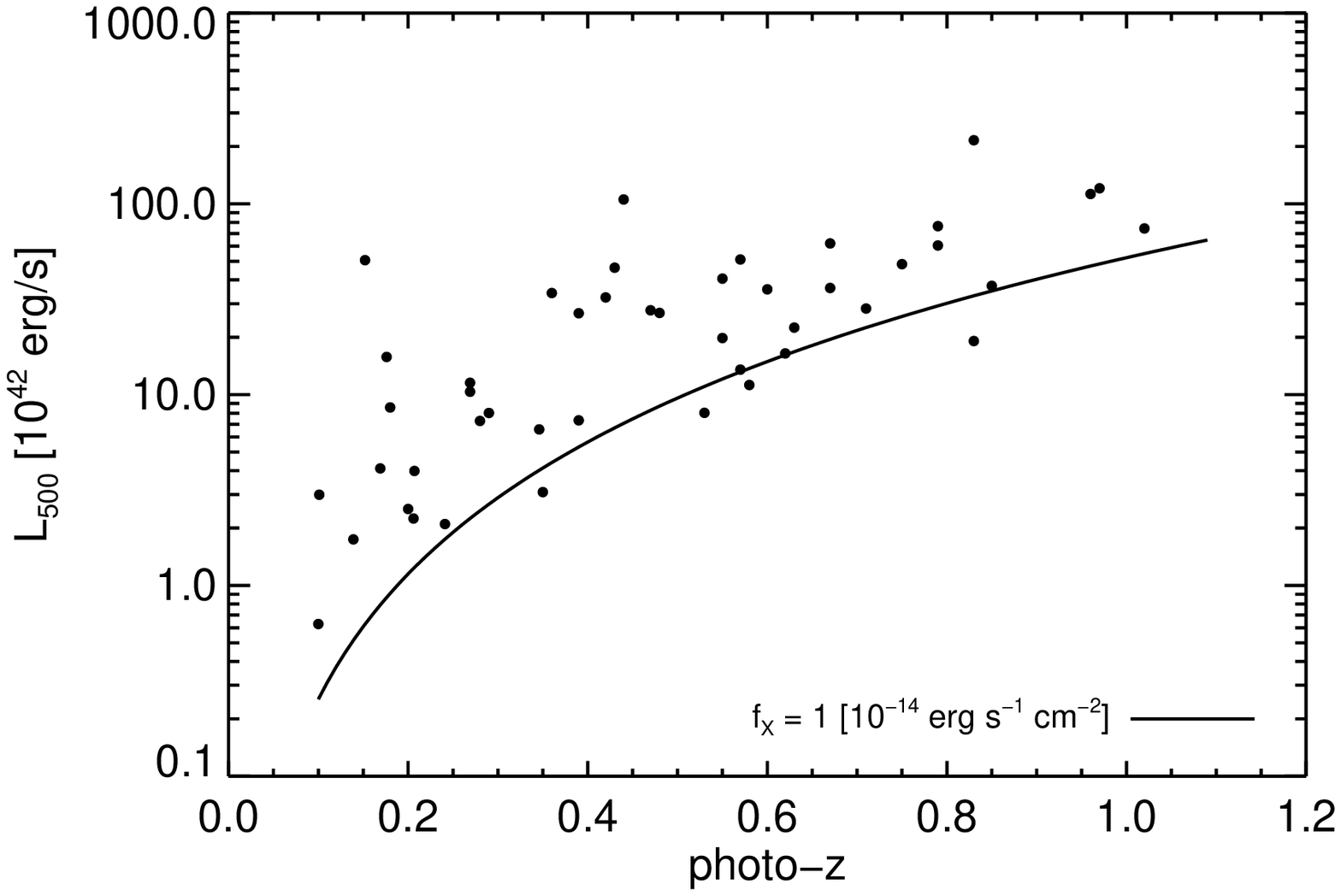}
\end{center}
\caption[Detection likelihoods and X-ray luminosities of the sample]{
  \textbf{Left:} Detection likelihood (det\_ml) as a function of total
  detected source counts (PN detector only) for point sources (open
  circles) and the detected clusters (full circles). Clusters are
  color coded by their extent (beta model core radius). The dashed red
  line shows the best fit linear relation in the \texttt{det\_ml} -
  counts plane for point sources. The solid line shows the same
  relation for extended sources (with slope fixed to the point-source
  fit, points weighted by their counts error). Typically, an extended
  source has to have 2.4 times more counts than a point source to be
  detected at the same \texttt{det\_ml} value.  See
  Sect.~\ref{sec:skycov} for details.  \textbf{Right:} Luminosity in
  the $0.5-2$~keV band (object rest-frame) for the present cluster
  sample as a function of redshift. The line shows the luminosity of a
  cluster with a measured flux of $1 \times 10^{14}$~erg s$^{-1}$
  cm$^{-2}$ (unabsorbed, observer rest-frame).  }
\label{fig:lxz}
\end{figure*}

The simplest statistical characteristics of a cluster survey are its
area coverage as a function of limiting flux (\emph{sky-coverage
  function}) and the cumulative surface density of the detected
objects above the given flux limit as a function of flux - the
so-called $\log N - \log S$ relation.\footnote{We use the standard
  notation of this relation, but keep writing $f_X$ as the source flux
  rather than $S$.}

In order to properly determine the survey's sky coverage, good
knowledge of the survey's selection function is necessary. For the
simple case when the selection function is the function of only flux,
the sky coverage is then the selection function of the survey scaled
by its geometric area.  Especially for the case of extended sources
the situation is more complex, since the selection function depends
also on other parameters (e.g the source extent and off-axis angle).
These effects can only be accounted for by Monte Carlo simulations.
At this moment, without the simulations at hand, we can still provide
a preliminary, empirically calibrated sky coverage calculation and
cluster $\log N - \log S$ relation. We will demonstrate that these
simple approaches show good agreement with the design aims for the
survey depth and previous measurements of the cluster $\log N - \log
S$ function.

While our source detection pipeline utilizes multiple detection bands
and likelihood thresholds (Sect.~\ref{sec:detectpipe}) we will for
simplicity (and ability to compare our results with published work)
characterize detections made in the standard $0.5-2$~keV band with a
$3\sigma$ detection threshold and a $5\sigma$ extent significance.

In order to obtain the survey sensitivity function for extended
sources, we first calculate the point source sensitivity for each
field. This is a simpler task since it does not require treatment of
the source extent.  We calculate the point source sensitivity function
by analytically inverting the detection likelihood calculation
(described in Sect.~\ref{sec:detect}) and obtaining the minimal
count-rate necessary for a point source to be detected at the required
detection threshold given the local background and exposure in the
detection cell.

The procedure is repeated for each survey field and the results are
combined for the whole survey area. In the areas where two or more
fields overlap, we compare the sensitivity maps pixel-by-pixel taking
the highest reached sensitivity (i.e. lowest local count-rate limit)
at the given position. This procedure is chosen because the present
catalog was derived from the detection pipeline that ran on each field
individually. An alternative approach is to combine the fields before
detection - reaching slightly deeper flux-limits in the overlapping
areas.\footnote{This was done for the ancillary catalog using the
  wavelet detection algorithm.}  This comes at the cost of losing the
information on the local PSF shape used by the maximum-likelihood
fitting algorithm, since the same sky location in two different
observations is imaged at different off-axis and position angles and
thus with different PSF.  Both approaches give comparable results and
we opt here to characterize the main scheme (i.e. detection on
individual fields).

\begin{figure}[t!]
\begin{center}
\includegraphics[width=0.5\textwidth]{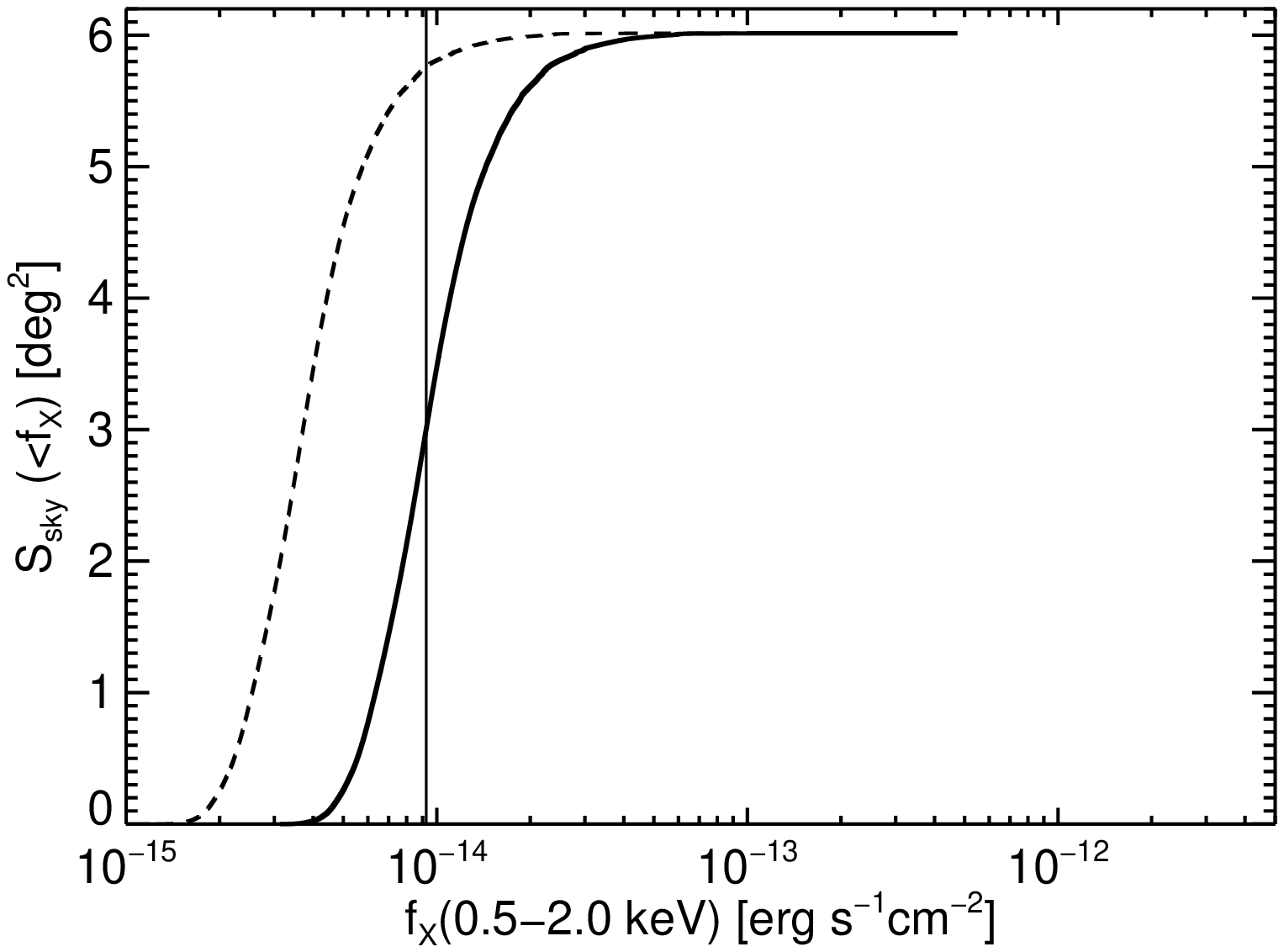}
\end{center}
\caption[The survey sky coverage]{The survey sky coverage. Dashed line
  shows the sky coverage as a function of limiting point source
  sensitivity in the $0.5-2.0$~keV band.  The empirically estimated
  extended source sensitivity is shown with a solid line. The median
  point source sensitivity of the survey is $3.7 \times 10^{-15}$~erg
  s$^{-1}$ cm$^{-2}$, the median sensitivity for the extended sources
  $9.3 \times 10^{-15}$~erg s$^{-1}$ cm$^{-2}$ (vertical line).}
\label{fig:skycov}
\end{figure}

The median \emph{point} source sensitivity calculated in this way for
the whole survey area is $3.7 \times 10^{-15}$~erg s$^{-1}$ cm$^{-2}$
for an energy-conversion factor\footnote{Assuming a power law spectrum
  with $\Gamma=1.7$ and $n_\mathrm{H}=1.25\times10^{20}$ cm$^{-2}$
  (median value of the galactic column density in the survey field)
  and using an on-axis PN response file.} of $1.5\times10^{-12}$~erg
s$^{-1}$ cm$^{-2}$. The corresponding sky coverage as a function of
flux is displayed in Fig.~\ref{fig:skycov}.

In the next step, we attempt to obtain a first order approximation to
the sky coverage function for the \emph{extended} sources by a simple
scaling to the point source function. In Fig.~\ref{fig:lxz} (left) we
show the dependence of the detection likelihood (i.e.  the
\texttt{det\_ml} parameter) on the total detected source counts for
point sources and the confirmed clusters from our sample (full
circles).

Photons from extended sources are distributed over a larger area and
thus require more counts to reach a given detection likelihood
compared to point sources.  For those, a simple linear relation in the
log-log plane is a good description of the counts-\texttt{det\_ml}
relation (dashed red line in Fig.~\ref{fig:lxz}).  Since the number of
our clusters is small a similar linear relation for them is only very
weakly constrained. We therefore fix the slope to the value from the
point source fit leaving only the intercept as a free parameter and
weighting the points by their counts error (solid red line).  The
offset of the extended-source best-fit line translates to a factor of
2.4 between the total required counts of point and extended sources at
any given \texttt{det\_ml}.  Fixing the slope has also the advantage
that the offset is independent of the selected detection
threshold. The best fit line roughly follows the locus of clusters
with extent (beta model core radius) close to the median value of
$\sim 20^{\prime\prime}$.  The solid red line in Fig.~\ref{fig:lxz}
(left) thus roughly gives the detected counts for a cluster with a
\emph{typical} extent detected with a given likelihood. We then use
this offset factor to scale up the point source sky coverage function
(see Fig.~\ref{fig:skycov}). The median flux limit for this
sky-coverage is $9.3 \times 10^{-15}$~erg s$^{-1}$ cm$^{-2}$ (using
the median ECF of our sample). In Fig.~\ref{fig:lxz} we display the
luminosity-redshift plane for our survey. The luminosity threshold for
a flux limited sample ($f_{\mathrm{min}}=1 \times 10^{-14}$~erg
s$^{-1}$ cm$^{-2}$) is also shown, demonstrating a rough agreement
with our calculation. Note that in the present sample we also include
fainter sources than this threshold (the lowest cluster flux is $\sim6
\times 10^{-15}$~erg s$^{-1}$ cm$^{-2}$).

This approach underestimates the effect of clusters with larger extent
- and thus overestimates the sky coverage at given flux.\footnote{The
  fit with a free slope gives an offset factor of $\sim 4$, the fit
  being skewed towards the locus of very extended clusters.} However,
since the detection probability itself is a strong function of source
extent, the only way to properly account for its effect is through
realistic simulations.

We present examples of preliminary sky coverage functions for extended
source detection on several (non-XMM-BCS) fields based on such Monte
Carlo simulations \citep{muehlegger10} in Fig.~\ref{fig:simsensf},
discussed in Sect.~\ref{sec:simsensf}.  These first results validate
our attempt to model to first approximation the extended source
sky-coverage by scaling the point source curve and also confirm that
the scaling factor between them is roughly $\sim 2.4$ (this scaling
factor is expected to hold only for observations with roughly same
depth as ours, $\sim10$~ks).

\subsection{Cluster $\log N - \log S$}
\label{sec:lognlogs}
We now use this empirical sky-coverage in order to calculate the
survey's $\log N - \log S$, defined in a standard way as:
\begin{equation}
\label{eq:lognlogs}
    N(>f_{X}) =
    \sum_{i=1}^{N_C}\frac{1}{\Omega(f_X^i)}\mathrm{deg}^{-2},
\end{equation}
where $N_C$ is the total number of clusters and $\Omega(f_X^i)$ is the
extended source sky-coverage corresponding to the flux of the $i$-th
cluster.  We characterized the survey sky coverage only for a
hypothetical single band ($0.5-2$~keV) detection scheme.  Since such a
detection scheme is not part of our pipeline, we opt to draw a
subsample from our cluster catalog derived from the three band scheme
(which includes the $0.5-2$~keV). For this calculation we consider
only clusters that would have also been detected in this hypothetical
single band run by setting the same detection- and extent likelihood
thresholds used for the sky-coverage calculation in the previous
section.

This requires us to recover the actual $0.5-2$~keV band detection
likelihoods from the total \texttt{det\_ml} parameter, which includes
contributions from all three detection bands.  As we described in
Sect.~\ref{sec:detectpipe}, \texttt{det\_ml} can be interpreted as
$\texttt{det\_ml} = -\ln P_{\mathrm{rand}}$, with $P_{\mathrm{rand}}$
being the probability of a false detection arising from pure
Poissonian fluctuations.  The actual definition of this parameter is
slightly more complex:
\begin{equation}
\label{eq:detml}
\texttt{det\_ml} = -\ln ( 1 - \Gamma(0.5\nu,L)),
\end{equation}
where $\Gamma$ is the incomplete gamma function and its arguments are
the number of degrees of freedom of the \texttt{emldetect} fit (for
extended sources $\nu = 3 +$ the number of detection bands times
number of instruments) and L is the sum of all the individual
likelihoods \citep[using the C statistics of][]{cash79}. This
definition effectively converts the joint likelihoods to two degrees
of freedom allowing to compare detections from different combination
of bands and instruments. However, for the conversion to a single band
detection likelihood, we need the original individual likelihoods
which we obtain by numerically inverting Eq.~\ref{eq:detml} for each
source.

We then calculate the joint detection likelihood from all three
instruments in the single, $0.5-2$~keV, band (all three detection
probabilities being independent) and subsequently calculate the new
\texttt{det\_ml} parameter normalized back to two degrees of freedom
using Eq.~\ref{eq:detml}. The number of clusters that have this new
single band \texttt{det\_ml} parameter above the required threshold
(i.e. equivalent to $\sim3\sigma$) is 40.

\begin{figure}[t!]
\begin{center}
\includegraphics[width=0.5\textwidth]{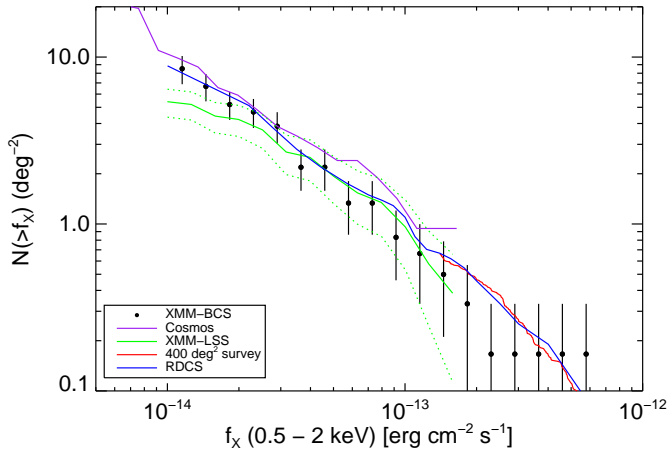}
 \end{center}
\caption[The $\log N-\log S$ relation]{The $\log N-\log S$ of the
  present sample in the $0.5 - 2.0$~keV band.  Fluxes are calculated
  in the $r_{500}$ aperture. Results from several surveys are also
  shown: COSMOS \citep{finoguenov07}, RDCS \citep[the ROSAT Deep
    Cluster Survey,][]{rosati98}, 400 deg$^2$ survey
  \citep{burenin07cccp1, vikhlinin09} and the XMM-LSS
  \citep{pacaud07}.  See Sect.~\ref{sec:lognlogs} for details.}
\label{fig:lognlogs}
\end{figure}

Finally, we calculate the $\log N - \log S$ according to
Eq.~\ref{eq:lognlogs} and the variance of the number counts as
$\sigma^2=\sum_{i=1}^{N_C}1/\Omega(f_X^i)^2$.  The recovered curve
(see Fig.~\ref{fig:lognlogs}) is in good agreement with the $\log N -
\log S$ of other surveys: e.g. COSMOS \citep{finoguenov07}, the RDCS
survey \citep[the ROSAT Deep Cluster Survey,][]{rosati98}, 400 deg$^2$
survey \citep{burenin07cccp1, vikhlinin09} and the XMM-LSS
survey\footnote{ Note that the XMM-LSS curve is only digitized from
  the figure in \citet{pacaud07} since the original curve is no longer
  available (Pacaud, private com.).} \citep{pacaud07}.  Since the area
and depth of the XMM-LSS survey match well the parameters of our
survey we discuss their comparison in more detail in
Sec.~\ref{sec:lsscomp}.

We note, that we used the fluxes in the $r_{500}$ aperture for our
calculation whereas the XMM-LSS uses a fixed physical aperture of
$0.5$ Mpc (typically very close to $r_{500}$), RDCS $\sim80-90\%$ of
the total flux (i.e. integrated out to infinity) and the 400 deg$^2$
survey the full total flux.  We have chosen $r_{500}$ because:
\textbf{a)} it requires less extrapolation based on a beta model whose
parameters are typically highly uncertain and is itself not
necessarily a good description of the surface brightness profiles and
\textbf{b)} it is the most natural choice when comparing to
theoretical predictions (the cluster is approximately virialized in
this radius and the scaling relations we employ are calibrated for
this overdensity).  However, assuming a typical cluster with $r_{500}
= 0.5$~Mpc well described by a beta model with $(\beta, r_{core}) =
(2/3, 180\, \mathrm{kpc})$ the flux extrapolated to infinity would be
higher by $\sim1/3$ moving our curve along the x-axis to higher fluxes
only very slightly - even closer to the RDCS and 400 deg$^2$ survey's
relations.

Uncertainties of the flux estimation (including the uncertainty on the
photo-zs) affect the $\log N - \log S$ only in a minor way. The main
source of uncertainty (not included in the error bars) is our current
lack of knowledge of the survey selection function (and thus only
tentative description of the sky-coverage). The good agreement with
previous work gives, however, support to our preliminary approach.

\subsection{Cross-correlation with known sources}
\label{sec:xcorrned}
The XMM-BCS field has an extensive multi-wavelength coverage and has
already been studied by the Southern Cosmology Survey \citep[M09,
  M10,][]{mcinnes09} who identified in optical data a number of
clusters in this area. Due to a significant overlap with our cluster
catalog we will address a more detailed comparison in
Sect.~\ref{sec:menan}.

In search of other known sources coincident with our clusters, we make
use of both the NASA/IPAC Extragalactic
Database\footnote{\texttt{nedwww.ipac.caltech.edu}} and the SIMBAD
Astronomical Database.\footnote{\texttt{simbad.u-strasbg.fr/simbad/}}

First we looked for associated known clusters. For this query a search
radius of $60^{\prime\prime}$ was selected, finding a single match -
the cluster 400d~J2325-5443 (alternative name: [BVH2007]~240)
identified in the 160 Square Degree ROSAT Survey \citep{vikhlinin98,
  mullis03} at spectroscopic redshift $z=0.102$. This cluster is
coincident with our cluster ID 476 with photometric redshift of 0.1
being in full agreement with the spectroscopic value.  The source is
also part of the 400 Square Degree ROSAT Survey. See
Appendix~\ref{sec:objectnotes} for more details on this source.

We also list galaxy matches, if they are within a 16 arcsec search
radius from the X-ray center in Table~\ref{tab:nedgals} (in the
appendix) with matches coming from the 2 Micron All Sky Survey
Extended objects catalog and the APM galaxy survey
\citep[][respectively]{2masx, maddox90}. Out of 13 matches, only two
galaxies have known spectroscopic redshifts, both obtained in the
frame of the 6dF Galaxy Survey \citep{jones046df}.  The first is 2MASX
J23254015-5444308 at $z=0.101$ coincident with the brightest galaxy in
cluster ID 476. The redshift value is concordant with the redshift
from the 160/400 Square Degree ROSAT surveys. The second match is the
brightest cluster galaxy of the system ID 150 at redshift $z=0.176$,
again in good agreement with our estimated photo-z of 0.2.

As can be seen, the survey field has a wealth of multi-wavelength
data, but very little spectroscopic measurements. This makes the
ongoing spectroscopic follow-up program very important, as redshifts
are essential for the full utilization of the available data sets.

Radio sources coincident with the X-ray detected clusters can bias the
SZE signal (filling the decrement). We checked for intervening radio
sources by cross-correlating our cluster catalog with the NED database
with a 1 arcmin search radius. We find 11 radio sources detected at
843 MHz by the Sydney University Molonglo Sky Survey
\citep[SUMS,][]{sumss_cat}. The source PMN~J2330-5436, $\sim30$~arcsec
from cluster ID 150, was detected by the Parkes-MIT-NRAO (PMN)
southern survey at 4.85 GHz \citep{pmn_cat}. The list of all
identified radio sources is given in Table~\ref{tab:radio}.

\begin{figure*}[ht!]
\begin{center}
\includegraphics[width=0.49\textwidth]{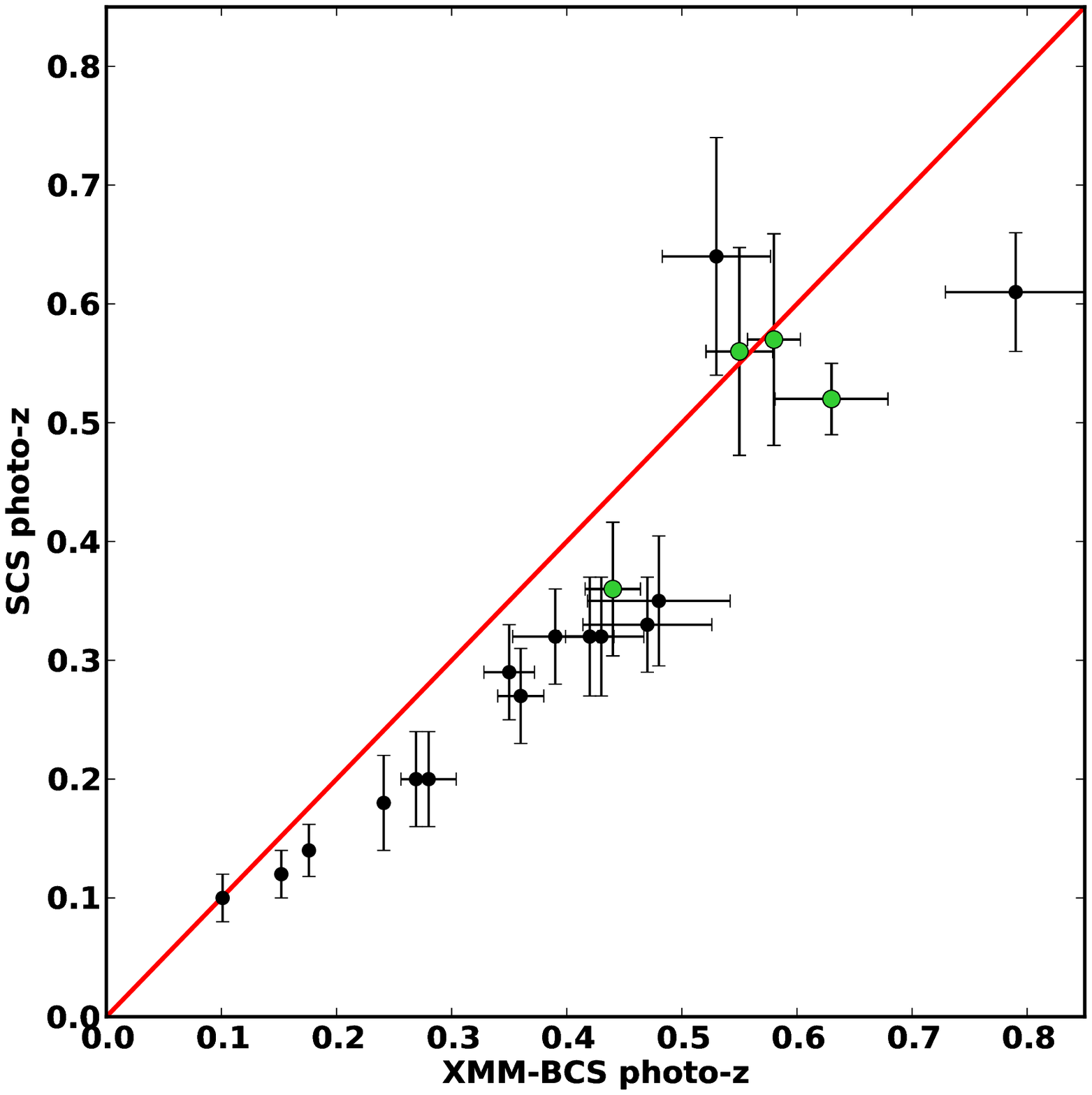}
\includegraphics[width=0.44\textwidth]{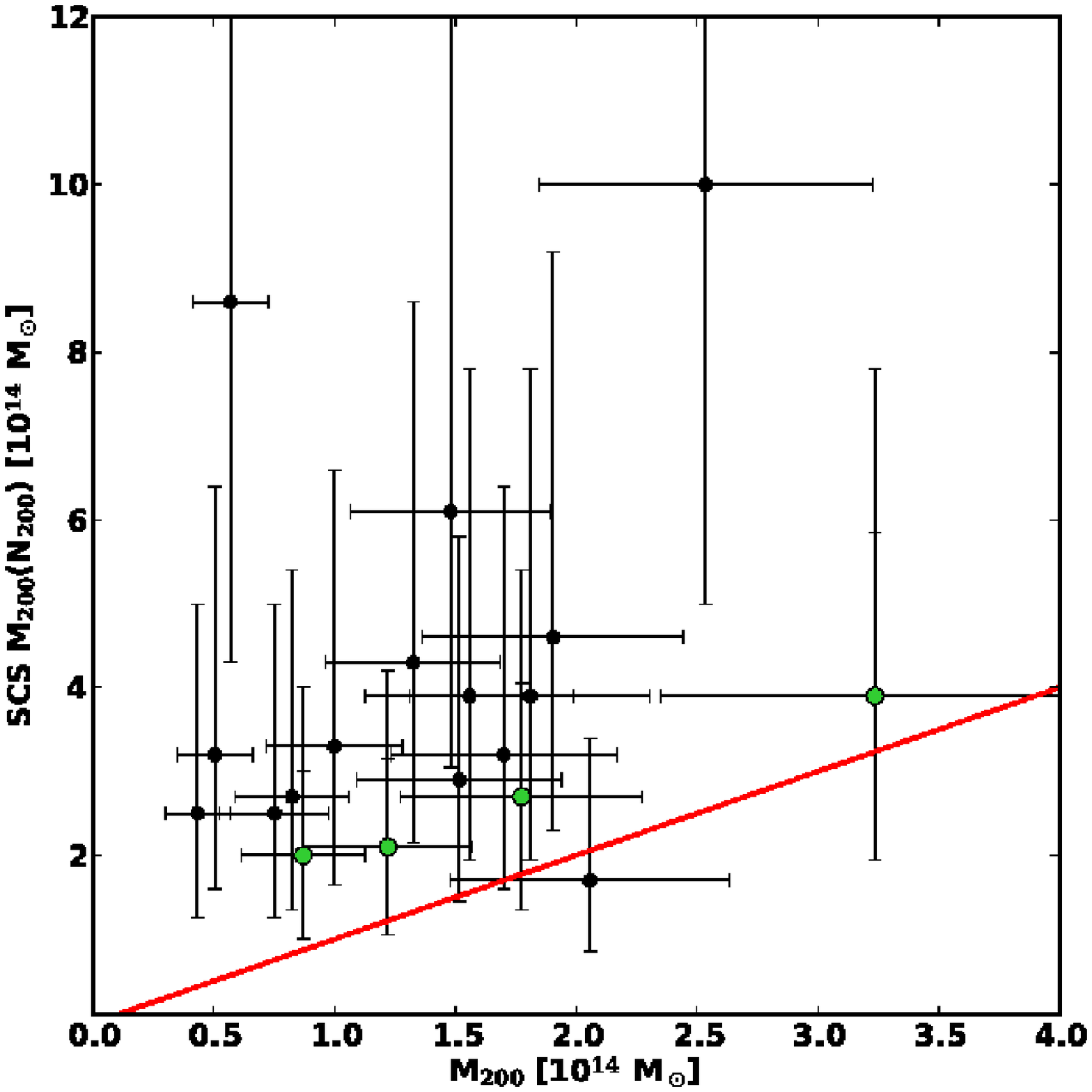}
\includegraphics[width=0.49\textwidth]{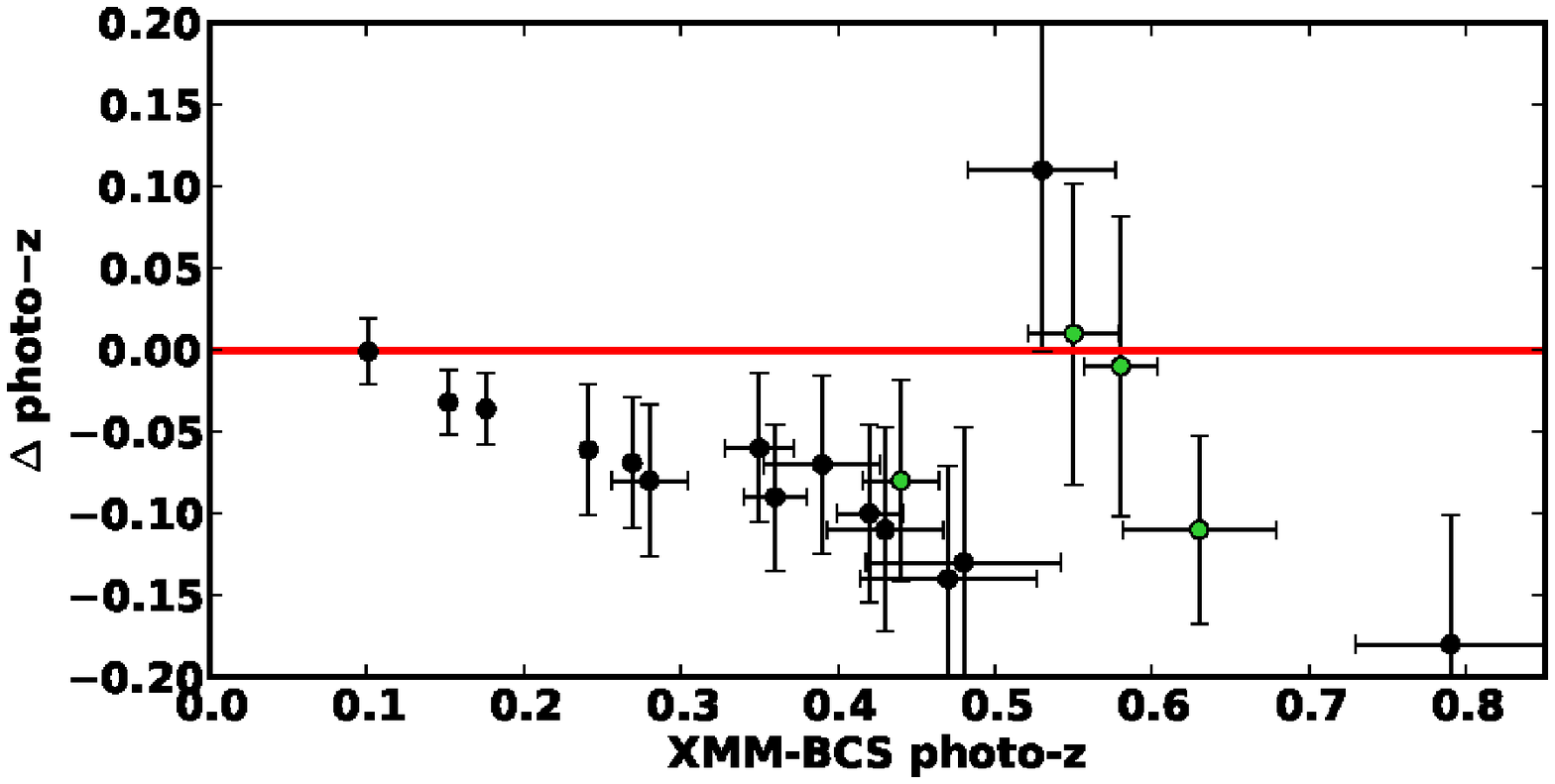}
\includegraphics[width=0.49\textwidth]{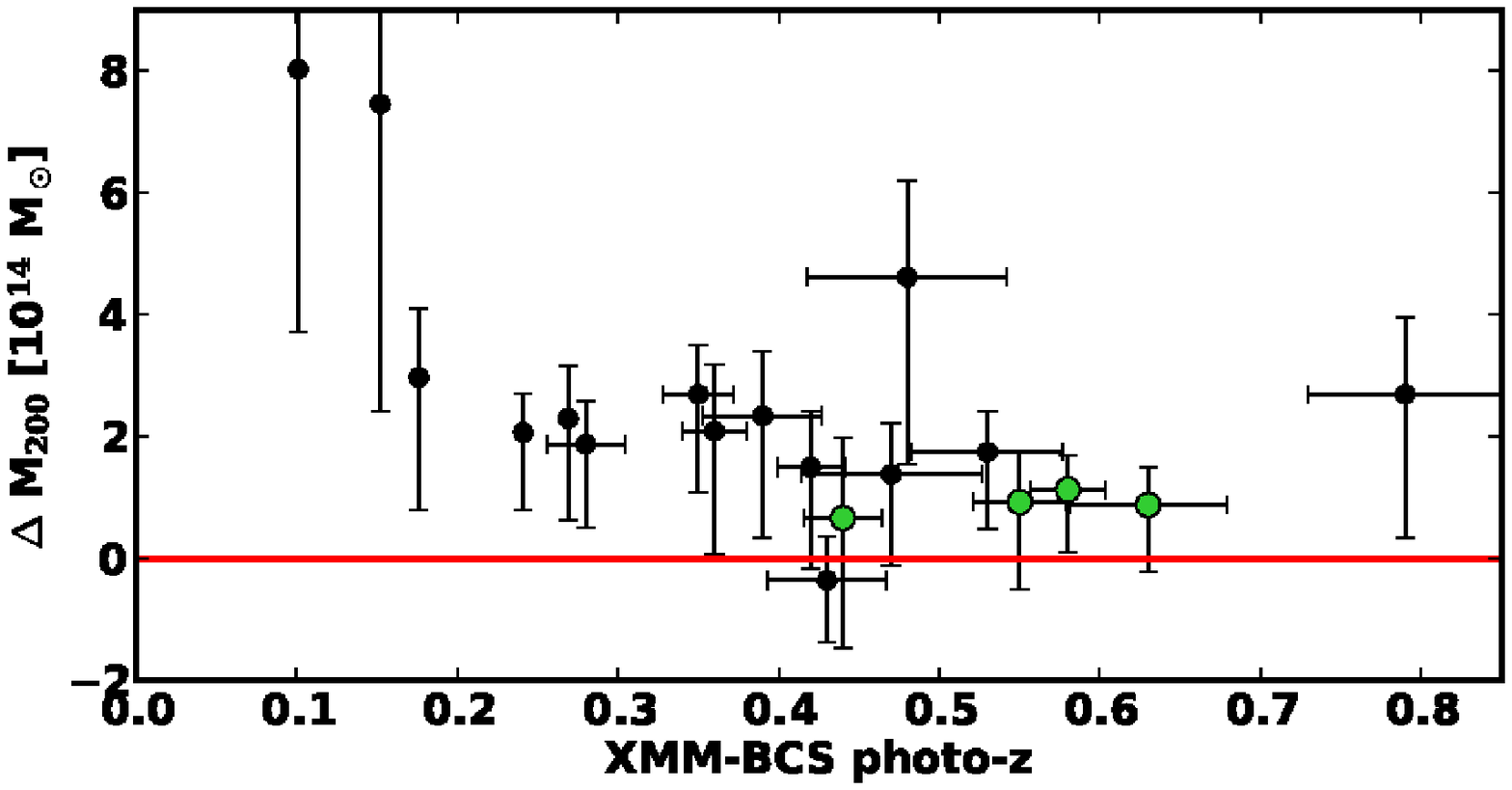}
\end{center}
\caption[Comparison of cluster parameters with the SCS
  sample]{\textbf{Top left:} Comparison of photometric redshifts for
  the 19 common clusters from our sample and the SCS cluster survey
  from \citet{menanteau09} and \citet{menanteau10}.  \textbf{Top
    right:} Comparison of masses for the same cluster sample in the
  r$_{200}$ aperture determined from the measured X-ray luminosity
  through scaling relations (x-axis) and the mean of the optically
  determined masses M(L$_{200}^{\mathrm{opt}}$) and M(N$_{200}$)
  (Table~\ref{tab:menan_xcounter}). The red line marks equality in
  both top panels.  \textbf{Bottom left:} Photo-z difference $\Delta=$
  photo-z(SCS) $-$ photo-z(XMM-BCS) as a function of our estimates of
  redshifts.  \textbf{Bottom right:} M$_{200}$ difference $\Delta=$
  M$_{200}$(SCS) $-$ M$_{200}$(XMM-BCS) as a function of our estimates
  of redshifts. The optical masses are significantly higher than the
  X-ray estimates especially at the low and high redshift ends. See
  text for discussion. Green points in all plots mark clusters from
  M10, black points those from M09.
}
\label{fig:scsphotoz}
\end{figure*}

\begin{table*}
 \centering
 \caption[Cross-matched radio sources]{Radio sources within
   $60^{\prime\prime}$ from the X-ray centers of the detected
   clusters. The quoted flux density S is at 843 Mhz (36 cm) for the
   SUMSS sources and at 4.85 GHz (6.2 cm) for the PMN detected object.
   The radio counterparts were obtained from the NASA/IPAC
   Extragalactic Database.}
\vspace{0.1cm}
 \label{tab:radio}
\begin{tabular}{rlcccc}
\hline
\hline
  \multicolumn{1}{c}{ID} &
  \multicolumn{1}{c}{Object Name} &
  \multicolumn{1}{c}{R.A. (deg)} &
  \multicolumn{1}{c}{Dec (deg)} &
  \multicolumn{1}{c}{S (mJy)} &
  \multicolumn{1}{c}{separation ($^{\prime\prime}$)} \\
\hline
018 & SUMSS J232952-560723 & 352.4677 & -56.1231 & $14.9 \pm 0.8$ & 56\\
035 & SUMSS J233345-553817 & 353.4416 & -55.6382 & $16.7 \pm 0.8$ & 6\\
044 & SUMSS J231654-545406 & 349.2274 & -54.9019 & $8.0  \pm 1.0$ & 14\\
109 & SUMSS J232737-541622 & 351.9047 & -54.2730 & $26.1 \pm 1.2$ & 9\\
110 & SUMSS J233003-541424 & 352.5146 & -54.2402 & $13.3 \pm 1.1$ & 6\\
189 & SUMSS J233044-560123 & 352.6860 & -56.0233 & $15.1 \pm 0.8$ & 36\\
210 & SUMSS J233406-554708 & 353.5253 & -55.7857 & $7.9  \pm 0.7$ & 3\\
288 & SUMSS J233459-545535 & 353.7495 & -54.9265 & $41.5 \pm 1.6$ & 37\\
426 & SUMSS J232138-541849 & 350.4092 & -54.3137 & $12.7 \pm 1.9$ & 14\\
534 & SUMSS J232446-552432 & 351.1951 & -55.4089 & $41.1 \pm 1.5$ & 17\\
546 & SUMSS J233113-543025 & 352.8076 & -54.5071 & $25.6 \pm 1.3$ & 26\\
150 & PMN J2330-5436       & 352.5075 & -54.6097 & $52.0 \pm 8.0$ & 34\\
\hline
\end{tabular}
\end{table*}

\subsection{Cross-correlation with the Southern Cosmology Survey clusters}
\label{sec:menan}

The Southern Cosmology Survey (SCS) carried out an optical cluster
search using the Blanco Cosmology Survey imaging data.
\citet[][hereafter M09]{menanteau09} provided a catalog of optically
selected clusters with photo-z $<0.8$ in a 8~deg$^2$ field partially
overlapping with the 6 deg$^2$ region presented in this
work. \citet[][M10]{menanteau10} then followed up this work by
creating a cluster catalog using the full $2005-2007$ BCS survey data
(i.e.  $\sim70$~deg$^2$, thus fully covering also the whole XMM-BCS
field), detecting 105 clusters with $M_{200}>3\times
10^{14}$~M$_{\odot}$ and photo-z$<0.8$.

Combining both these catalogs, we find in total 30 SCS clusters whose
optical coordinates lie inside our 6~deg$^2$ region.\footnote{ In the
  present work we thus do not consider SCS clusters that lie only
  partially in our 6~deg$^2$ region or in the 8~deg$^2$ mosaic
  extension.}  Out of these 30 systems, 26 come from the M09 catalog
(which contains clusters also below the mass limit applied in M10),
two are included in both M09 and M10 and two additional clusters are
from M10.  For the two clusters in both M09 and M10 we will use the
updated parameters from M10.

The SCS catalog provides the BCG coordinates for each system while our
catalog lists the X-ray centroids. For cross-correlation of the two
catalogs we take a conservative $60^{\prime\prime}$ matching radius,
which yields 19 clusters.  We summarize the properties of matched
clusters in Table~\ref{tab:menan_xcounter} and provide a more detailed
comparison of their parameters in the next two sections.

\subsubsection{Comparison of photometric redshifts}
First, we turn to the comparison of the photometric redshifts for the
19 matches. M09 and M10 utilize the BPZ code \citep{benitez00} to
estimate photo-zs while our method is based on the red-sequence method
as described in Sect.~\ref{sec:photoz}. For the SCS clusters we use
the photo-z errors published in \citet{mcinnes09} where possible (the
M09 and M10 catalogs do not provide error bars). For the remaining
cases we assume a $15\%$ error, which is the mean precision of the
photo-zs where errors are available.

As can be seen in Fig.~\ref{fig:scsphotoz} (top left) there is no case
of catastrophic disagreement. We find a gap in the SCS photo-z
distribution in the $0.35-0.5$ photo-z range that is not present in
our redshift distribution.  The most important feature is, however,
the systematic offset between the photo-z estimates.  The SCS photo-zs
are on average $\sim20\%$~lower than our values. This trend roughly
holds in the whole redshift range, as can be seen from the photo-z
residuals plotted against redshift in Fig.~\ref{fig:scsphotoz}, bottom
left). We found a similar bias when comparing the SCS photo-zs to the
spectroscopic subsample in Sect.~\ref{sec:speczcompare}.  For the five
systems with spectroscopic redshifts the photo-zs were on average
underestimated by $\sim19\%$.

In order to investigate potential sources of the discrepancy we check
whether the photo-z residuals depend on any of the available
parameters, most importantly the richness parameter $N_{200}$,
integrated optical luminosity $L_{200}^{\mathrm{opt}}$ and the
BCG-X-ray centroid offset. However, we do not find any statistically
significant dependence.

\subsubsection{X-ray - optical mass comparison}

M09 and M10 provide rough mass estimates for their clusters based on
the optical proxies $N_{200}$ and $L_{200}^{\mathrm{opt}}$ using the
scaling relations from \citet{reyes08}. The richness estimator
$N_{200}$ is defined as the number of E/S0 ridgeline cluster members
brighter than $0.4L^*$. The integrated cluster luminosity
$L_{200}^{\mathrm{opt}}$ is the summed $r$ band luminosity of the
member galaxies included in the $N_{200}$ calculation. Both parameters
are calculated within an aperture where the \emph{galaxy} density
equals $200/\Omega_{\mathrm{M}}$ times the \emph{mean} density of
galaxies in the Universe. Fortunately, \citet{johnston07} found that
this aperture is an unbiased estimate of the radius where the
\emph{matter} density is 200 times the \emph{critical} density of the
Universe, i.e. the optical masses and our X-ray estimates come from
roughly the same apertures and can thus be directly compared.

\begin{table*}
\centering
\caption[Comparison with the SCS sample]{The 19 SCS clusters recovered
  in the XMM-BCS survey. SCS References: M09 - \citet{menanteau09},
  M10 - \citet{menanteau10}. The optical masses
  $M(L_{200}^{\mathrm{opt}})$ and $M(N_{200})$ are taken from M09 and
  M10, the weak lensing mass measurements are provided by
  \citet{mcinnes09}.  The X-ray mass estimates obtained in the present
  work are taken from Table~\ref{tab:physpar}. The ID of the X-ray
  counterpart and its distance from the BCG are listed in the last two
  columns. The masses are in units of $10^{13}$~M$_{\odot}$.}
\vspace{0.1cm}
\label{tab:menan_xcounter}
\begin{tabular}{ccccccccc}
\hline
\hline
   \multicolumn{1}{c}{SCS ID} &
   \multicolumn{1}{c}{Ref.} &
   \multicolumn{1}{c}{photo-z (SCS)} &
   \multicolumn{1}{c}{M(L$_{200}^{\mathrm{opt}}$)} &
   \multicolumn{1}{c}{M(N$_{200}$)} &
   \multicolumn{1}{c}{M$_{200}^{\mathrm{wl}}$} &
   \multicolumn{1}{c}{M$_{200}^{\mathrm{X}}$} &
   \multicolumn{1}{c}{XMM-BCS ID} &
   \multicolumn{1}{c}{separation ($^{\prime\prime}$)} \\
 \hline
  SCSO J233430.2-543647.5 & M09  & 0.35 & 36  &  61   & $-                   $ &           $ 14.8 \pm 4.1 $ & 357  & $ 26.6  $ \\
  SCSO J232211.0-561847.4 & M09  & 0.61 & 56  &  46   & $4.7_{-4.7}^{+26.1}     $ &        $ 19.0 \pm 5.4 $ & 527  & $ 1.6  $ \\
  SCSO J232540.2-544430.9 & M09  & 0.10 & 21  &  86   & $2.3_{-2.3}^{+8.9}      $ &        $ 5.7 \pm 1.6 $ & 476  & $ 3.2  $ \\
  SCSO J232230.9-541608.3 & M09  & 0.12 & 16  &  100  & $8.5_{-5.9}^{+9.2}      $ &        $ 25.4 \pm 6.9 $ & 070  & $ 0.6  $ \\
  SCSO J233000.4-543707.7 & M09  & 0.14 & 12  &  43   & $-                   $ &           $ 13.3 \pm 3.6 $ & 150  & $ 1.7  $ \\
  SCSO J232419.6-552548.9 & M09  & 0.18 & 12  &  25   & $<2.6                $ &           $ 4.3 \pm 1.3 $ & 547  & $ 1.0  $ \\
  SCSO J233215.5-544211.6 & M09  & 0.20 & 17  &  33   & $10.2_{-6.1}^{+8.4 }    $ &        $ 10.0 \pm 2.8 $ & 511  & $ 10.0  $ \\
  SCSO J233037.1-554338.8 & M09  & 0.20 & 10  &  27   & $16.2_{-7.7}^{+10.7}    $ &        $ 8.2 \pm 2.3 $ & 034  & $ 2.3  $ \\
  SCSO J232200.4-544459.7 & M09  & 0.27 & 17  &  39   & $2.6_{-2.6}^{+8.6 }     $ &        $ 18.1 \pm 5.0 $ & 136  & $ 3.9  $ \\
  SCSO J233522.6-553237.0 & M09  & 0.29 & 22  &  32   & $8.5_{-8.5}^{+16.0}     $ &        $ 5.1 \pm 1.6 $ & 528  & $ 17.4  $ \\
  SCSO J232956.0-560808.3 & M09  & 0.32 & 20  &  39   & $21.3_{-17.3}^{+27.5}$    &        $ 15.6 \pm 4.3 $ & 018  & $ 1.7  $ \\
  SCSO J232839.5-551353.8 & M09  & 0.32 & 10  &  17   & $16.9_{-13.2}^{+20.1}$    &        $ 20.6 \pm 5.8 $ & 088  & $ 36.2  $ \\
  SCSO J232633.6-550111.5 & M09  & 0.32 & 28  &  32   & $<4.8               $  &           $ 17.0 \pm 4.7 $ & 126  & $ 3.1  $ \\
  SCSO J233003.6-541426.7 & M09  & 0.33 & 9   &  29   & $28.1_{-14.7}^{+20.7}$    &        $ 15.2 \pm 4.2 $ & 110  & $ 7.4  $ \\
  SCSO J232619.8-552308.8 & M09  & 0.52 & 12  &  21   & $28.1_{-22.2}^{+33.4}$    &        $ 12.2 \pm 3.5 $ & 082  & $ 9.5  $ \\
  SCSO J231651.0-545356.0 & M10  & 0.36 & 27  &  39   & $-                  $  &           $ 32.4 \pm 8.9 $ & 044  & $ 24.7  $ \\
  SCSO J232856.0-552428.0 & M10  & 0.57 & 35  &  20   & $-                  $  &           $ 8.7 \pm 2.6 $ & 090  & $ 6.7  $ \\
  SCSO J233420.0-542732.0 & M10  & 0.56 & 36  &  27   & $-                  $  &           $ 17.7 \pm 5.0 $ & 158  & $ 41.6  $ \\
  SCSO J233556.0-560602.0 & M10  & 0.64 & 47  &  25   & $-                  $  &           $ 7.5 \pm 2.3 $ & 386  & $ 31.3  $ \\
  \hline
\end{tabular}
\end{table*}

In Fig.~\ref{fig:scsphotoz} (top right) we compare our X-ray masses
with the optical masses $M(N_{200})$ calculated from the $N_{200}$
parameter.  The optical masses are estimated to be accurate within a
factor of two (M09), where this factor should include also the
uncertainty in extrapolating the \citet{reyes08} scaling relations to
higher redshifts (the scaling relations were calibrated for redshifts
$z<0.3$).  We used the factor two uncertainty to calculate the
$M(N_{200})$ error bars in Fig.~\ref{fig:scsphotoz}.  We find that the
optical masses are significantly higher than the X-ray mass estimates
$M_{200}$ by a factor of $\sim2.6$ (median value).

\citet[][]{reichert11} investigate X-ray luminosity based
scaling relations on a large compilation of cluster samples from the
literature. They find only very few systems deviating from the mean
$L-M$ relation by more than a factor two (i.e. with actual mass two or
more times higher than the luminosity prediction). We thus do not
expect our X-ray masses to be underestimated by similar factors even
in individual cases. The observed bias in mass goes in the opposite
direction as that found in the photo-zs (i.e. photo-zs were
underestimated while masses overestimated).  The photo-zs, however
influence the mass estimates and therefore it is not straightforward
to disentangle all the factors contributing to this discrepancy.  The
influence of the redshift uncertainty is likely more important for
nearby systems, where it translates to larger differences in the
angular size of the aperture.  The discrepancy for the
$M(L_{200}^{\mathrm{opt}})$ masses is similar. We note here, however,
that the $M(L_{200}^{\mathrm{opt}})$ masses in M09 were obtained from
the scaling relations of \citet{reyes08} prior to their
erratum-correction.\footnote{Scaling relations with the updated
  coefficients are available at: \texttt{arxiv.org/abs/0802.2365}.}

The bottom right panel of Fig.~\ref{fig:scsphotoz} displays the mass
residuals versus our photometric redshifts. We find an anticorrelation
between redshift and the mass residuals (using a Spearman's rank
correlation coefficient, neglecting the error bars). Given that the
\citet{reyes08} relations are calibrated only out to $z\lesssim0.3$ it
is however impossible to either confirm the presence of such an
anticorrelation or its potential causes.We also note that the four
clusters from M10 agree with our measurements better than all but one
cluster from M09.

\begin{figure}[t!]
\begin{center}
    \includegraphics[width=0.5\textwidth]{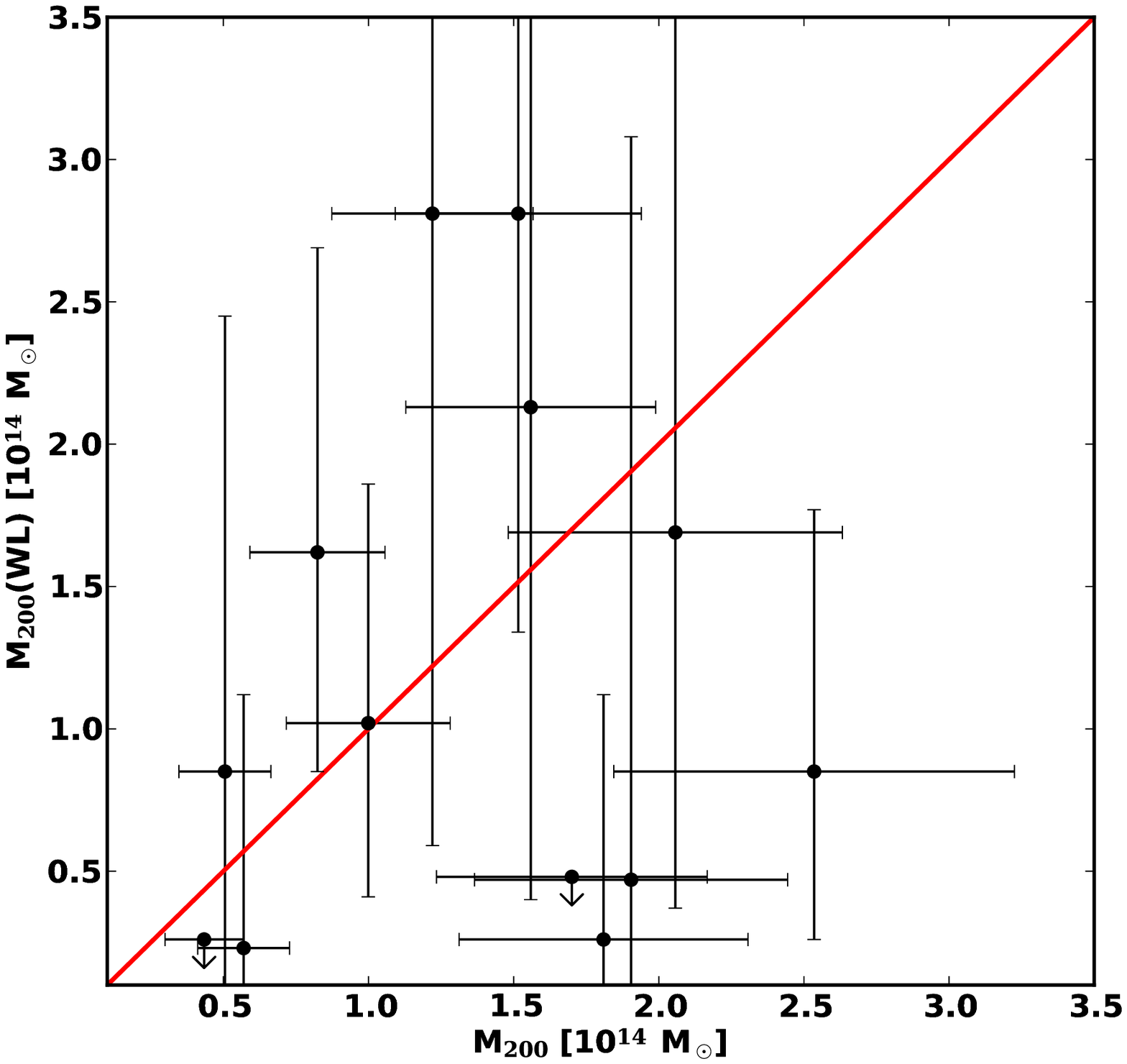}
\end{center}
\caption[Comparison of X-ray and weak lensing masses]{Comparison of
  X-ray masses ($M_{200}$, x-axis) with the weak lensing measurements
  ($M(\mathrm{WL})$, y-axis) from \citet{mcinnes09} for 13 clusters in
  our sample. Although the scatter and uncertainties are large, the
  agreement is considerably better than with the optical masses
  $M(N_{200})$ displayed in Fig.~\ref{fig:scsphotoz}.  }
\label{fig:scswl}
\end{figure}

\citet{mcinnes09} provides weak lensing mass measurements for the
clusters from M09 among which 13 clusters are also in our sample (for
two of these systems only upper limits could be set).  We compare the
weak lensing masses with our X-ray estimates in
Fig.~\ref{fig:scswl}. The agreement is significantly better than for
$M(N_{200})$ masses, although the scatter and uncertainty in the weak
lensing mass estimates are large (mostly due to the limited depth of
the optical data, but also influenced by the uncertainty in the
photo-zs). From their full sample, \citet{mcinnes09} also noted that
the $M(L_{200}^{\mathrm{opt}})$ seems to overestimate the total mass
compared to the weak lensing estimates.

An in-depth comparison of optical and X-ray masses will be addressed
in an upcoming work, where we will provide also our own measurements
of $N_{200}$ and $L_{200}^{\mathrm{opt}}$ (Song et al., in
prep.). This will allow us to properly investigate the presence of
potential biases in the different mass estimators methods and
calibrate our own relations.

\subsubsection{Parameter upper limits for X-ray non-detections}
\begin{table*}
\centering
\caption[X-ray upper limits for the SCS non-detections]{The 11 SCS
  clusters from \citet{menanteau09} that lie in the XMM-BCS core
  survey but have no X-ray cluster detection, for which we provide
  flux and mass upper limits.}
\vspace{0.1cm}
\label{tab:menan_fluxlim}
\begin{tabular}{ccccccc}
\hline
\hline
  \multicolumn{1}{c}{SCS ID} &
  \multicolumn{1}{c}{photo-z} &
  \multicolumn{1}{c}{f$^{lim}_X$} &
  \multicolumn{1}{c}{L$^{lim}_{500}$} &
  \multicolumn{1}{c}{M$^{lim}_{500}$} \\
& (SCS) &($10^{-14}$~erg s$^{-1}$ cm$^{-2}$) &  ($10^{43}$~erg s$^{-1})$ & $(10^{13}$ M$_{\odot}$) \\
\hline
SCSO J232829.7-544255.4  & 0.68 &   0.79 &   1.6 &   5.5  \\
SCSO J233106.9-555119.5  & 0.19 &   0.69 &   0.1 &   1.3  \\
SCSO J233550.6-552820.4  & 0.22 &   0.68 &   0.1 &   1.6  \\
SCSO J232156.4-541428.8  & 0.33 &   0.88 &   0.3 &   2.8  \\
SCSO J233231.4-540135.8  & 0.33 &   1.59 &   0.6 &   4.0  \\
SCSO J233110.6-555213.5  & 0.39 &   0.73 &   0.4 &   3.1  \\
SCSO J233618.3-555440.3  & 0.49 &   1.51 &   1.4 &   5.9  \\
SCSO J232215.9-555045.6  & 0.56 &   0.84 &   1.1 &   4.8  \\
SCSO J232247.6-541110.1  & 0.57 &   0.72 &   0.9 &   4.4  \\
SCSO J232342.3-551915.1  & 0.67 &   1.22 &   2.4 &   7.0  \\
SCSO J233403.7-555250.7  & 0.71 &   0.56 &   1.3 &   4.7  \\

\hline
\end{tabular}
\end{table*}

For the 11 SCS clusters that lie in the core area of our survey but
have no X-ray counterparts we provide X-ray flux upper limits in
Table~\ref{tab:menan_fluxlim} and mass limits using the SCS photo-z
values.

The flux limits were calculated using the same procedure as we used
for the survey sky coverage calculation (Sect.~\ref{sec:skycov}, i.e.
calculating the minimal flux needed for the source to be detected at
the given position and our detection threshold).

The flux was then converted to luminosity using the photometric
redshift from either M09 or M10. We calculated the mass upper limits
from the $L-M$ scaling relation as detailed in
Sect.~\ref{sec:physpar}. The obtained upper limits on the mass are
considerably lower than the M09/M10 estimates.

We also check for possible miss-classification (or confusion if a
central AGN is present) by cross-correlating the positions of these 11
clusters with our X-ray \emph{point} source catalog with a threshold
of $16^{\prime\prime}$. In this aperture we find no matches. The
non-detection of these sources is due to either their low X-ray fluxes
or a spurious detection.

\section{Discussion}
\label{sec:discuss}
In this section we discuss the additional effects that influence the
precision of the physical parameters provided in our catalog. We also
give an outlook on the upcoming work in the context of the XMM-BCS
survey.

\subsection{Error budget of the X-ray analysis}
\label{sec:systematics}

For the present catalog, we restricted ourselves to provide only
formal statistical errors for the estimated parameters
(Table~\ref{tab:physpar}) that include the Poisson errors of the flux
measurement, a $5\%$ systematic error from the background modeling and
the intrinsic scatter of the scaling relations. Although we used the
bolometric luminosity to calculate further physical parameters, here
we assumed the intrinsic scatter found in the $0.5-2$~keV luminosity
relations. This scatter is slightly larger than the bolometric one and
it gives a more realistic error estimate since the band luminosity
is, in fact, our only direct observable, while the temperature
required for the bolometric correction is not. We determine physical
parameters with the following precision (mean across the whole
redshift and flux ranges): flux and luminosity to $\sim16\%$, T$_{500}$
and M$_{500}$ to $\sim30\%$, and Y$_{500}$ to $\sim60\%$.

In this section we discuss several additional sources of systematic
errors and their impact on the estimated fluxes and other parameters.
All below reported relative errors are obtained by averaging over the
whole cluster sample. Several of the considered effects are redshift
dependent, but we typically allow broad parameter ranges and thus our
uncertainty estimates are rather conservative.

\textbf{1)} Good precision photometric redshifts are crucial for the
determination of each physical parameter. Photo-z estimates in the
present work have a mean error of $\sim10\%$ and show good agreement
with the available spectroscopic measurements
(Sect.~\ref{sec:speczcompare}). In order to estimate the impact of the
photo-z uncertainty on the measured physical parameters we offset the
redshifts (Table~\ref{tab:physpar}) by their $1\sigma$ errors to both
sides and rerun the iterative physical parameter estimation procedures
(see Sect.~\ref{sec:physpar}).

\begin{figure*}
\begin{center}
\includegraphics[width=0.49\textwidth]{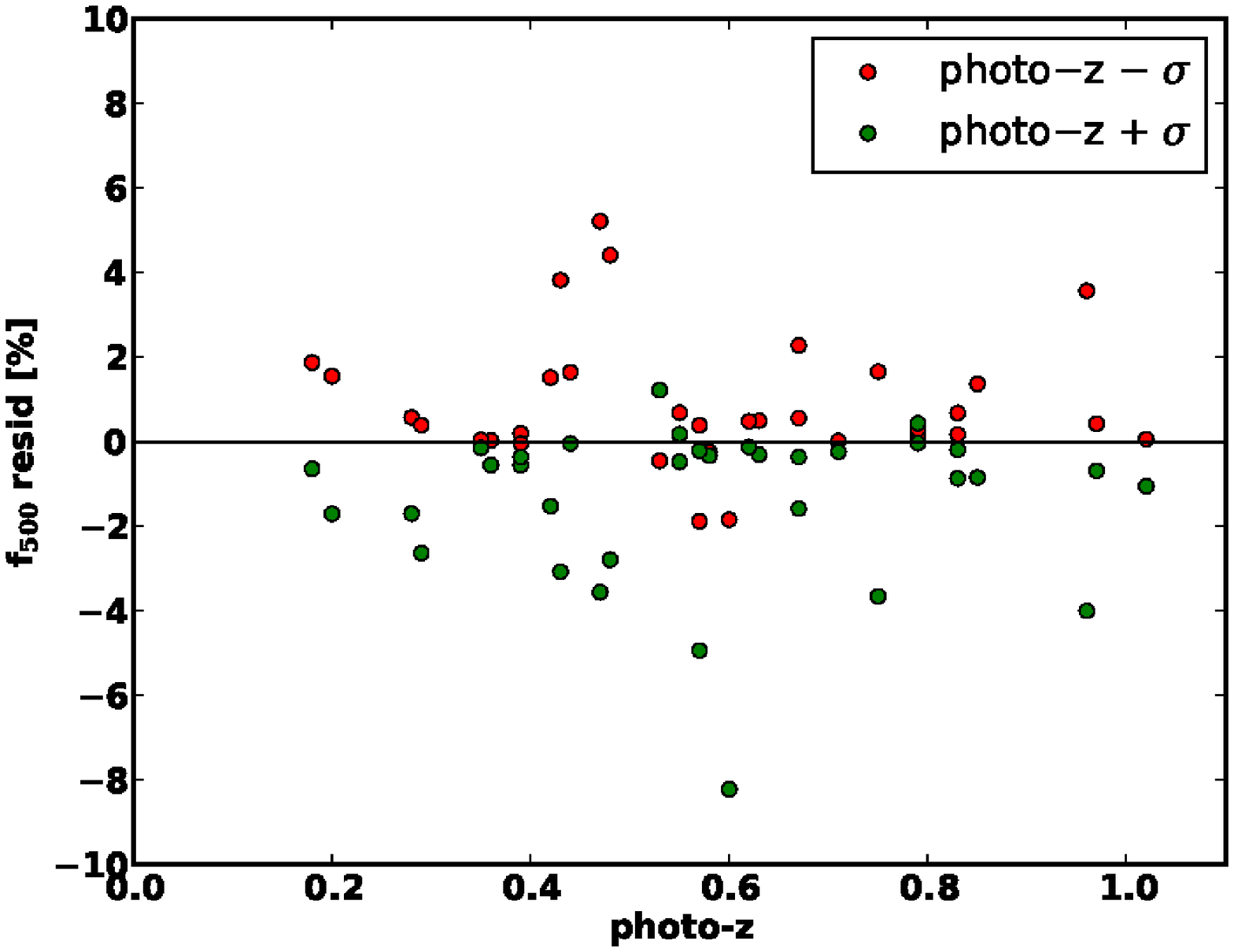}
\includegraphics[width=0.49\textwidth]{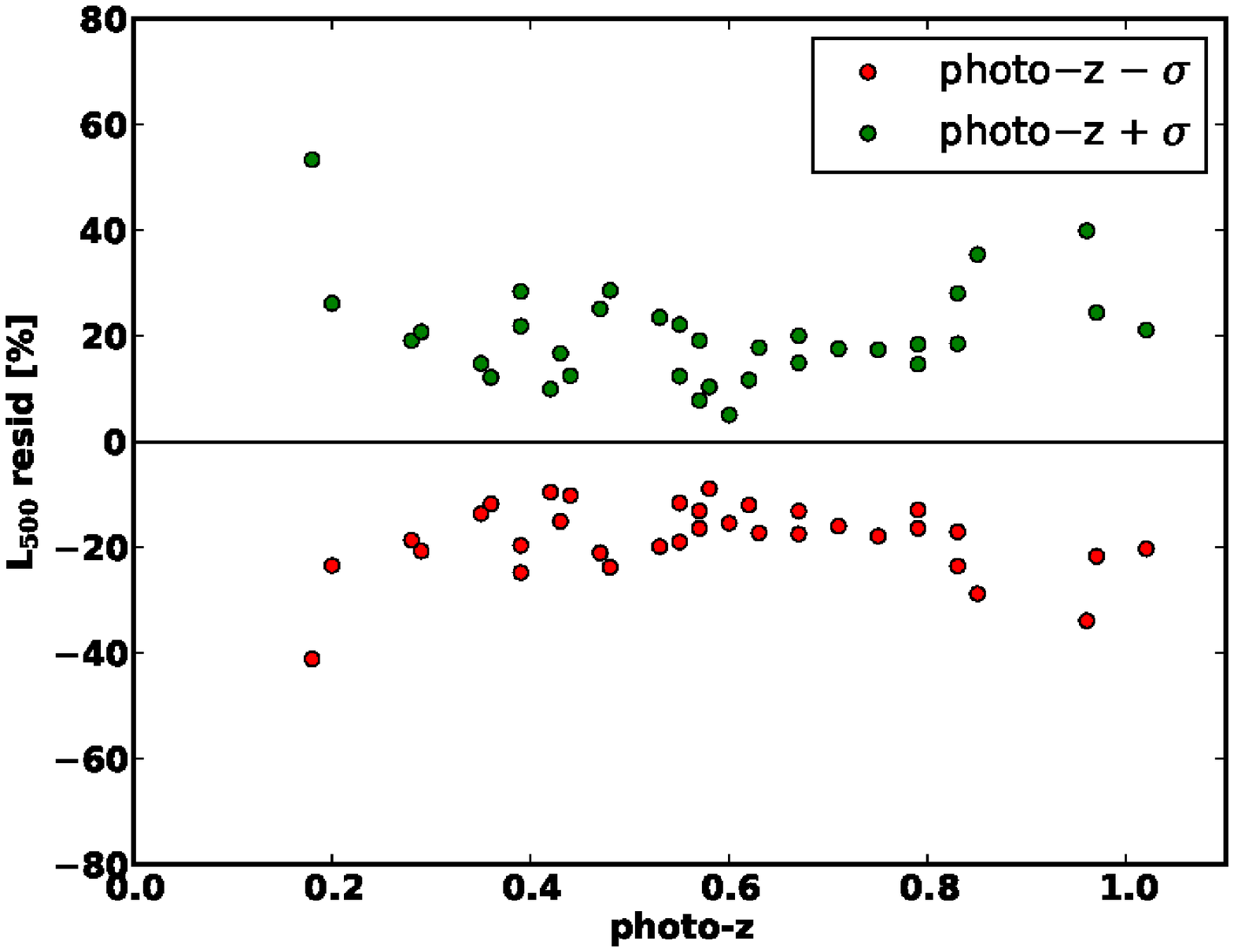}
\end{center}


\caption[Flux sensitivity on photo-z uncertainty]{\textbf{Left:} The
  effect of the photometric redshift uncertainty on the determined
  flux in the $r_{500}$ aperture. \textbf{Right:} The effect of the
  photometric redshift uncertainty on the determined luminosity in the
  $r_{500}$ aperture.  Green points mark the relative difference of
  flux (luminosity) for photo-z increased by $1\,\sigma$ compared to
  the mean value.  Red points are for the case when we decrease the
  photo-zs by the same amount.}
\label{fig:photozuncertain}
\end{figure*}

We find that, for the flux f$_X(<r_{500})$, all values are consistent
within their $1\,\sigma$ uncertainty and for most clusters the
relative difference is below the $\sim2\%$ level
(Fig.~\ref{fig:photozuncertain}, left). Change in the photo-z affects
the flux in a complex way - it is entering directly the
energy-conversion factor (ECF) calculation (lower redshift leads to a
lower ECF), and also during the iterative process through the scaling
relations, which then feed back into the aperture size itself as well
as the temperature which again affects the ECF value.  This complex
dependence explains the scatter of the flux residuals in
Fig.~\ref{fig:photozuncertain}, leading to different convergence
points for different input photo-zs. Interestingly, a lower photo-z
value leads to a \emph{higher} flux in the $r_{500}$ aperture. The
reason is that the direct effect of decreasing the photo-z would be to
lower temperature and mass and thus also reduce the $r_{500}$
value. However, the redshift dependence of the angular distance is
stronger and thus the \emph{angular} size of the $r_{500}$ aperture is
actually larger for lower redshifts, which leads to the increase in
the f$_X(<r_{500})$ values (we confirm this explanation by checking
the flux in fixed sky apertures).

For luminosities the photo-z errors translate into a $\sim20\%$
uncertainty (Fig.~\ref{fig:photozuncertain}, right). Here the
dependence is dominated by the cosmological redshift dimming and thus
higher redshifts yield also higher luminosities.  If we now use the
perturbed redshift and luminosity values to recalculate temperatures
and masses, we find that the $T_{500}$ values vary on the
$\sim5\%$~level, while for $M_{500}$ the uncertainty is on the
$\sim7\%$~level.

\textbf{2)} In the present work, we have utilized the bolometric
luminosity scaling relations of \citet{pratt09} based on the REXCESS
cluster sample \citep[][]{boehringer07}.  These scaling relations are
best suited for our purposes for several reasons. They were derived
from XMM-\emph{Newton} observations (removing possible calibration
issues between relations derived from different instruments) of a
representative cluster sample \citep[REXCESS,][]{boehringer07}. The
sample covers a great range of the cluster luminosity function without
a bias towards a morphological structure type (like e.g. presence of a
cooling core or merging activity). By using the bolometric relations
we can also utilize the results of \citet[][]{reichert11} to
estimate the effect of non-self-similar evolution on the estimated
parameters (see below).  Additionally, the $L-M$ relation is based on
the $L-Y_{\mathrm{X}}$ and $Y_{\mathrm{X}}-M$ scaling relation, which
is found to be more robust than previous direct $L-M$ calibrations
\citep{arnaud07}.

The direct application of these scaling relations, however requires
extrapolations both to higher redshifts (the REXCESS cluster sample
includes only local clusters with $z\lesssim0.2$) and to the low-mass
regime of groups of galaxies. The physical parameters provided in
Table~\ref{tab:physpar} were obtained by assuming the redshift
evolution of the cluster scaling relations to be self-similar. This is
a standard assumption supported by predictions of a purely
gravitationally driven cluster growth
\citep[e.g.][]{kaiser86}. However, there is increasing evidence, that
the evolution of the luminosity scaling relations is slower than the
self-similar expectation (see \citet[][]{reichert11} and
references therein). We test the influence of this assumption by using
the simplified approach proposed by \citet{fassbender11a} by removing
a part of the self-similar evolution factor from the relations (the
$L-T$ "no evolution" scenario). This approach is consistent with the
more detailed analysis of Reichert et al. In this picture, the
predicted temperatures are on average higher by $9\%$ and masses by
$15\%$ compared to the self-similar scaling relations. At the high
redshift end ($z>0.8$) this effect is even more important ($\sim20\%$
and $30\%$ increase, respectively), while for $z<0.2$ the effect is
less than $5\%$.

The cluster sample of \citet{vikhlinin09} covers a similar redshift
range as our sample ($0-0.9$) and extends in mass down to roughly the
median mass of our sample ($\sim1\times10^{14}$~M$_{\odot}$). Using
the $L-M$ relation from this samples gives masses only slightly higher
(by $\sim5\%$, standard deviation $\sim13\%$) than our results. This
difference grows with redshift due to the deviation of the evolution
index from the self-similar value (\citet{vikhlinin09} obtained
$E(z)^{1.61}$).

Pertaining the assumptions that the REXCESS scaling relations hold
also bellow M$_{500}\approx1\times10^{14}$~M$_{\odot}$,
\citet[][]{sun11} found very good agreement between the pressure
profiles of their sample of groups and those derived from
REXCESS. \citet{leauthaud10} used the groups detected in the COSMOS
survey \citep{finoguenov07} to calibrate the $L-M$ scaling relation
with weak lensing mass measurements. Their evolution factor is
consistent with the one found by \citet[][]{reichert11},
i.e. giving only slightly higher masses than the self-similar
scenario. These findings give us an indication that the scaling
relations employed in the present work are indeed adequate also in the
low-mass clusters/groups regime.

\textbf{3)} The Galactic hydrogen column densities reported by the LAB
HI survey \citep{lab} are systematically lower by $\sim27\%$ than the
\citet{dl} values in the whole survey area.  The effect on the derived
luminosities is, however, only marginal ($\sim1.6\%$).

\textbf{4)} In order to quantify the effect of possible deviations of
the cluster metallicity from the mean value of 0.3 solar, we bracket
the possible metallicities in the very conservative range of (0.1,
0.6) solar. The corresponding range of $L_{\mathrm{X}}(<r_{500})$
deviations from the fiducial value (for 0.3 solar metallicity) is
\protect{$(1.6\%,-1.5\%)$}, i.e. lower metallicities lead to higher
luminosities and vice versa.

\textbf{5)} We also test the quality of the flux extrapolation
correction described in Sect.~\ref{sec:physpar}. The correction
coefficients are calculated by integrating a beta model between
($r_{plat}$, $r_{500}$), if $r_{plat} < r_{500}$. We use
Eq.~\ref{eq:beta} to estimate the $\beta$ and $r_{C}$
parameters. Alternatively, we can use the canonical value $\beta=2/3$
and $r_{C}$ obtained from the maximum likelihood fit in the source
detection step. The two extrapolation methods give fluxes
(luminosities) differing on average for the whole sample by
$~2\%$. For individual objects the relative difference of fluxes is
clearly correlated with the amount of extrapolation needed and is
roughly of the size of the correction itself. This means that the
extrapolation is currently only very weakly constrained and thus
highly uncertain. Fortunately, for most of our sample the required
extrapolation factor is small.

\textbf{6)} The combined MOS1 and MOS2 counts are converted to flux
and luminosities using the MOS2 response matrix (see
Sect.~\ref{sec:physpar}). We have chosen the MOS2 response matrix over
the MOS1, because some sources lie on the missing MOS1 CCD\#6, where
no meaningful response matrix can be calculated.  If the MOS1 response
is used instead, the luminosities obtained purely from the combined
MOS detectors are on average lower by $2\%$ (excluding clusters
detected on the position of the missing MOS1 chip). The final
$L_{\mathrm{X}}(<r_{500})$ calculated as the weighted average of the
individual PN and MOS luminosities is affected by less than $1.3\%$.

\textbf{7)} The response matrices used in our analysis are calculated
for a fixed radius of $150$~arcsec. This range is roughly the average
extraction radius of our clusters (i.e. from which growth curves are
extracted and local background estimated). We calculate response
matrices for two additional radii - 60 arcsec and 240 arcsec, to check
how the spatial averaging of the spectral response impacts the derived
ECFs and thus flux and luminosity. In this very conservative range of
extraction radii we found the average effect to be of the order of
$~2.5\%$.

\textbf{8)} The uncertainties in the absolute normalization of the
effective area of the detectors decrease the flux measurement
precision.  \citet{nevalainen10} found an agreement between
$0.5-2$~keV fluxes measured by PN and both MOS cameras to be better
than $\sim5-7\%$ and for ACIS on \emph{Chandra} and the PN found the
fluxes to differ only by $2\%$.

\textbf{9)} We have tested the impact of using the up-to-date version
of the SAS package (SAS v. 11.0.0), with respect to the SAS v. 7.1.0
adopted for the analysis in this paper. In an end-to-end reprocessing
of a subsample of six clusters fully sampling our flux and off-axis
angle distributions we find on average only very small differences
($<4\%$ in flux, $<1.5\%$ in temperatures and $<2.5\%$ in
masses). These differences can get larger for low quality candidates
(e.g. quality flag 3 in Table~\ref{tab:flags}), but in all test cases
they were consistent within the error bars.

\textbf{10)} We also tried to run the physical parameter estimation
procedure (Sect.~\ref{sec:physpar}) from several initial values of
$T_{500}/r_{500}$. The iteration procedure always converged to the
same solution, confirming its independence from the starting values.

We will provide tests of photometric accuracy of the growth curve
method in a subsequent publication based on simulations using
realistic backgrounds (i.e. using our survey fields as background for
the simulated clusters). Presently we provide a comparison of the
X-ray photometry obtained by our pipeline in comparison with the
XMM-LSS project in Appendix~\ref{sec:lsscomp}.  In addition, our
algorithm was also applied to the clusters SPT-CL~J2332-5358 and
SPT-CL~J2342-5411 in \citet{suhada10} detected in the extension of the
XMM-BCS survey.  These sources have been independently analysed by
\citet{andersson10} using deeper pointed data (from a $19$~ks long
XMM-\emph{Newton} observation for the first cluster and from a
$134$~ks long \emph{Chandra} exposure for the
second). \citet{andersson10} find for SPT-CL~J2332-5358
($z_{\mathrm{photo}}=0.32$) an X-ray luminosity L$_{500} = 3.0 \pm 0.1
\times 10^{44}$~erg s$^{-1}$ and for SPT-CL~2342-5411
($z_{\mathrm{photo}}=1.08$) L$_{500} = 2.9 \pm 0.3 \times 10^{44}$~erg
s$^{-1}$ in good agreement with our values: L$_{500} = 2.7 \pm 0.2
\times 10^{44}$~erg s$^{-1}$ for SPT-CL~2332-5358 and L$_{500} = 2.9
\pm 0.3 \times 10^{44}$~erg s$^{-1}$ for SPT-CL~2342-5411
(luminosities in the $0.5-2$~keV band). Estimated temperatures and
masses are also consistent within error bars although the
uncertainties on these parameters are significant given the exposure
time in the \citet{suhada10} analysis was $\lesssim 3$~ks.

A proper understanding of a realistic error budget of a cluster sample
is crucial for its modelling in the cosmological context. From our
analysis we find that most effects are typically on the $\sim2\%$
level (under conservative assumptions) and the major contributing
factors are the uncertainty of the photo-z measurements and the
required extrapolations of the scaling relations (both in the range of
$5-30\%$ depending on the parameter and the redshift of the
system). For a few clusters an additional significant source of
uncertainty is connected with the flux extrapolation.  A full
self-consistent treatment of the error propagation (including their
full covariance matrices) and its impact on the cosmological modeling
of the sample will be addressed in subsequent work.

\subsection{Project outlook}
The present sample establishes the observational base of the X-ray
part of the XMM-BCS survey. In upcoming work we will use the available
multi-wavelength data to follow several lines of investigations, some
of which have already been initiated:

\begin{itemize}
\item The X-ray cluster catalog will be extended to cover the whole
  14~deg$^2$ area. The preliminary source catalog is already available
  and we will follow up this work by estimating the photometric
  redshifts and physical parameters for the clusters in the same way
  as presented in this work. The full cluster catalog is expected to
  comprise $\sim100$ clusters and groups of galaxies.

\item We will calculate the selection function based on Monte Carlo
  simulations developed by \citet{muehlegger10}. This analysis will
  allow us to construct a well controlled subsample from the full
  cluster catalog that will be suitable for cosmological modelling.

\item A more detailed analysis of optical properties of the clusters
  presented in this sample will be provided in Song et al., in
  prep. We will provide here measurements of the $N_{200}$ and
  $L_{200}^{\mathrm{opt}}$ parameters and investigate their mass
  scaling relations.

\item A detailed comparison of the X-ray, optical and mid-infrared
  cluster samples will allow us to gain good understanding of the
  selection function of each method. We will study the cluster/group
  population in this field and establish its multi-wavelength
  properties. The \emph{Spitzer} imaging data will also be used to
  improve the photometric redshift estimates, especially for distant
  systems with redshift $z\gtrsim0.8$.

\item We have initiated further X-ray-SZE studies based on a
  cooperation with the SPT collaboration. The current SPT cluster
  sample \citep{vanderlinde10,williamson11} includes only sources with
  minimal detection significance of $5\sigma \,(7\sigma)$,
  respectively. There are only two clusters in the 14~deg$^2$ above
  this threshold, SPT-CL~J2332-5358 and SPT-CL~J2342-5411, which are
  also independently detected in our survey \citep{suhada10}. Using
  our X-ray selected cluster catalog we can also safely investigate
  lower significance SPT detections.  As a first example, cluster ID
  044, (XBCS~231653.1-545413) was found to have a direct SPT detection
  at the $4.2\sigma$~level (B. Benson, private communication). Another
  approach is a stacking analysis of the SZE data for the X-ray
  selected clusters.  Here, a preliminary analysis of the top eleven
  clusters ranked by their X-ray predicted SPT detection significance
  yields a $\gtrsim6\sigma$ detection.  We will explore both
  approaches in more depth in upcoming work, but already now it is
  clear, that with a joint SZE and X-ray analysis we are able to
  explore a completely new mass regime in the SPT survey.

\item The multi-wavelength coverage of the field provides
  opportunities also for non-cluster science.  As an example, we have
  detected a total of 3065 X-ray point sources in the survey (1639 in
  the core region and 1426 in the extension). Most of these point
  sources are AGN and using the available multi-wavelength data we
  will be able to carry out a study with a focus on the obscured AGN
  population.
\end{itemize}

\section{Summary  and conclusions}
\label{sec:conclusions}
\begin{itemize}
\item
We have provided the analysis of the 6~deg$^2$ XMM-\emph{Newton} field
in the framework of the XMM-BCS survey. We have carried out X-ray
source detection and constructed a catalog of 46 clusters and groups
of galaxies.

\item
Based on four band optical imaging provided by the Blanco Cosmology
Survey we have confirmed that these X-ray detections are coincident
with overdensities of red galaxies. Using the red sequence method we
have measured the photometric redshifts of these systems.

\item
We have initiated a spectroscopic follow-up program by carrying out
long-slit spectroscopy observations using the EFOSC2 instrument at the
3.6~m NTT telescope at La Silla, Chile. We have obtained spectroscopic
redshifts for BCG galaxies in 12 clusters and in four cases also for
one additional member galaxy. This sample covers the redshift range
$0<z<0.4$ (i.e. roughly up to the median redshift of the sample) and
constitutes the first spectroscopic information for the field. We find
good agreement between our photometric estimates and the spectroscopic
values, but the spectroscopic sample has to be extended in redshift,
in order to be able to provide a rigorous calibration of the photo-zs.

\item
Using the redshift information we measured the X-ray luminosities for
our cluster sample. From luminosity scaling relations we estimate
their most important physical parameters, e.g. mass, temperature and
the $Y_{\mathrm{X}}$ parameter.  We discuss the influence of several
factors on the precision of the provided estimates. The uncertainty of
the photometric redshift estimates and the extrapolation of the
scaling relations to high redshift systems and into the group regime
are identified as the most important factors that determine the
overall errors in the physical parameters.  We verify our X-ray
parameter estimation method by analyzing the C1 sample of the XMM-LSS
survey \citep{pacaud07}. We find good agreement between the parameters
provided by both pipelines.

\item
The present sample of clusters and groups of galaxies covers the
redshift range from $z=0.1$ to redshift $z\approx1$ with a median
redshift of $z=0.47$.  The median temperature of the clusters is $\sim
2$~keV, and the median $M_{500}$ mass $9 \times 10^{13}$~M$_{\odot}$
(based on luminosity scaling relations). With our $\sim10$~ks
XMM-\emph{Newton} observations we are thus able to effectively probe
the cluster/group transition regime practically at all redshifts up to
$z\approx1$.

\item
We provide a preliminary, simplified calculation of the survey sky
coverage which does not require extensive Monte Carlo
simulations. Using this calculation we characterize our cluster sample
by its $\log N - \log S$ relation. We find good agreement with the
relations established by the RDCS survey \citep{rosati98}, 400deg$^2$
survey \citep{burenin07cccp1,vikhlinin09} and the XMM-LSS project
\citep{pacaud07}.
\item
We carried out first comparisons with optical studies available from
the Southern Cosmology Survey
\citep[SCS,][]{menanteau09,menanteau10}. In this preliminary
investigation we find the SCS photometric redshifts to be biased low
by $\sim20\%$ with respect to our estimates (both photometric and
spectroscopic, where available). We find a discrepancy between the
X-ray and optical mass estimates, with optical masses being
significantly higher.  We compare our masses to weak lensing mass
measurements available for 13 clusters in our sample from
\citet{mcinnes09}. Although the weak lensing mass uncertainties are
large, there is no statistical inconsistency between the two mass
estimators.
\end{itemize}

The presented results illustrate the potential of medium-deep, X-ray
surveys to deliver cluster samples for cosmological modelling. These
samples then in combination with available multi-wavelength data
(particularly in optical, near-infrared and SZE) will allow us to
probe the dependence of the selection functions on relevant cluster
observables and provide thus an important input for upcoming
large-area multi-wavelength cluster surveys.

\begin{acknowledgements}
We thank the referee for detailed comments on the manuscript. We are
thankful to Bradford Benson for providing the preliminary SPT
analysis.  We thank Stefania Giodini and Veronica Biffi for carrying
out GROND observations for several XMM-BCS clusters.  We thank Rodion
Burenin for providing the 400 deg$^2$ survey $\log N - \log S$
relation and Hermann Brunner for useful discussions.  We are thankful
to Martin Pan\v{c}i\v{s}in and Alexandra Wei\ss mann for their comments
on the manuscript.  RS acknowledges support by the DfG in the program
SPP1177.  HB acknowledges support for the research group through The
Cluster of Excellence 'Origin and Structure of the Universe', funded
by the Excellence Initiative of the Federal Government of Germany, EXC
project number 153. This research has been partially supported through
a NASA grant NNX07AT95G to UMBC.  DP acknowledges the kind hospitality
of the Max-Planck-Institute for extraterrestrial Physik.  This
research has made use of the NASA/IPAC Extragalactic Database (NED)
which is operated by the Jet Propulsion Laboratory, California
Institute of Technology, under contract with the National Aeronautics
and Space Administration.
\end{acknowledgements}

\bibliographystyle{aa}
\bibliography{clusters}

\begin{thebibliography}{93}
\expandafter\ifx\csname natexlab\endcsname\relax\def\natexlab#1{#1}\fi

\bibitem[{{Akritas} \& {Bershady}(1996)}]{bces}
{Akritas}, M.~G. \& {Bershady}, M.~A. 1996, \apj, 470, 706

\bibitem[{{Andersson} {et~al.}(2010){Andersson}, {Benson}, {Ade}, {Aird},
  {Armstrong}, {Bautz}, {Bleem}, {Brodwin}, {Carlstrom}, {Chang}, {Crawford},
  {Crites}, {de Haan}, {Desai}, {Dobbs}, {Dudley}, {Foley}, {Forman},
  {Garmire}, {George}, {Gladders}, {Halverson}, {High}, {Holder}, {Holzapfel},
  {Hrubes}, {Jones}, {Joy}, {Keisler}, {Knox}, {Lee}, {Leitch}, {Lueker},
  {Marrone}, {McMahon}, {Mehl}, {Meyer}, {Mohr}, {Montroy}, {Murray}, {Padin},
  {Plagge}, {Pryke}, {Reichardt}, {Rest}, {Ruel}, {Ruhl}, {Schaffer}, {Shaw},
  {Shirokoff}, {Song}, {Spieler}, {Stalder}, {Staniszewski}, {Stark}, {Stubbs},
  {Vanderlinde}, {Vieira}, {Vikhlinin}, {Williamson}, {Yang}, \&
  {Zahn}}]{andersson10}
{Andersson}, K., {Benson}, B.~A., {Ade}, P.~A.~R., {et~al.} 2010, arXiv:
  1006.3068

\bibitem[{{Armstrong} {et~al.}(2010){Armstrong}, {Mohr}, {Adams}, {Beldica},
  {Cai}, {Darnell}, {Daues}, {Desai}, {Gower}, {Mossessian}, {Ngeow}, {Lin},
  {Neilson}, {Tucker}, {Bertin}, \& {BCS Collaboration}}]{armstrong10}
{Armstrong}, B., {Mohr}, J., {Adams}, D., {et~al.} 2010, in Bulletin of the
  American Astronomical Society, Vol.~42, American Astronomical Society Meeting
  Abstracts 215, 438.07--+

\bibitem[{{Arnaud} {et~al.}(2007){Arnaud}, {Pointecouteau}, \&
  {Pratt}}]{arnaud07}
{Arnaud}, M., {Pointecouteau}, E., \& {Pratt}, G.~W. 2007, \aap, 474, L37

\bibitem[{{Arnaud} {et~al.}(2010){Arnaud}, {Pratt}, {Piffaretti},
  {B{\"o}hringer}, {Croston}, \& {Pointecouteau}}]{arnaud10}
{Arnaud}, M., {Pratt}, G.~W., {Piffaretti}, R., {et~al.} 2010, \aap, 517, A92+

\bibitem[{{Ben{\'{\i}}tez}(2000)}]{benitez00}
{Ben{\'{\i}}tez}, N. 2000, \apj, 536, 571

\bibitem[{{Bertin}(2006)}]{bertin06}
{Bertin}, E. 2006, in Astronomical Society of the Pacific Conference Series,
  Vol. 351, Astronomical Data Analysis Software and Systems XV, ed.
  {C.~Gabriel, C.~Arviset, D.~Ponz, \& S.~Enrique}, 112--+

\bibitem[{{Bertin} \& {Arnouts}(1996)}]{bertin96}
{Bertin}, E. \& {Arnouts}, S. 1996, \aaps, 117, 393

\bibitem[{{Bielby} {et~al.}(2010){Bielby}, {Finoguenov}, {Tanaka}, {McCracken},
  {Daddi}, {Hudelot}, {Ilbert}, {Kneib}, {Le F{\`e}vre}, {Mellier}, {Nandra},
  {Petitjean}, {Srianand}, {Stalin}, \& {Willott}}]{bielby10}
{Bielby}, R.~M., {Finoguenov}, A., {Tanaka}, M., {et~al.} 2010, \aap, 523, A66+

\bibitem[{{B\"ohringer} {et~al.}(2005){B\"ohringer}, {Mullis}, {Rosati},
  {Lamer}, {Fassbender}, {Schwope}, \& {Schuecker}}]{boehringer05}
{B\"ohringer}, H., {Mullis}, C., {Rosati}, P., {et~al.} 2005, The Messenger,
  120, 33

\bibitem[{{B{\"o}hringer} {et~al.}(2010){B{\"o}hringer}, {Pratt}, {Arnaud},
  {Borgani}, {Croston}, {Ponman}, {Ameglio}, {Temple}, \&
  {Dolag}}]{boehringer10b}
{B{\"o}hringer}, H., {Pratt}, G.~W., {Arnaud}, M., {et~al.} 2010, \aap, 514,
  A32+

\bibitem[{{B{\"o}hringer} {et~al.}(2007){B{\"o}hringer}, {Schuecker}, {Pratt},
  {Arnaud}, {Ponman}, {Croston}, {Borgani}, {Bower}, {Briel}, {Collins},
  {Donahue}, {Forman}, {Finoguenov}, {Geller}, {Guzzo}, {Henry}, {Kneissl},
  {Mohr}, {Matsushita}, {Mullis}, {Ohashi}, {Pedersen}, {Pierini}, {Quintana},
  {Raychaudhury}, {Reiprich}, {Romer}, {Rosati}, {Sabirli}, {Temple}, {Viana},
  {Vikhlinin}, {Voit}, \& {Zhang}}]{boehringer07}
{B{\"o}hringer}, H., {Schuecker}, P., {Pratt}, G.~W., {et~al.} 2007, \aap, 469,
  363

\bibitem[{{B{\"o}hringer} {et~al.}(2000){B{\"o}hringer}, {Voges}, {Huchra},
  {McLean}, {Giacconi}, {Rosati}, {Burg}, {Mader}, {Schuecker}, {Simi{\c c}},
  {Komossa}, {Reiprich}, {Retzlaff}, \& {Tr{\"u}mper}}]{boehringer00}
{B{\"o}hringer}, H., {Voges}, W., {Huchra}, J.~P., {et~al.} 2000, \apjs, 129,
  435

\bibitem[{{Bonamente} {et~al.}(2008){Bonamente}, {Joy}, {LaRoque}, {Carlstrom},
  {Nagai}, \& {Marrone}}]{bonamente08}
{Bonamente}, M., {Joy}, M., {LaRoque}, S.~J., {et~al.} 2008, \apj, 675, 106

\bibitem[{{Bower} {et~al.}(1992){Bower}, {Lucey}, \& {Ellis}}]{bower92}
{Bower}, R.~G., {Lucey}, J.~R., \& {Ellis}, R.~S. 1992, \mnras, 254, 601

\bibitem[{{Bruzual} \& {Charlot}(2003)}]{bruzual03}
{Bruzual}, G. \& {Charlot}, S. 2003, \mnras, 344, 1000

\bibitem[{{Bullock} {et~al.}(2001){Bullock}, {Kolatt}, {Sigad}, {Somerville},
  {Kravtsov}, {Klypin}, {Primack}, \& {Dekel}}]{bullock01}
{Bullock}, J.~S., {Kolatt}, T.~S., {Sigad}, Y., {et~al.} 2001, \mnras, 321, 559

\bibitem[{{Burenin} {et~al.}(2007){Burenin}, {Vikhlinin}, {Hornstrup},
  {Ebeling}, {Quintana}, \& {Mescheryakov}}]{burenin07cccp1}
{Burenin}, R.~A., {Vikhlinin}, A., {Hornstrup}, A., {et~al.} 2007, \apjs, 172,
  561

\bibitem[{{Cash}(1979)}]{cash79}
{Cash}, W. 1979, \apj, 228, 939

\bibitem[{{Cavaliere} \& {Fusco-Femiano}(1976)}]{cavaliere76}
{Cavaliere}, A. \& {Fusco-Femiano}, R. 1976, \aap, 49, 137

\bibitem[{{da Silva} {et~al.}(2004){da Silva}, {Kay}, {Liddle}, \&
  {Thomas}}]{dasilva04}
{da Silva}, A.~C., {Kay}, S.~T., {Liddle}, A.~R., \& {Thomas}, P.~A. 2004,
  \mnras, 348, 1401

\bibitem[{{De Luca} \& {Molendi}(2004)}]{deluca04}
{De Luca}, A. \& {Molendi}, S. 2004, \aap, 419, 837

\bibitem[{{Dickey} \& {Lockman}(1990)}]{dl}
{Dickey}, J.~M. \& {Lockman}, F.~J. 1990, \araa, 28, 215

\bibitem[{{Ettori}(2000)}]{ettori00}
{Ettori}, S. 2000, \mnras, 311, 313

\bibitem[{{Fassbender}(2008)}]{fassbender-thesis}
{Fassbender}, R. 2008, arXiv: 0806.0861

\bibitem[{{Fassbender} {et~al.}(2011{\natexlab{a}}){Fassbender},
  {B{\"o}hringer}, {Nastasi}, {\v{S}uhada}, {M{\"u}hlegger}, {de Hoon},
  {Kohnert}, {Lamer}, {Mohr}, {Pierini}, {Pratt}, {Quintana}, {Rosati},
  {Santos}, \& {Schwope}}]{fassbender11c}
{Fassbender}, R., {B{\"o}hringer}, H., {Nastasi}, A., {et~al.}
  2011{\natexlab{a}}, New Journal of Physics (submitted)

\bibitem[{{Fassbender} {et~al.}(2011{\natexlab{b}}){Fassbender},
  {B{\"o}hringer}, {Santos}, {Pratt}, {{\v S}uhada}, {Kohnert}, {Lerchster},
  {Rovilos}, {Pierini}, {Chon}, {Schwope}, {Lamer}, {M{\"u}hlegger}, {Rosati},
  {Quintana}, {Nastasi}, {de Hoon}, {Seitz}, \& {Mohr}}]{fassbender11a}
{Fassbender}, R., {B{\"o}hringer}, H., {Santos}, J.~S., {et~al.}
  2011{\natexlab{b}}, \aap, 527, A78+

\bibitem[{{Finoguenov} {et~al.}(2009){Finoguenov}, {Connelly}, {Parker},
  {Wilman}, {Mulchaey}, {Saglia}, {Balogh}, {Bower}, \& {McGee}}]{finoguenov09}
{Finoguenov}, A., {Connelly}, J.~L., {Parker}, L.~C., {et~al.} 2009, \apj, 704,
  564

\bibitem[{{Finoguenov} {et~al.}(2007){Finoguenov}, {Guzzo}, {Hasinger},
  {Scoville}, {Aussel}, {B{\"o}hringer}, {Brusa}, {Capak}, {Cappelluti},
  {Comastri}, {Giodini}, {Griffiths}, {Impey}, {Koekemoer}, {Kneib},
  {Leauthaud}, {Le F{\`e}vre}, {Lilly}, {Mainieri}, {Massey}, {McCracken},
  {Mobasher}, {Murayama}, {Peacock}, {Sakelliou}, {Schinnerer}, {Silverman},
  {Smol{\v c}i{\'c}}, {Taniguchi}, {Tasca}, {Taylor}, {Trump}, \&
  {Zamorani}}]{finoguenov07}
{Finoguenov}, A., {Guzzo}, L., {Hasinger}, G., {et~al.} 2007, \apjs, 172, 182

\bibitem[{{Gregory} {et~al.}(1994){Gregory}, {Vavasour}, {Scott}, \&
  {Condon}}]{pmn_cat}
{Gregory}, P.~C., {Vavasour}, J.~D., {Scott}, W.~K., \& {Condon}, J.~J. 1994,
  \apjs, 90, 173

\bibitem[{{Haiman} {et~al.}(2005){Haiman}, {Allen}, {Bahcall}, {Bautz},
  {B\"ohringer}, {Borgani}, {Bryan}, {Cabrera}, {Canizares}, {Citterio},
  {Evrard}, {Finoguenov}, {Griffiths}, {Hasinger}, {Henry}, {Jahoda},
  {Jernigan}, {Kahn}, {Lamb}, {Majumdar}, {Mohr}, {Molendi}, {Mushotzky},
  {Pareschi}, {Peterson}, {Petre}, {Predehl}, {Rasmussen}, {Ricker}, {Ricker},
  {Rosati}, {Sanderson}, {Stanford}, {Voit}, {Wang}, {White}, \&
  {White}}]{haiman05}
{Haiman}, Z., {Allen}, S., {Bahcall}, N., {et~al.} 2005, arXiv: 0507013

\bibitem[{{Haiman} {et~al.}(2001){Haiman}, {Mohr}, \& {Holder}}]{haiman01}
{Haiman}, Z., {Mohr}, J.~J., \& {Holder}, G.~P. 2001, \apj, 553, 545

\bibitem[{{High} {et~al.}(2010){High}, {Stalder}, {Song}, {Ade}, {Aird},
  {Allam}, {Armstrong}, {Barkhouse}, {Benson}, {Bertin}, {Bhattacharya},
  {Bleem}, {Brodwin}, {Buckley-Geer}, {Carlstrom}, {Challis}, {Chang},
  {Crawford}, {Crites}, {de Haan}, {Desai}, {Dobbs}, {Dudley}, {Foley},
  {George}, {Gladders}, {Halverson}, {Hamuy}, {Hansen}, {Holder}, {Holzapfel},
  {Hrubes}, {Joy}, {Keisler}, {Lee}, {Leitch}, {Lin}, {Lin}, {Loehr}, {Lueker},
  {Marrone}, {McMahon}, {Mehl}, {Meyer}, {Mohr}, {Montroy}, {Morell}, {Ngeow},
  {Padin}, {Plagge}, {Pryke}, {Reichardt}, {Rest}, {Ruel}, {Ruhl}, {Schaffer},
  {Shaw}, {Shirokoff}, {Smith}, {Spieler}, {Staniszewski}, {Stark}, {Stubbs},
  {Tucker}, {Vanderlinde}, {Vieira}, {Williamson}, {Wood-Vasey}, {Yang},
  {Zahn}, \& {Zenteno}}]{high10}
{High}, F.~W., {Stalder}, B., {Song}, J., {et~al.} 2010, \apj, 723, 1736

\bibitem[{{Hincks} {et~al.}(2010){Hincks}, {Acquaviva}, {Ade}, {Aguirre},
  {Amiri}, {Appel}, {Barrientos}, {Battistelli}, {Bond}, {Brown}, {Burger},
  {Chervenak}, {Das}, {Devlin}, {Dicker}, {Doriese}, {Dunkley}, {D{\"u}nner},
  {Essinger-Hileman}, {Fisher}, {Fowler}, {Hajian}, {Halpern}, {Hasselfield},
  {Hern{\'a}ndez-Monteagudo}, {Hilton}, {Hilton}, {Hlozek}, {Huffenberger},
  {Hughes}, {Hughes}, {Infante}, {Irwin}, {Jimenez}, {Juin}, {Kaul}, {Klein},
  {Kosowsky}, {Lau}, {Limon}, {Lin}, {Lupton}, {Marriage}, {Marsden},
  {Martocci}, {Mauskopf}, {Menanteau}, {Moodley}, {Moseley}, {Netterfield},
  {Niemack}, {Nolta}, {Page}, {Parker}, {Partridge}, {Quintana}, {Reid},
  {Sehgal}, {Sievers}, {Spergel}, {Staggs}, {Stryzak}, {Swetz}, {Switzer},
  {Thornton}, {Trac}, {Tucker}, {Verde}, {Warne}, {Wilson}, {Wollack}, \&
  {Zhao}}]{hincks10}
{Hincks}, A.~D., {Acquaviva}, V., {Ade}, P.~A.~R., {et~al.} 2010, \apjs, 191,
  423

\bibitem[{{Hu} \& {Kravtsov}(2003)}]{hukravtsov03}
{Hu}, W. \& {Kravtsov}, A.~V. 2003, \apj, 584, 702

\bibitem[{{Jeltema} {et~al.}(2008){Jeltema}, {Hallman}, {Burns}, \&
  {Motl}}]{jeltema08}
{Jeltema}, T.~E., {Hallman}, E.~J., {Burns}, J.~O., \& {Motl}, P.~M. 2008,
  \apj, 681, 167

\bibitem[{{Johnston} {et~al.}(2007){Johnston}, {Sheldon}, {Wechsler}, {Rozo},
  {Koester}, {Frieman}, {McKay}, {Evrard}, {Becker}, \& {Annis}}]{johnston07}
{Johnston}, D.~E., {Sheldon}, E.~S., {Wechsler}, R.~H., {et~al.} 2007, arXiv:
  0709.1159

\bibitem[{{Jones} {et~al.}(2004){Jones}, {Saunders}, {Colless}, {Read},
  {Parker}, {Watson}, {Campbell}, {Burkey}, {Mauch}, {Moore}, {Hartley},
  {Cass}, {James}, {Russell}, {Fiegert}, {Dawe}, {Huchra}, {Jarrett}, {Lahav},
  {Lucey}, {Mamon}, {Proust}, {Sadler}, \& {Wakamatsu}}]{jones046df}
{Jones}, D.~H., {Saunders}, W., {Colless}, M., {et~al.} 2004, \mnras, 355, 747

\bibitem[{{Kaastra}(1992)}]{kaastra92}
{Kaastra}, J.~S. 1992, Internal SRON-Leiden Report, Updated Version 2.0, 1

\bibitem[{{Kaiser}(1986)}]{kaiser86}
{Kaiser}, N. 1986, \mnras, 222, 323

\bibitem[{{Kalberla} {et~al.}(2005){Kalberla}, {Burton}, {Hartmann}, {Arnal},
  {Bajaja}, {Morras}, \& {P{\"o}ppel}}]{lab}
{Kalberla}, P.~M.~W., {Burton}, W.~B., {Hartmann}, D., {et~al.} 2005, \aap,
  440, 775

\bibitem[{{Kravtsov} {et~al.}(2006){Kravtsov}, {Vikhlinin}, \&
  {Nagai}}]{kravtsov06}
{Kravtsov}, A.~V., {Vikhlinin}, A., \& {Nagai}, D. 2006, \apj, 650, 128

\bibitem[{{Kuntz} \& {Snowden}(2008)}]{kuntzsnowden08}
{Kuntz}, K.~D. \& {Snowden}, S.~L. 2008, \aap, 478, 575

\bibitem[{{Leauthaud} {et~al.}(2010){Leauthaud}, {Finoguenov}, {Kneib},
  {Taylor}, {Massey}, {Rhodes}, {Ilbert}, {Bundy}, {Tinker}, {George}, {Capak},
  {Koekemoer}, {Johnston}, {Zhang}, {Cappelluti}, {Ellis}, {Elvis}, {Giodini},
  {Heymans}, {Le F{\`e}vre}, {Lilly}, {McCracken}, {Mellier},
  {R{\'e}fr{\'e}gier}, {Salvato}, {Scoville}, {Smoot}, {Tanaka}, {Van
  Waerbeke}, \& {Wolk}}]{leauthaud10}
{Leauthaud}, A., {Finoguenov}, A., {Kneib}, J., {et~al.} 2010, \apj, 709, 97

\bibitem[{{Liedahl} {et~al.}(1995){Liedahl}, {Osterheld}, \&
  {Goldstein}}]{liedahl95}
{Liedahl}, D.~A., {Osterheld}, A.~L., \& {Goldstein}, W.~H. 1995, \apjl, 438,
  L115

\bibitem[{{Maddox} {et~al.}(1990){Maddox}, {Efstathiou}, {Sutherland}, \&
  {Loveday}}]{maddox90}
{Maddox}, S.~J., {Efstathiou}, G., {Sutherland}, W.~J., \& {Loveday}, J. 1990,
  \mnras, 243, 692

\bibitem[{{Majumdar} \& {Mohr}(2003)}]{majumdar03}
{Majumdar}, S. \& {Mohr}, J.~J. 2003, \apj, 585, 603

\bibitem[{{Mantz} {et~al.}(2010{\natexlab{a}}){Mantz}, {Allen}, {Ebeling},
  {Rapetti}, \& {Drlica-Wagner}}]{mantz10}
{Mantz}, A., {Allen}, S.~W., {Ebeling}, H., {Rapetti}, D., \& {Drlica-Wagner},
  A. 2010{\natexlab{a}}, \mnras, 406, 1773

\bibitem[{{Mantz} {et~al.}(2010{\natexlab{b}}){Mantz}, {Allen}, {Rapetti}, \&
  {Ebeling}}]{mantz10a}
{Mantz}, A., {Allen}, S.~W., {Rapetti}, D., \& {Ebeling}, H.
  2010{\natexlab{b}}, \mnras, 406, 1759

\bibitem[{{Markevitch}(1998)}]{markevitch98}
{Markevitch}, M. 1998, \apj, 504, 27

\bibitem[{{Marriage} {et~al.}(2010){Marriage}, {Acquaviva}, {Ade}, {Aguirre},
  {Amiri}, {Appel}, {Barrientos}, {Battistelli}, {Bond}, {Brown}, {Burger},
  {Chervenak}, {Das}, {Devlin}, {Dicker}, {Doriese}, {Dunkley}, {Dunner},
  {Essinger-Hileman}, {Fisher}, {Fowler}, {Hajian}, {Halpern}, {Hasselfield},
  {Hern'andez-Monteagudo}, {Hilton}, {Hilton}, {Hincks}, {Hlozek},
  {Huffenberger}, {Hughes}, {Hughes}, {Infante}, {Irwin}, {Juin}, {Kaul},
  {Klein}, {Kosowsky}, {Lau}, {Limon}, {Lin}, {Lupton}, {Marsden}, {Martocci},
  {Mauskopf}, {Menanteau}, {Moodley}, {Moseley}, {Netterfield}, {Niemack},
  {Nolta}, {Page}, {Parker}, {Partridge}, {Quintana}, {Reese}, {Reid},
  {Sehgal}, {Sherwin}, {Sievers}, {Spergel}, {Staggs}, {Swetz}, {Switzer},
  {Thornton}, {Trac}, {Tucker}, {Warne}, {Wilson}, {Wollack}, \&
  {Zhao}}]{marriage10}
{Marriage}, T.~A., {Acquaviva}, V., {Ade}, P.~A.~R., {et~al.} 2010, arXiv:
  1010.1065

\bibitem[{{Marrone} {et~al.}(2009){Marrone}, {Smith}, {Richard}, {Joy},
  {Bonamente}, {Hasler}, {Hamilton-Morris}, {Kneib}, {Culverhouse},
  {Carlstrom}, {Greer}, {Hawkins}, {Hennessy}, {Lamb}, {Leitch}, {Loh},
  {Miller}, {Mroczkowski}, {Muchovej}, {Pryke}, {Sharp}, \&
  {Woody}}]{marrone09}
{Marrone}, D.~P., {Smith}, G.~P., {Richard}, J., {et~al.} 2009, \apjl, 701,
  L114

\bibitem[{{Mauch} {et~al.}(2003){Mauch}, {Murphy}, {Buttery}, {Curran},
  {Hunstead}, {Piestrzynski}, {Robertson}, \& {Sadler}}]{sumss_cat}
{Mauch}, T., {Murphy}, T., {Buttery}, H.~J., {et~al.} 2003, \mnras, 342, 1117

\bibitem[{{McInnes} {et~al.}(2009){McInnes}, {Menanteau}, {Heavens}, {Hughes},
  {Jimenez}, {Massey}, {Simon}, \& {Taylor}}]{mcinnes09}
{McInnes}, R.~N., {Menanteau}, F., {Heavens}, A.~F., {et~al.} 2009, \mnras,
  399, L84

\bibitem[{{Melin} {et~al.}(2011){Melin}, {Bartlett}, {Delabrouille}, {Arnaud},
  {Piffaretti}, \& {Pratt}}]{melin11}
{Melin}, J., {Bartlett}, J.~G., {Delabrouille}, J., {et~al.} 2011, \aap, 525,
  A139+

\bibitem[{{Menanteau} {et~al.}(2010){Menanteau}, {Hughes}, {Barrientos},
  {Deshpande}, {Hilton}, {Infante}, {Jimenez}, {Kosowsky}, {Moodley},
  {Spergel}, \& {Verde}}]{menanteau10}
{Menanteau}, F., {Hughes}, J.~P., {Barrientos}, L.~F., {et~al.} 2010, \apjs,
  191, 340

\bibitem[{{Menanteau} {et~al.}(2009){Menanteau}, {Hughes}, {Jimenez},
  {Hernandez-Monteagudo}, {Verde}, {Kosowsky}, {Moodley}, {Infante}, \&
  {Roche}}]{menanteau09}
{Menanteau}, F., {Hughes}, J.~P., {Jimenez}, R., {et~al.} 2009, \apj, 698, 1221

\bibitem[{{Mewe} {et~al.}(1985){Mewe}, {Gronenschild}, \& {van den
  Oord}}]{mewe85}
{Mewe}, R., {Gronenschild}, E.~H.~B.~M., \& {van den Oord}, G.~H.~J. 1985,
  \aaps, 62, 197

\bibitem[{{Mohr} {et~al.}(2008){Mohr}, {Adams}, {Barkhouse}, {Beldica},
  {Bertin}, {Cai}, {da Costa}, {Darnell}, {Daues}, {Jarvis}, {Gower}, {Lin},
  {Martelli}, {Neilsen}, {Ngeow}, {Ogando}, {Parga}, {Sheldon}, {Tucker},
  {Kuropatkin}, \& {Stoughton}}]{mohr08}
{Mohr}, J.~J., {Adams}, D., {Barkhouse}, W., {et~al.} 2008, in Presented at the
  Society of Photo-Optical Instrumentation Engineers (SPIE) Conference, Vol.
  7016, Society of Photo-Optical Instrumentation Engineers (SPIE) Conference
  Series

\bibitem[{{Motl} {et~al.}(2005){Motl}, {Hallman}, {Burns}, \&
  {Norman}}]{motl05}
{Motl}, P.~M., {Hallman}, E.~J., {Burns}, J.~O., \& {Norman}, M.~L. 2005,
  \apjl, 623, L63

\bibitem[{{M\"uhlegger}(2010)}]{muehlegger10}
{M\"uhlegger}, M. 2010, PhD thesis, Technische Universit\"at M\"unchen

\bibitem[{{Mullis} {et~al.}(2003){Mullis}, {McNamara}, {Quintana}, {Vikhlinin},
  {Henry}, {Gioia}, {Hornstrup}, {Forman}, \& {Jones}}]{mullis03}
{Mullis}, C.~R., {McNamara}, B.~R., {Quintana}, H., {et~al.} 2003, \apj, 594,
  154

\bibitem[{{Nagai}(2006)}]{nagai06}
{Nagai}, D. 2006, \apj, 650, 538

\bibitem[{{Navarro} {et~al.}(1997){Navarro}, {Frenk}, \& {White}}]{navarro97}
{Navarro}, J.~F., {Frenk}, C.~S., \& {White}, S.~D.~M. 1997, \apj, 490, 493

\bibitem[{{Nevalainen} {et~al.}(2010){Nevalainen}, {David}, \&
  {Guainazzi}}]{nevalainen10}
{Nevalainen}, J., {David}, L., \& {Guainazzi}, M. 2010, arXiv:1008.2102

\bibitem[{{O'Hara} {et~al.}(2006){O'Hara}, {Mohr}, {Bialek}, \&
  {Evrard}}]{ohara06}
{O'Hara}, T.~B., {Mohr}, J.~J., {Bialek}, J.~J., \& {Evrard}, A.~E. 2006, \apj,
  639, 64

\bibitem[{{Pacaud} {et~al.}(2007){Pacaud}, {Pierre}, {Adami}, {Altieri},
  {Andreon}, {Chiappetti}, {Detal}, {Duc}, {Galaz}, {Gueguen}, {Le F{\`e}vre},
  {Hertling}, {Libbrecht}, {Melin}, {Ponman}, {Quintana}, {Refregier},
  {Sprimont}, {Surdej}, {Valtchanov}, {Willis}, {Alloin}, {Birkinshaw},
  {Bremer}, {Garcet}, {Jean}, {Jones}, {Le F{\`e}vre}, {Maccagni}, {Mazure},
  {Proust}, {R{\"o}ttgering}, \& {Trinchieri}}]{pacaud07}
{Pacaud}, F., {Pierre}, M., {Adami}, C., {et~al.} 2007, \mnras, 382, 1289

\bibitem[{{Pacaud} {et~al.}(2006){Pacaud}, {Pierre}, {Refregier}, {Gueguen},
  {Starck}, {Valtchanov}, {Read}, {Altieri}, {Chiappetti}, {Gandhi}, {Garcet},
  {Gosset}, {Ponman}, \& {Surdej}}]{pacaud06}
{Pacaud}, F., {Pierre}, M., {Refregier}, A., {et~al.} 2006, \mnras, 372, 578

\bibitem[{{Pierre} {et~al.}(2007){Pierre}, {Chiappetti}, {Pacaud}, {Gueguen},
  {Libbrecht}, {Altieri}, {Aussel}, {Gandhi}, {Garcet}, {Gosset}, {Paioro},
  {Ponman}, {Read}, {Refregier}, {Starck}, {Surdej}, {Valtchanov}, {Adami},
  {Alloin}, {Alshino}, {Andreon}, {Birkinshaw}, {Bremer}, {Detal}, {Duc},
  {Galaz}, {Jones}, {Le F{\`e}vre}, {Le F{\`e}vre}, {Maccagni}, {Mazure},
  {Quintana}, {R{\"o}ttgering}, {Sprimont}, {Tasse}, {Trinchieri}, \&
  {Willis}}]{pierre07}
{Pierre}, M., {Chiappetti}, L., {Pacaud}, F., {et~al.} 2007, \mnras, 382, 279

\bibitem[{{Plagge} {et~al.}(2010){Plagge}, {Benson}, {Ade}, {Aird}, {Bleem},
  {Carlstrom}, {Chang}, {Cho}, {Crawford}, {Crites}, {de Haan}, {Dobbs},
  {George}, {Hall}, {Halverson}, {Holder}, {Holzapfel}, {Hrubes}, {Joy},
  {Keisler}, {Knox}, {Lee}, {Leitch}, {Lueker}, {Marrone}, {McMahon}, {Mehl},
  {Meyer}, {Mohr}, {Montroy}, {Padin}, {Pryke}, {Reichardt}, {Ruhl},
  {Schaffer}, {Shaw}, {Shirokoff}, {Spieler}, {Stalder}, {Staniszewski},
  {Stark}, {Vanderlinde}, {Vieira}, {Williamson}, \& {Zahn}}]{plagge10}
{Plagge}, T., {Benson}, B.~A., {Ade}, P.~A.~R., {et~al.} 2010, \apj, 716, 1118

\bibitem[{{Planck Collaboration}(2011{\natexlab{a}})}]{planck11}
{Planck Collaboration}. 2011{\natexlab{a}}, {Planck early results 08: The
  all-sky early Sunyaev-Zeldovich cluster sample} ({Submitted to \aap,
  [arXiv:astro-ph/1101.2024]})

\bibitem[{{Planck Collaboration}(2011{\natexlab{b}})}]{planck11b}
{Planck Collaboration}. 2011{\natexlab{b}}, {Planck early results 10:
  Statistical analysis of Sunyaev-Zeldovich scaling relations for X-ray galaxy
  clusters} ({Submitted to \aap, [arXiv:astro-ph/1101.2043]})

\bibitem[{{Planck Collaboration}(2011{\natexlab{c}})}]{planck11c}
{Planck Collaboration}. 2011{\natexlab{c}}, {Planck early results 11:
  Calibration of the local galaxy cluster Sunyaev-Zeldovich scaling relations}
  ({Submitted to \aap, [arXiv:astro-ph/1101.2026]})

\bibitem[{{Pratt} \& {Arnaud}(2003)}]{pratt03}
{Pratt}, G.~W. \& {Arnaud}, M. 2003, \aap, 408, 1

\bibitem[{{Pratt} {et~al.}(2009){Pratt}, {Croston}, {Arnaud}, \&
  {B{\"o}hringer}}]{pratt09}
{Pratt}, G.~W., {Croston}, J.~H., {Arnaud}, M., \& {B{\"o}hringer}, H. 2009,
  \aap, 498, 361

\bibitem[{{Reichert} {et~al.}(2011){Reichert}, {B{\"o}hringer}, {Fassbender},
  \& {M{\"u}hlegger}}]{reichert11}
{Reichert}, A., {B{\"o}hringer}, H., {Fassbender}, R., \& {M{\"u}hlegger}, M.
  2011, \aap, 535, A4+

\bibitem[{{Reiprich}(2001)}]{reiprich01}
{Reiprich}, T.~H. 2001, PhD thesis, Max-Planck-Institut f{\"u}r
  extraterrestrische Physik, P.O.~Box 1312, 85741 Garching, Germany

\bibitem[{{Reiprich} \& {B{\"o}hringer}(2002)}]{reiprich02}
{Reiprich}, T.~H. \& {B{\"o}hringer}, H. 2002, \apj, 567, 716

\bibitem[{{Reyes} {et~al.}(2008){Reyes}, {Mandelbaum}, {Hirata}, {Bahcall}, \&
  {Seljak}}]{reyes08}
{Reyes}, R., {Mandelbaum}, R., {Hirata}, C., {Bahcall}, N., \& {Seljak}, U.
  2008, \mnras, 390, 1157

\bibitem[{{Rosati} {et~al.}(1998){Rosati}, {della Ceca}, {Norman}, \&
  {Giacconi}}]{rosati98}
{Rosati}, P., {della Ceca}, R., {Norman}, C., \& {Giacconi}, R. 1998, \apjl,
  492, L21+

\bibitem[{{Scharf}(2002)}]{scharf02}
{Scharf}, C. 2002, \apj, 572, 157

\bibitem[{{Skrutskie} {et~al.}(2006){Skrutskie}, {Cutri}, {Stiening},
  {Weinberg}, {Schneider}, {Carpenter}, {Beichman}, {Capps}, {Chester},
  {Elias}, {Huchra}, {Liebert}, {Lonsdale}, {Monet}, {Price}, {Seitzer},
  {Jarrett}, {Kirkpatrick}, {Gizis}, {Howard}, {Evans}, {Fowler}, {Fullmer},
  {Hurt}, {Light}, {Kopan}, {Marsh}, {McCallon}, {Tam}, {Van Dyk}, \&
  {Wheelock}}]{2masx}
{Skrutskie}, M.~F., {Cutri}, R.~M., {Stiening}, R., {et~al.} 2006, \aj, 131,
  1163

\bibitem[{{Skrutskie} {et~al.}(2000){Skrutskie}, {Schneider}, {Stiening},
  {Strom}, {Weinberg}, {Beichman}, {Chester}, {Cutri}, {Lonsdale}, {Elias},
  {Elston}, {Capps}, {Carpenter}, {Huchra}, {Liebert}, {Monet}, {Price}, \&
  {Seitzer}}]{skrutskie00}
{Skrutskie}, M.~F., {Schneider}, S.~E., {Stiening}, R., {et~al.} 2000, VizieR
  Online Data Catalog, 1, 2003

\bibitem[{{Song} {et~al.}(2011){Song}, {Mohr}, {Barkhouse}, {Warren}, \&
  {Rude}}]{song11}
{Song}, J., {Mohr}, J.~J., {Barkhouse}, W.~A., {Warren}, M.~S., \& {Rude}, C.
  2011, arXiv: 1104.2332

\bibitem[{{Staniszewski} {et~al.}(2009){Staniszewski}, {Ade}, {Aird}, {Benson},
  {Bleem}, {Carlstrom}, {Chang}, {Cho}, {Crawford}, {Crites}, {de Haan},
  {Dobbs}, {Halverson}, {Holder}, {Holzapfel}, {Hrubes}, {Joy}, {Keisler},
  {Lanting}, {Lee}, {Leitch}, {Loehr}, {Lueker}, {McMahon}, {Mehl}, {Meyer},
  {Mohr}, {Montroy}, {Ngeow}, {Padin}, {Plagge}, {Pryke}, {Reichardt}, {Ruhl},
  {Schaffer}, {Shaw}, {Shirokoff}, {Spieler}, {Stalder}, {Stark},
  {Vanderlinde}, {Vieira}, {Zahn}, \& {Zenteno}}]{staniszewski09}
{Staniszewski}, Z., {Ade}, P.~A.~R., {Aird}, K.~A., {et~al.} 2009, \apj, 701,
  32

\bibitem[{{Sun} {et~al.}(2011){Sun}, {Sehgal}, {Voit}, {Donahue}, {Jones},
  {Forman}, {Vikhlinin}, \& {Sarazin}}]{sun11}
{Sun}, M., {Sehgal}, N., {Voit}, G.~M., {et~al.} 2011, \apjl, 727, L49+

\bibitem[{{Sunyaev} \& {Zel'dovich}(1972)}]{sunyaev72}
{Sunyaev}, R.~A. \& {Zel'dovich}, Y.~B. 1972, Comments on Astrophysics and
  Space Physics, 4, 173

\bibitem[{{{\v S}uhada} {et~al.}(2010){{\v S}uhada}, {Song}, {B{\"o}hringer},
  {Benson}, {Mohr}, {Fassbender}, {Finoguenov}, {Pierini}, {Pratt},
  {Andersson}, {Armstrong}, \& {Desai}}]{suhada10}
{{\v S}uhada}, R., {Song}, J., {B{\"o}hringer}, H., {et~al.} 2010, \aap, 514,
  L3+

\bibitem[{{Vanderlinde} {et~al.}(2010){Vanderlinde}, {Crawford}, {de Haan},
  {Dudley}, {Shaw}, {Ade}, {Aird}, {Benson}, {Bleem}, {Brodwin}, {Carlstrom},
  {Chang}, {Crites}, {Desai}, {Dobbs}, {Foley}, {George}, {Gladders}, {Hall},
  {Halverson}, {High}, {Holder}, {Holzapfel}, {Hrubes}, {Joy}, {Keisler},
  {Knox}, {Lee}, {Leitch}, {Loehr}, {Lueker}, {Marrone}, {McMahon}, {Mehl},
  {Meyer}, {Mohr}, {Montroy}, {Ngeow}, {Padin}, {Plagge}, {Pryke}, {Reichardt},
  {Rest}, {Ruel}, {Ruhl}, {Schaffer}, {Shirokoff}, {Song}, {Spieler},
  {Stalder}, {Staniszewski}, {Stark}, {Stubbs}, {van Engelen}, {Vieira},
  {Williamson}, {Yang}, {Zahn}, \& {Zenteno}}]{vanderlinde10}
{Vanderlinde}, K., {Crawford}, T.~M., {de Haan}, T., {et~al.} 2010, \apj, 722,
  1180

\bibitem[{{Vikhlinin} {et~al.}(2009){Vikhlinin}, {Burenin}, {Ebeling},
  {Forman}, {Hornstrup}, {Jones}, {Kravtsov}, {Murray}, {Nagai}, {Quintana}, \&
  {Voevodkin}}]{vikhlinin09}
{Vikhlinin}, A., {Burenin}, R.~A., {Ebeling}, H., {et~al.} 2009, \apj, 692,
  1033

\bibitem[{{Vikhlinin} {et~al.}(1998){Vikhlinin}, {McNamara}, {Forman}, {Jones},
  {Quintana}, \& {Hornstrup}}]{vikhlinin98}
{Vikhlinin}, A., {McNamara}, B.~R., {Forman}, W., {et~al.} 1998, \apj, 502, 558

\bibitem[{{Williamson} {et~al.}(2011){Williamson}, {Benson}, {High},
  {Vanderlinde}, {Ade}, {Aird}, {Andersson}, {Armstrong}, {Ashby}, {Bautz},
  {Bazin}, {Bertin}, {Bleem}, {Bonamente}, {Brodwin}, {Carlstrom}, {Chang},
  {Clocchiatti}, {Crawford}, {Crites}, {de Haan}, {Desai}, {Dobbs}, {Dudley},
  {Fazio}, {Foley}, {Forman}, {Garmire}, {George}, {Gladders}, {Gonzalez},
  {Halverson}, {Holder}, {Holzapfel}, {Hoover}, {Hrubes}, {Jones}, {Joy},
  {Keisler}, {Knox}, {Lee}, {Leitch}, {Lueker}, {Luong-Van}, {Marrone},
  {McMahon}, {Mehl}, {Meyer}, {Mohr}, {Montroy}, {Murray}, {Padin}, {Plagge},
  {Pryke}, {Reichardt}, {Rest}, {Ruel}, {Ruhl}, {Saliwanchik}, {Saro},
  {Schaffer}, {Shaw}, {Shirokoff}, {Song}, {Spieler}, {Stalder}, {Stanford},
  {Staniszewski}, {Stark}, {Story}, {Stubbs}, {Vieira}, {Vikhlinin}, \&
  {Zenteno}}]{williamson11}
{Williamson}, R., {Benson}, B.~A., {High}, F.~W., {et~al.} 2011, arXiv:
  1101.1290

\bibitem[{{Zenteno} {et~al.}(2011){Zenteno}, {Song}, {Desai}, {Armstrong},
  {Mohr}, {Ngeow}, {Barkhouse}, {Allam}, {Andersson}, {Bazin}, {Benson},
  {Bertin}, {Brodwin}, {Buckley-Geer}, {Hansen}, {High}, {Lin}, {Lin}, {Liu},
  {Rest}, {Smith}, {Stalder}, {Stark}, {Tucker}, \& {Yang}}]{zenteno11}
{Zenteno}, A., {Song}, J., {Desai}, S., {et~al.} 2011, arXiv: 1103.4612

\end{thebibliography}

\Online

\begin{appendix}

\section{Ancillary information}
\label{sec:flags}
\subsection{Quality flags}
In this section we provide useful ancillary data for our clusters
described using several X-ray quality flags and diagnostic parameters
compiled in Table~\ref{tab:flags}. Here is the description of the
table's columns:

\begin{itemize}
\item \textbf{ID:} the cluster identification number.
\item \textbf{BCS field:} the identification number of the BCS field,
  on which the cluster is lying. Some clusters lie on two or more
  tiles, in those cases we provide the name of the tile with the
  largest overlap region.

\item \textbf{XMM OBSID:} The official identification number of the
  XMM-\emph{Newton} pointing containing the cluster. If the cluster
  lies in two (or three) adjacent observations we provide the OBSID of
  the pointing which provides the best constraint on the cluster flux
  (typically the one where the cluster is at the smallest off-axis
  angle).

\item \textbf{Internal field ID: } XMM-BCS internal field ID used in
  the text as shorthand for the OBSID. See also Table~\ref{tab:obslog}
  for the full cross-listing of field IDs.

\item \textbf{flag$_{\mathrm{\mathbf{HC}}}$:} The hot chip flag is a four
character string, with the characters being either T for "true" or F for
"false". The significance of the characters:\\
1. character: Does the observation have a hot MOS2 CCD\#5? \\
2. character: Does the cluster lie on the MOS2 CCD\#5?\\
3. character: Does the observation have a hot MOS1 CCD\#4?\\
4. character: Does the cluster lie on the MOS1 CCD\#4?\\
For the problematics of the hot chips see Sect.~\ref{sec:hc}.

\item \textbf{SNR$_{\mathrm{\mathbf{Xflux}}}$:} The flux estimation
  significance determined as $F_{500}/\sigma_{F_{500}}$, where
  $F_{500}$ is the source flux in the $r_{500}$ aperture, and
  $\sigma_{F_{500}}$ is its error (including shot noise and $5\%$
  background modelling uncertainty, Sect.~\ref{sec:gca}).

\item \textbf{flag$_{\mathrm{\mathbf{inst}}}$:} The instrument flag
  equals 0 if the physical parameters of the source were obtained
  using the PN and both MOS cameras. If flag$_{\mathrm{inst}}=1$ only
  PN could be used and if flag$_{\mathrm{inst}}=2$ only the
  combination of the two MOS cameras was utilized.

\item \textbf{Q$^{\mathrm{\mathbf{PN}}}_{\mathrm{\mathbf{plat}}}$:}
  The automatic plateau fit quality flag for the PN growth
  curves. These flags have the same meaning as in
  \citet{boehringer00}. In summary, as described in
  Sect.~\ref{sec:gca} we fit a line to the growth curve between
  r$_{plat}$ and the outer extraction radius. The flag describes the
  quality of this fit by calculating the ratio of the predicted count
  rate from the linear fit to the expectation, if the plateau was
  constant and equal to the estimated plateau flux.
  Q$^{\mathrm{PN}}_{\mathrm{plat}}=1$: the growth curve shows neither
  significant increase nor decrease outside r$_{plat}$. This value is
  assigned if the linear extrapolation does not differ by more than
  $0.8\%$ per bin from the constant value.
  Q$^{\mathrm{PN}}_{\mathrm{plat}}=2$: marks a declining curve
  (decline $>0.8\%$/bin).  A decline can occur if the background model
  (determined from a fit to the whole field) slightly overestimates
  the local background. In this case we attempt to estimate the
  plateau level from the 3 bins closest to r$_{plat}$. If the final
  fit is acceptable (no significant residual decline), the plateau is
  accepted and assigned this quality flag.
  Q$^{\mathrm{PN}}_{\mathrm{plat}}=3$ and
  Q$^{\mathrm{PN}}_{\mathrm{plat}}=4$: in case that the plateau is
  rising an attempt is made to iteratively exclude the outermost bins
  and in a second step also the innermost bins. This procedure helps
  in correcting an outer rise of the growth curve due to a neighboring
  source and if necessary by skipping over a few bins if the curve
  fluctuates in the radial range close to r$_{plat}$. If this
  procedure converges, after the exclusion of the outermost bins the
  plateau is accepted and flagged with
  Q$^{\mathrm{PN}}_{\mathrm{plat}}=3$. If the procedure converges, but
  it required also the second step of excluding the innermost bins we
  assign Q$^{\mathrm{PN}}_{\mathrm{plat}}=4$.  If
  Q$^{\mathrm{PN}}_{\mathrm{plat}}=5$, the plateau is rising and the
  increase could not be corrected for by the above described
  procedure.  If there are only two or less radial bins outside the
  plateau radius can not be established and we assign
  Q$^{\mathrm{PN}}_{\mathrm{plat}}=9$.

Q$^{\mathrm{PN}}_{\mathrm{plat}}\leq4$ mark generally good quality
plateaus (naturally, the lower the flag the
better). Q$^{\mathrm{PN}}_{\mathrm{plat}}=5$ is a serious warning and
Q$^{\mathrm{PN}}_{\mathrm{plat}}=9$ is not recommended to be used at
all. In fact, for the parameters in Table~\ref{tab:physpar} we do not
use plateaus with this flag with the exception for the systems ID 476
and 139 where an alternative solution is not available due to their
significant blending (Appendix~\ref{sec:objectnotes}). \\

\item \textbf{Q$^{\mathrm{\mathbf{MOS}}}_{\mathrm{\mathbf{plat}}}$:} The same as
Q$^{\mathrm{PN}}_{\mathrm{plat}}$ but applied to the MOS1+MOS2 growth curve.
\item \textbf{Q$^{\mathrm{\mathbf{PN}}}_{\mathrm{\mathbf{gca}}}$:}
  Visual flag set considering the overall quality of the PN growth
  curve solution (taking into account the presence of chip gaps,
  anomalous background, potential contamination etc.). Value equal to
  1 is the best (no problems), equal to 3 the worse (to be considered
  as a warning). If the given detector was not included in the
  analysis (as marked by flag$_{\mathrm{\mathbf{inst}}}$)
  Q$^{\mathrm{\mathbf{PN}}}_{\mathrm{\mathbf{gca}}}$ is set to 9.

\item \textbf{Q$^{\mathrm{\mathbf{MOS}}}_{\mathrm{\mathbf{gca}}}$:}
  The same as Q$^{\mathrm{\mathbf{PN}}}_{\mathrm{\mathbf{gca}}}$ but
  applied to the MOS1+MOS2 growth curve.

\item \textbf{Q$_{\mathrm{\mathbf{TOT}}}$:} This flag is a
  non-quantitative ancillary X-ray flag. It is assigned during a
  visual inspection of each source, but takes into consideration also
  all above flags. Sources with this flag equal to 1 (best) and 2
  (only mild warnings) have high quality X-ray photometry
  measurements. Flag equal to 3 signifies problems with the X-ray
  photometry e.g. due to problematic locations on the detectors, bad
  quality plateaus, source blending etc. In the present work, we have
  included also Q$_{\mathrm{\mathbf{TOT}}}=3$ sources in all analyses
  since they constitute only about $\sim10\%$ of the sample.
\end{itemize}

\begin{landscape}
\begin{table}[ht!]
\rowcolors{1}{tableShade}{white}
 \centering
 \begin{small}
 \caption[Ancillary X-ray quality flags]{Cluster ID and proper name cross-reference table. We also
list the BCS field name for each cluster and additional ancillary X-ray quality
flags (see text for details).}
\vspace{0.1cm}
 \label{tab:flags}
\begin{tabular}{ccccccccccccc}
\hline
\hline
  \multicolumn{1}{c}{ID} &
  \multicolumn{1}{c}{Name} &
  \multicolumn{1}{c}{BCS field} &
  \multicolumn{1}{c}{XMM OBSID} &
  \multicolumn{1}{c}{Internal field ID} &
  \multicolumn{1}{c}{flag$_{\mathrm{HC}}$} &
  \multicolumn{1}{c}{SNR$_{\mathrm{Xflux}}$} &
  \multicolumn{1}{c}{flag$_{\mathrm{inst}}$} &
  \multicolumn{1}{c}{Q$^{\mathrm{PN}}_{\mathrm{plat}}$} &
  \multicolumn{1}{c}{Q$^{\mathrm{MOS}}_{\mathrm{plat}}$} &
  \multicolumn{1}{c}{Q$^{\mathrm{PN}}_{\mathrm{gca}}$} &
  \multicolumn{1}{c}{Q$^{\mathrm{MOS}}_{\mathrm{gca}}$} &
  \multicolumn{1}{c}{Q$_{\mathrm{TOT}}$} \\
\hline
  011 & XBCS J232713.7-560341 & BCS2327-5602 & 0505380301 & F03 & TFFF & 6.7 & 0 & 1 & 1 & 2 & 2 & 1\\
  018 & XBCS J232955.9-560810 & BCS2327-5602 & 0505380401 & F04 & TFFF & 10.2 & 0 & 3 & 1 & 1 & 1 & 1\\
  032 & XBCS J232842.7-553358 & BCS2327-5529 & 0505381001 & F10 & FFFF & 11.9 & 0 & 1 & 1 & 1 & 1 & 1\\
  033 & XBCS J232810.7-555024 & BCS2327-5602 & 0505381001 & F10 & FFFF & 11.9 & 0 & 1 & 1 & 2 & 2 & 1\\
  034 & XBCS J233036.9-554337 & BCS2332-5529 & 0505381101 & F11 & TTFF & 6.3 & 0 & 1 & 1 & 1 & 2 & 1\\
  035 & XBCS J233345.3-553819 & BCS2332-5529 & 0505381201 & F12 & FFFF & 6.3 & 0 & 1 & 1 & 3 & 3 & 1\\
  038 & XBCS J233403.1-554856 & BCS2332-5602 & 0505381201 & F12 & FFFF & 6.6 & 0 & 1 & 1 & 2 & 1 & 2\\
  039 & XBCS J231917.1-551928 & BCS2319-5529 & 0505381401 & F14 & FFFF & 16.2 & 0 & 1 & 1 & 1 & 1 & 2\\
  044 & XBCS J231653.1-545413 & BCS2316-5455 & 0505382201 & F22 & TTFF & 14.9 & 0 & 1 & 1 & 1 & 1 & 1\\
  069 & XBCS J232351.1-545332 & BCS2324-5455 & 0505382401 & F24 & TFTF & 5.2 & 0 & 1 & 1 & 2 & 2 & 1\\
  070 & XBCS J232230.9-541609 & BCS2324-5421 & 0505383801 & F38 & TFTT & 49.1 & 0 & 1 & 1 & 1 & 1 & 1\\
  081 & XBCS J232723.3-551545 & BCS2327-5529 & 0505381701 & F17 & TFFF & 6.0 & 0 & 3 & 1 & 2 & 3 & 2\\
  082 & XBCS J232618.7-552309 & BCS2327-5529 & 0505381701 & F17 & TFFF & 5.8 & 0 & 1 & 1 & 1 & 2 & 1\\
  088 & XBCS J232842.0-551324 & BCS2327-5529 & 0505381801 & F18 & FFFF & 6.7 & 1 & 1 & 9 & 2 & 9 & 2\\
  090 & XBCS J232856.8-552429 & BCS2327-5529 & 0505381801 & F18 & FFFF & 4.3 & 0 & 1 & 1 & 1 & 3 & 1\\
  094 & XBCS J233204.4-551243 & BCS2332-5529 & 0505381901 & F19 & TTFF & 9.2 & 0 & 5 & 1 & 2 & 2 & 1\\
  109 & XBCS J232737.4-541614 & BCS2328-5421 & 0505384001 & F40 & FFFF & 6.8 & 0 & 1 & 1 & 2 & 2 & 2\\
  110 & XBCS J233003.9-541420 & BCS2331-5421 & 0505384001 & F40 & FFFF & 7.7 & 0 & 1 & 1 & 1 & 2 & 1\\
  126 & XBCS J232633.4-550114 & BCS2328-5455 & 0505382501 & F25 & TFTF & 13.3 & 0 & 1 & 1 & 1 & 1 & 1\\
  127 & XBCS J232723.8-550353 & BCS2328-5455 & 0505382501 & F25 & TFTF & 8.8 & 0 & 1 & 1 & 1 & 1 & 1\\
  132 & XBCS J232802.0-545545 & BCS2328-5455 & 0505382601 & F26 & FFFF & 8.6 & 0 & 1 & 1 & 1 & 2 & 2\\
  136 & XBCS J232200.9-544500 & BCS2320-5455 & 0505383101 & F31 & FFTF & 10.9 & 0 & 1 & 1 & 1 & 1 & 1\\
  139 & XBCS J232534.9-544316 & BCS2324-5455 & 0505383201 & F32 & TFTF & 7.5 & 0 & 9 & 9 & 3 & 3 & 3\\
  150 & XBCS J233000.4-543706 & BCS2331-5421 & 0505383401 & F34 & FFFF & 20.4 & 0 & 1 & 1 & 1 & 1 & 1\\
  152 & XBCS J232940.0-544719 & BCS2328-5455 & 0505383401 & F34 & FFFF & 6.1 & 0 & 4 & 1 & 3 & 2 & 2\\
  156 & XBCS J233531.6-543511 & BCS2335-5421 & 0505383601 & F36 & FFFF & 14.2 & 0 & 1 & 1 & 2 & 2 & 1\\
  158 & XBCS J233424.8-542731 & BCS2335-5421 & 0505383601 & F36 & FFFF & 6.7 & 0 & 1 & 1 & 2 & 2 & 1\\
  210 & XBCS J233405.8-554709 & BCS2336-5602 & 0505381201 & F12 & FFFF & 5.0 & 0 & 1 & 1 & 2 & 2 & 1\\
  227 & XBCS J232210.2-552512 & BCS2323-5529 & 0505381501 & F15 & FFFF & 7.9 & 0 & 1 & 1 & 2 & 2 & 1\\
  245 & XBCS J232403.8-550121 & BCS2324-5455 & 0505382401 & F24 & TFTF & 5.4 & 0 & 1 & 1 & 1 & 2 & 1\\
  275 & XBCS J233447.8-551625 & BCS2336-5529 & 0505382001 & F20 & TTFF & 6.6 & 0 & 1 & 1 & 2 & 2 & 1\\
  287 & XBCS J233650.9-551756 & BCS2336-5529 & 0505382101 & F21 & TTFF & 2.8 & 0 & 4 & 4 & 3 & 3 & 3\\
  288 & XBCS J233500.5-545459 & BCS2335-5455 & 0505382801 & F28 & TFFF & 4.4 & 0 & 5 & 1 & 1 & 2 & 1\\
  357 & XBCS J233428.8-543624 & BCS2335-5421 & 0505383601 & F36 & FFFF & 7.5 & 0 & 1 & 1 & 1 & 2 & 1\\
  386 & XBCS J233554.3-560534 & BCS2336-5602 & 0505380601 & F06 & TFFF & 3.9 & 0 & 1 & 3 & 1 & 3 & 1\\
  430 & XBCS J232533.4-554358 & BCS2323-5529 & 0505384801 & F09b & FFFF & 4.8 & 0 & 1 & 1 & 1 & 2 & 1\\
  444 & XBCS J233620.1-553108 & BCS2336-5529 & 0505382001 & F20 & TFFF & 5.0 & 0 & 1 & 1 & 1 & 3 & 2\\
  457 & XBCS J232828.2-541450 & BCS2328-5421 & 0505384001 & F40 & FFFF & 3.8 & 2 & 9 & 1 & 9 & 2 & 3\\
  476 & XBCS J232540.0-544428 & BCS2324-5455 & 0505383201 & F32 & TFTF & 8.7 & 0 & 9 & 5 & 3 & 3 & 3\\
  502 & XBCS J231944.0-543824 & BCS2320-5455 & 0505383001 & F30 & TFFF & 11.2 & 0 & 1 & 1 & 2 & 3 & 1\\
  511 & XBCS J233215.1-544202 & BCS2331-5455 & 0505383501 & F35 & TFFF & 6.3 & 0 & 3 & 1 & 2 & 2 & 1\\
  527 & XBCS J232210.9-561846 & BCS2323-5602 & 0554561001 & F01b & TFFF & 6.1 & 0 & 1 & 3 & 1 & 3 & 1\\
  528 & XBCS J233520.6-553239 & BCS2336-5529 & 0505382001 & F20 & TFFF & 3.1 & 0 & 1 & 1 & 3 & 3 & 1\\
  538 & XBCS J233406.2-544352 & BCS2335-5455 & 0505383601 & F36 & FFFF & 2.4 & 2 & 9 & 3 & 9 & 2 & 3\\
  543 & XBCS J233243.3-544947 & BCS2331-5455 & 0554560601 & F35b & TFTF & 3.7 & 1 & 1 & 1 & 3 & 3 & 1\\
  547 & XBCS J232419.6-552550 & BCS2323-5529 & 0505381601 & F16 & TFTF & 3.2 & 2 & 9 & 1 & 9 & 1 & 3\\

\hline
\end{tabular}
\end{small}
\end{table}
\end{landscape}

\begin{table*}[ht!]
\centering
\caption[Cross-matched galaxies]{Galaxies identified in the NED
  database to be within $16^{\prime\prime}$ from the X-ray center. We
  list spectroscopic redshifts where available - in both cases the
  identified galaxies are BCGs.  Redshift reference: $^a$
  \citet{jones046df}.}
 \vspace{0.1cm}
\label{tab:nedgals}
\begin{tabular}{clcccc}
\hline
\hline
  \multicolumn{1}{c}{ID} &
  \multicolumn{1}{c}{Object Name} &
  \multicolumn{1}{c}{R.A. (deg)} &
  \multicolumn{1}{c}{DEC (deg)} &
  \multicolumn{1}{c}{redshift} &
  \multicolumn{1}{c}{separation} \\
\hline
  034 & APMUKS(BJ) B232750.10-560012.1 & 352.6544 & -55.7274 &  & $1.9^{\prime\prime}$\\
  039 & 2MASX J23191712-5519284 & 349.8214 & -55.3245 &  & $0.5^{\prime\prime}$\\
  041 & 2MASX J23190212-5523195 & 349.7588 & -55.3888 &  & $1.3^{\prime\prime}$\\
  070 & 2MASX J23223092-5416086 & 350.6289 & -54.2691 &  & $0.8^{\prime\prime}$\\
  094 & APMUKS(BJ) B232918.62-552918.2 & 353.0196 & -55.2122 &  & $2.5^{\prime\prime}$\\
  127 & 2MASX J23272468-5503589 & 351.8528 & -55.0664 &  & $9.6^{\prime\prime}$\\
  150 & 2MASX J23300047-5437069 & 352.5019 & -54.6187 & 0.177$^a$ & $1.5^{\prime\prime}$\\
  152 & 2MASX J23294006-5447220 & 352.4168 & -54.7895 &  & $3.1^{\prime\prime}$\\
  227 & APMUKS(BJ) B231920.30-554137.6 & 350.5444 & -55.4194 &  & $4.4^{\prime\prime}$\\
  268 & APMUKS(BJ) B232326.14-554657.3 & 351.5618 & -55.5074 &  & $14.0^{\prime\prime}$\\
  476 & 2MASX J23254015-5444308 & 351.4173 & -54.7419 & 0.101$^a$ & $3.1^{\prime\prime}$\\
  511 & APMUKS(BJ) B232929.68-545847.0 & 353.0645 & -54.7035 &  & $10.8^{\prime\prime}$\\
  547 & 2MASX J23241957-5525494 & 351.0816 & -55.4303 &  & $0.6^{\prime\prime}$\\
\hline
\end{tabular}
\end{table*}

\subsection{Notes on individual sources}

\label{sec:objectnotes}
Some of the identified clusters required individual treatment and in
this section we provide notes for these cases:
\begin{itemize}
\item \textbf{ID 011}: this high redshift cluster lies on the heavily
  flared field F03 with no quiescent period. Therefore, the field was
  not used for the sensitivity function calculation and the $\log N -
  \log S$. The double component background model accounts in principle
  in the first approximation for the enhanced background and therefore
  we provide the basic X-ray parameters for this cluster. The
  diagnostic flags (Table~\ref{tab:flags}) indicate that the growth
  curve solution is quite reliable, but due to the flaring all
  physical parameters should be treated with caution.

\item \textbf{ID 038}: This source consists of two completely
  overlapping systems, one with photometric redshift $z=0.39\pm0.05$
  and the second with $z=0.74\pm0.07$. Since there is no direct way to
  disentangle the contribution of the two sources, we will assume that
  all the flux comes from the more nearby system.  In this case, the
  estimated physical parameters are upper limits.

\item \textbf{ID 070}: is a nearby cluster with large extent and
  measured flux. It lies on a hot MOS2 CCD$\#5$ and due to its extent
  it is impossible to obtain a background area on this chip
  uncontaminated by the source emission. Therefore we can not use the
  procedure described in Sect.~\ref{sec:hc}, where we fit a double
  component model to the hot chip independently from the rest of the
  field. Instead we discard the data from this chip completely.
  Consistently, we use the MOS1 response matrix instead of the MOS2
  one for further analysis.

\item \textbf{ID 081}: in this distant cluster (photo-z$=0.85$)
the BCG is offset by $\sim19\arcsec$ from the X-ray centroid. The
emission from this system might be contaminated by an intervening AGN
(the X-ray center is coincident with a bright galaxy which could
harbor one).

\item \textbf{ID 109}: due to the limited depth of the available
  optical data we can provide only very tentative redshift estimate
  for this system.

\item \textbf{ID 139 and 476}: We detect two nearby, high significance
  extended sources in this region ($\sim1.3$~arcmin apart). The
  systems are confirmed as independent also in redshift space by our
  spectroscopic measurements (ID 476 at $z=0.102$ and ID 139 at
  $z=0.169$). In order to measure the flux of each cluster we excise
  the other source. Due to their proximity, however, full deblending
  is not possible and therefore both fluxes are likely overestimated.
  The analysis of the sources is further complicated by the presence
  of a very bright X-ray point source at $\sim2$~arcmin distance from
  the clusters and high quiescent soft proton contamination in the PN
  camera ($39\%$ background increase, Sect.~\ref{sec:pipe}).

The cluster catalog of \citet[][]{burenin07cccp1} based on ROSAT data
includes a source with a center roughly between the two systems
(i.e. very likely misclassified as a single cluster due to the limited
resolution of ROSAT).

\item \textbf{ID 275}: also lies on a hot MOS2 CCD$\#5$. Its detection
  likelihood is completely dominated by the MOS2 detection, however
  the source is not flagged as spurious based on the criteria
  described in Sect.~\ref{sec:hc}, because it would be above the
  detection threshold even without the MOS2 data. In this case, the
  background modeling of the hot chip was possible and this background
  was used in the subsequent growth curve analysis.

\end{itemize}

\subsection{Galactic counterparts}
All NED galactic counterparts within 16$^{\prime\prime}$ from the
X-ray centroids of our clusters are summarized in
Table~\ref{tab:nedgals} (see also Sect.~\ref{sec:xcorrned}).

\subsection{Wavelet detections}
We have also carried out a detection using a wavelet detection
algorithm developed for the COSMOS project by \citet{finoguenov07,
  finoguenov09}. Every cluster presented in our current sample has
also been confirmed by this approach.  In addition, we have identified
five systems (Table~\ref{tab:physpar2}) with a wavelet detection but
no SAS-based detection in any setup. All five systems are coincident
with significant galaxy overdensities. We find many interloping X-ray
point-sources in these systems (a potential source of
misclassification in the SAS detections).  Even after conservative
point source removal, residual contamination makes all the estimated
X-ray parameters for these systems highly uncertain. These detections
are not included in the statistical description of the sample
(e.g. the $\log N-\log S$ relation) and are listed here only for
completeness (in Table~\ref{tab:physpar2}).

\begin{landscape}
\begin{table}[ht!]
\rowcolors{1}{tableShade}{white}
 \centering
\caption[Physical parameters for the lower quality detections]{\small{Physical
parameters for the lower quality detections.}}
\vspace{0.1cm}
 \label{tab:physpar2}
 \begin{scriptsize}
\begin{tabular}{lcccccccccccc}
\hline
\hline
   \multicolumn{1}{c}{ID} &
   \multicolumn{1}{c}{R.A. (J2000)} &
   \multicolumn{1}{c}{Dec (J2000)} &
   \multicolumn{1}{c}{z} &
   \multicolumn{1}{c}{$r_{\mathrm{plat}}$} &
   \multicolumn{1}{c}{$F_{plat}$} &
   \multicolumn{1}{c}{$r_{500}$} &
   \multicolumn{1}{c}{$F_{500}$} &
   \multicolumn{1}{c}{$L_{500}$} &
   \multicolumn{1}{c}{$T_{500}$} &
   \multicolumn{1}{c}{$M_{500}$} &
   \multicolumn{1}{c}{$Y_{500}$} &
   \multicolumn{1}{c}{M$_{200}$}  \\
&(deg) &(deg) & (photo.) &(arcmin/$r_{500}^{-1}$) &($10^{-14}$~erg s$^{-1}$ cm$^{-2}$)
 & (kpc) & ($10^{-14}$~erg s$^{-1}$cm$^{-2}$)  & ($10^{43}$~erg s$^{-1})$
 & (keV) & $(10^{13}$ M$_{\odot}$) & ($10^{13}$ M$_{\odot}$~keV) & $(10^{13}$ M$_{\odot}$) \\
\hline
     355 &$ 352.8205 $&$ -54.7064 $&$ 0.31 \pm 0.05 $&$ 2.2/1.1 $&$ 3.20 \pm 0.94 $&$ 541 \pm 55 $&$ 2.66 \pm 0.79 $&$ 0.8 \pm 0.2 $&$ 1.6 \pm 0.5 $&$ 6.2 \pm 1.9 $&$ 0.6 \pm 0.4 $&$ 8.5 \pm 2.6 $ \\
 534 &$ 351.1993 $&$ -55.4128 $&$ 0.31 \pm 0.02 $&$ 1.4/0.7 $&$ 2.18 \pm 1.11 $&$ 523 \pm 64 $&$ 2.22 \pm 1.13 $&$ 0.7 \pm 0.4 $&$ 1.5 \pm 0.5 $&$ 5.6 \pm 2.0 $&$ 0.5 \pm 0.4 $&$ 7.6 \pm 2.8 $ \\
 536 &$ 354.2994 $&$ -55.3514 $&$ 0.69 \pm 0.07 $&$ 0.8/0.7 $&$ 0.83 \pm 0.23 $&$ 487 \pm 49 $&$ 0.85 \pm 0.24 $&$ 1.8 \pm 0.5 $&$ 1.9 \pm 0.6 $&$ 7.1 \pm 2.1 $&$ 0.8 \pm 0.5 $&$ 10.0 \pm 3.0 $ \\
 540 &$ 350.1461 $&$ -54.4628 $&$ 0.24 \pm 0.02 $&$ 1.3/0.6 $&$ 1.66 \pm 0.46 $&$ 474 \pm 48 $&$ 1.75 \pm 0.48 $&$ 0.3 \pm 0.1 $&$ 1.2 \pm 0.3 $&$ 3.8 \pm 1.2 $&$ 0.2 \pm 0.1 $&$ 5.2 \pm 1.6 $ \\
 541 &$ 350.5048 $&$ -54.3296 $&$ 0.59 \pm 0.06 $&$ 0.9/0.7 $&$ 1.06 \pm 0.30 $&$ 506 \pm 51 $&$ 1.08 \pm 0.30 $&$ 1.5 \pm 0.4 $&$ 1.8 \pm 0.5 $&$ 7.0 \pm 2.1 $&$ 0.8 \pm 0.5 $&$ 9.9 \pm 3.0 $ \\

\hline
\end{tabular}
\end{scriptsize}

\end{table}
\end{landscape}

\section{Test of the sensitivity function from preliminary Monte Carlo simulations}
\label{sec:simsensf}
The determination of the survey  selection function is a crucial
requirement for the cosmological modeling of the cluster sample,
scaling relation studies etc. Due to the complex nature of
the extended source detection, this question can be properly addressed only
by detailed Monte Carlo simulations. In the present work
we utilized a simplifying approach that allowed us to get a
first estimate of the sensitivity functions and the recovered
$\log N - \log S$ relation (Sect.~\ref{sec:skycov} and \ref{sec:lognlogs}).

The software for the Monte Carlo simulations \citep{muehlegger10} is
in an advanced development stage which allows us to carry out a
preliminary test of our simplified approach.

The simulation pipeline uses the survey fields themselves and injects
mock beta model clusters into the observations at random positions
across the field-of-view. The field is then processed with the
detection pipeline. The process is repeated on a grid of cluster
fluxes and core radii and the cluster detection probability is derived
as a function of these parameters.  The use of real observations
instead of model backgrounds allows us to derive a realistic selection
function.  The simulation software is described in detail in
\citet{muehlegger10}.

\begin{figure}[t!]
\begin{center}
\includegraphics[width=0.24\textwidth]{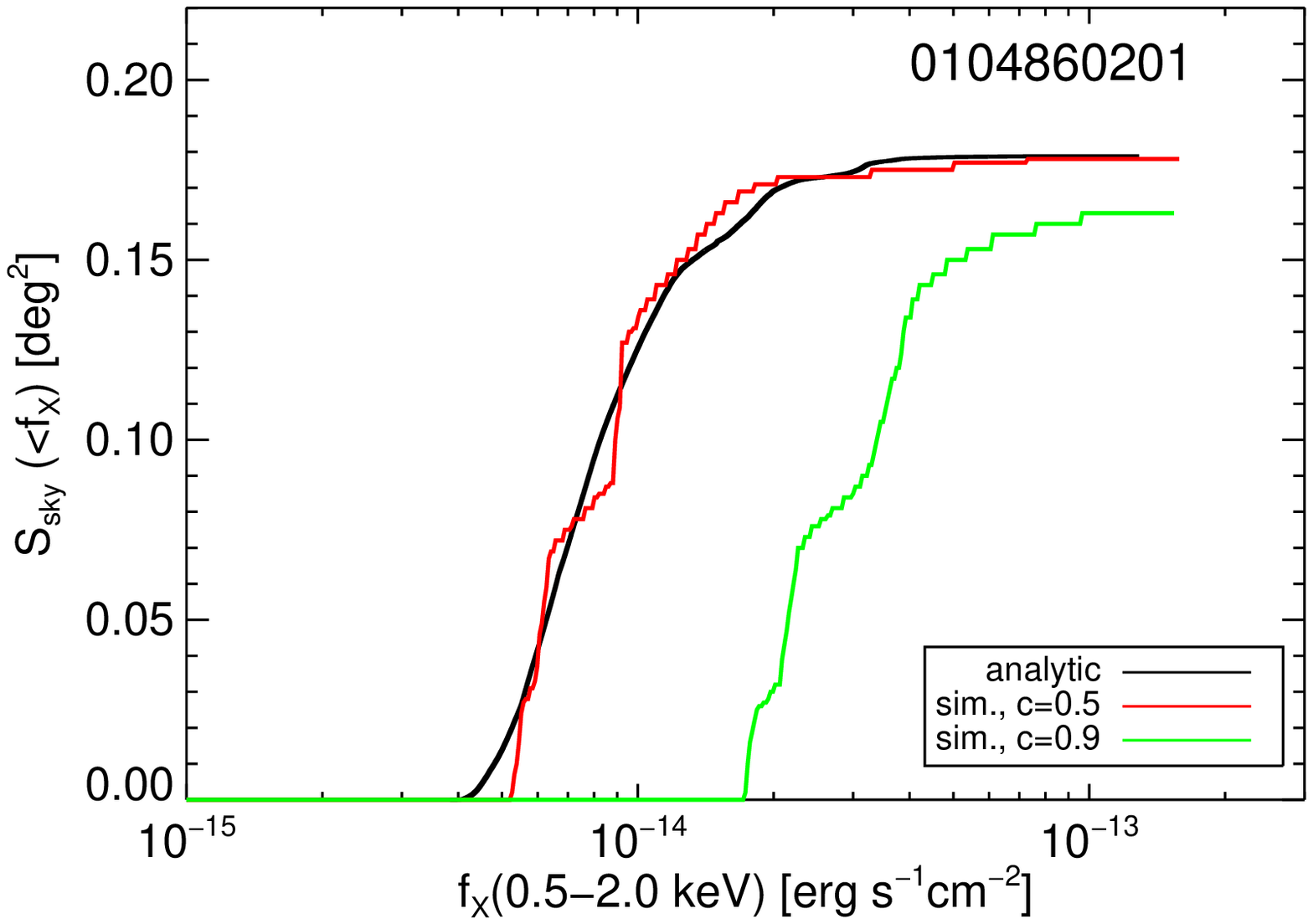}
\includegraphics[width=0.24\textwidth]{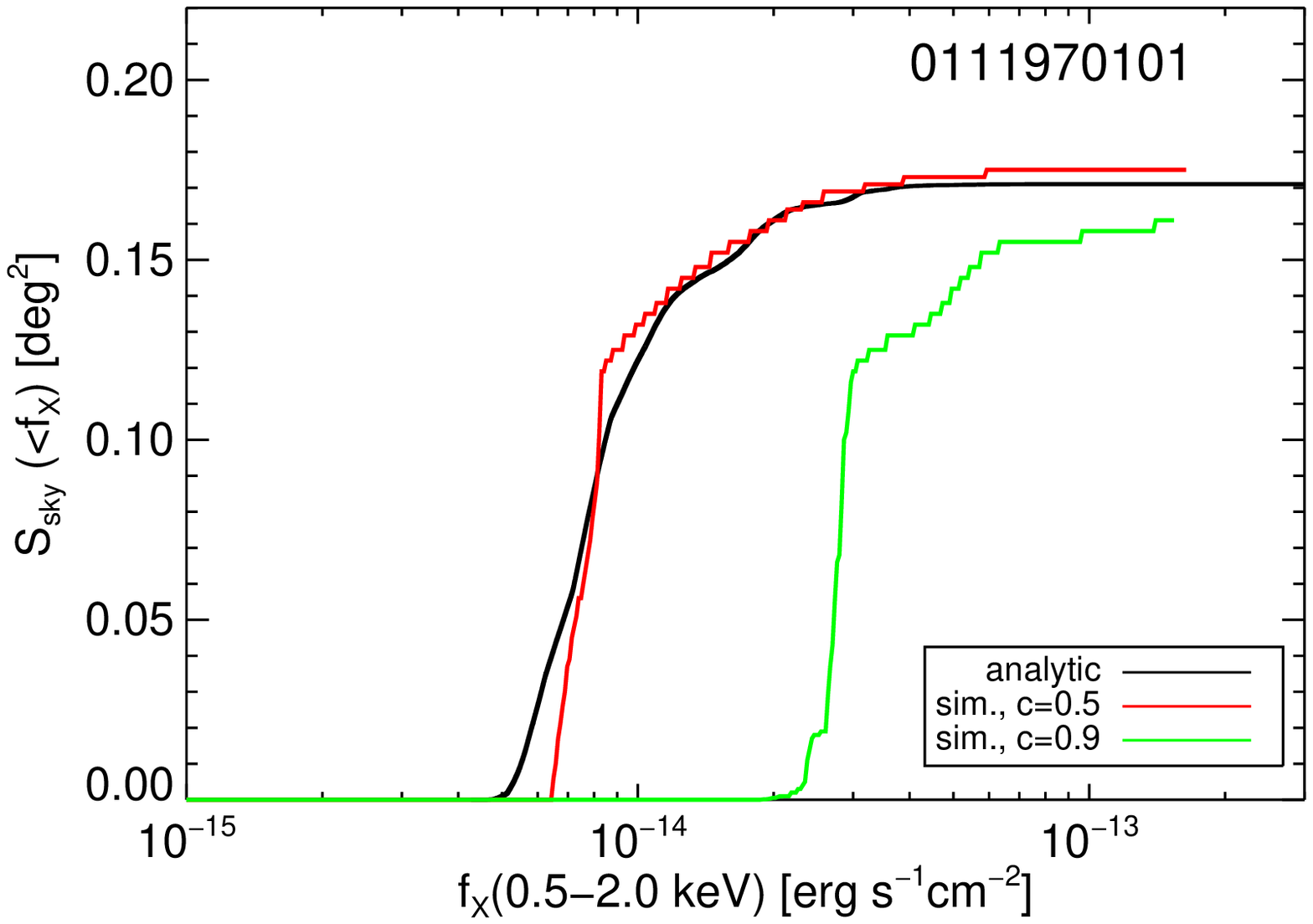}
\includegraphics[width=0.24\textwidth]{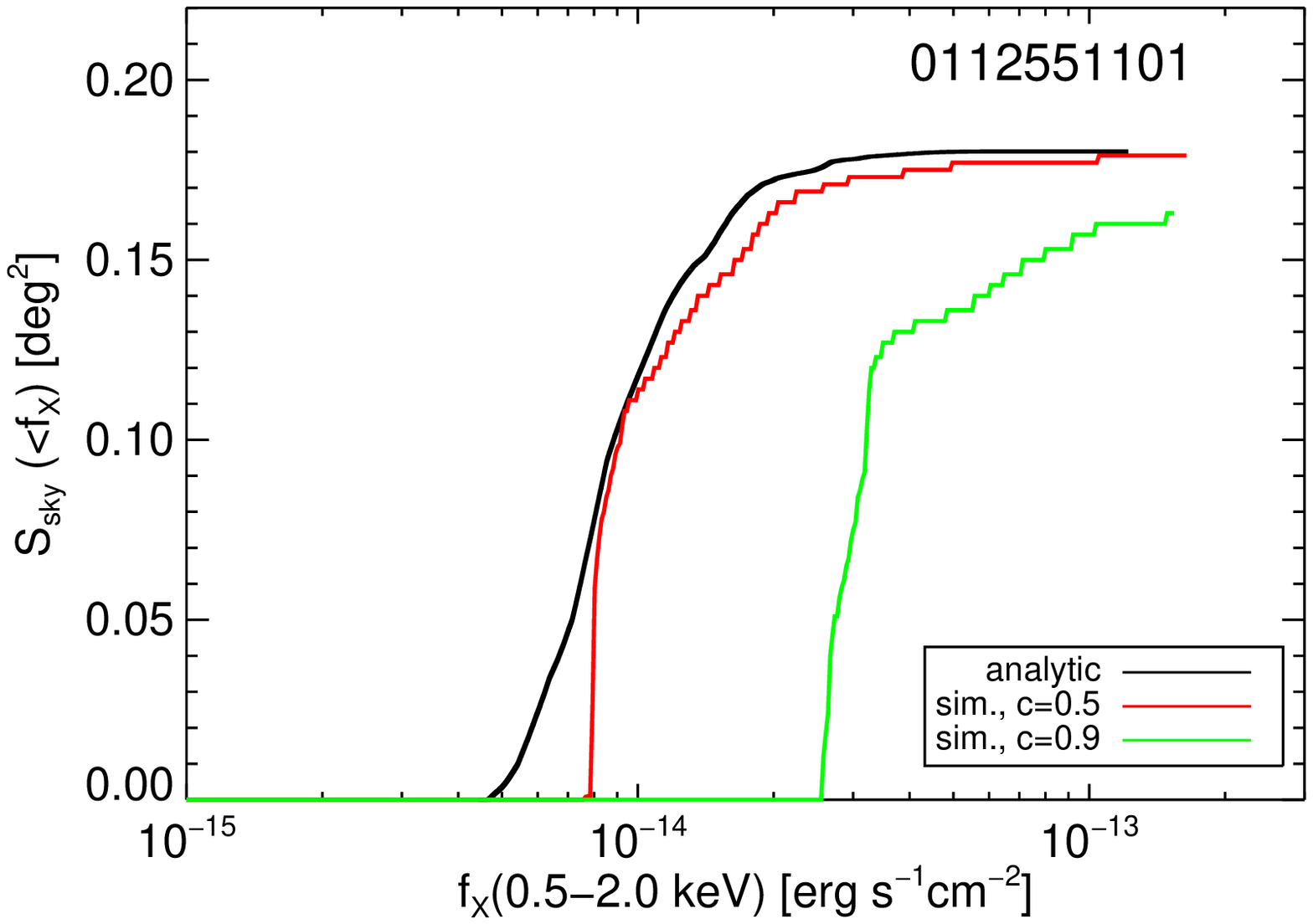}
\end{center}
\caption[Sky coverage comparison with simulations]{Sky-coverage for
  extended sources in three XDCP fields from Monte Carlo simulations
  at the $50\%$ (red curve) and $90\%$ (green) completeness level.
  The black solid curve shows the sky-coverage calculated by scaling
  the point source curve (dashed) with an offset factor of 2.4 (see
  Sect.~\ref{sec:skycov}). The simple scaling is shown to be a good
  first order description of the extended source sensitivity
  function. See Appendix~\ref{sec:simsensf} and Sect.~\ref{sec:skycov}
  for details.}
\label{fig:simsensf}
\end{figure}

Simulations are currently available for a subset of the
XMM-\emph{Newton} Distant Cluster Project
\citep[XDCP,][submitted]{boehringer05, fassbender-thesis,
  fassbender11c} fields.  From these fields we selected 3 observations
(XMM OBSIDs 0104860201, 0111970101, 0112551101) which have similar
depth to our survey fields (e.g. cleaned exposure times $\sim 10$~ks
and enough area unaffected by the central source to safely assess the
background). We processed these fields with our detection pipeline and
calculated the point source sensitivity function and the scaled
extended sensitivity function as described in
Sect.~\ref{sec:skycov}. The comparison with the sensitivity function
derived from the simulations are displayed in Fig.~\ref{fig:simsensf}.
The simple calculation (black curves) already matches the realistic
calculation (red curves) very well, capturing also the transition
parts of the curve.  The curves from simulations include the effect of
incompleteness of the output catalogs. The red curves in
Fig.~\ref{fig:simsensf} are calculated for a $50\%$ completeness level
(c=0.5). The completeness of our cluster catalog can be assessed only
by simulations, but is certainly higher than $50\%$.  This means that
the preliminary analytic sky coverage function overestimates the sky
coverage. The use of the true sky-coverage function would lead to an
increase of the weighting factor in Eq.~\ref{eq:lognlogs} and would
move the points in Fig.~\ref{fig:lognlogs} in the relevant flux range
slightly higher.  The sensitivity function for a $90\%$ completeness
scenario is plotted in green and as expected yields a much smaller
area for the given flux.

Additional subtle effects slightly influence this comparison,
e.g. leading to different normalizations of the two curves in the
saturated high-end part: \textbf{1)} All the fields have a bright
source in the center of the field-of-view, which has to be
excised. The excision is treated slightly differently in the
simulations and in the simple calculations leading to slightly
different total \emph{geometric} areas.  \textbf{2)} The simulated
curves were calculated for the single band detection scheme in the
$0.35 - 2.4$~keV while our analytic solution is for the $0.5 - 2$~keV
band. The fluxes were converted to the $0.5 - 2$~keV band, but
detection in these different energy ranges could cause slightly
different completeness and contamination fractions.

We conclude, that our first-order approach yields a good description
of the sensitivity function for a $50\%$ completeness level. To
estimate the completeness of our sample Monte Carlo simulations are
needed.  The sensitivity functions from Sec.~\ref{sec:skycov} provides
sufficient precision for present applications and the preliminary
$\log N - \log S$ is already in good agreement with previous
findings. The described simulation pipeline will be applied to the
whole XMM-BCS survey in subsequent work and the realistic selection
function will be utilized for further analysis and modelling of the
final cluster sample.

\section{Comparison with the XMM-LSS survey}
The first part of the XMM-LSS survey \citep[the initial 5
  deg$^2$,][]{pierre07, pacaud06, pacaud07} offers an excellent match
to our survey not only with respect to the area, but also to the
typical depth (having only slightly higher average exposure
times). Since the XMM-LSS project has already carried out Monte Carlo
simulations to calibrate its detection and source-characterization
pipeline, we make here an effort to compare results derived from our
XMM-BCS pipeline with results published by XMM-LSS.

\subsection{Cluster detection comparison}
\label{sec:lsscomp}
A full comparison of the source detection pipelines would be only of
limited use and is currently impossible since only a small part of the
XMM-LSS extended sources have been thus far spectroscopically
confirmed \citep[the so-called C1 sample
  of][]{pacaud07}.\footnote{Catalog available at :
  \\ \texttt{heasarc.gsfc.nasa.gov/W3Browse/all/xmmlssoid.html}}
Therefore, we restrict ourselves to the reanalysis of the C1 sample.

We downloaded all the XMM-LSS fields with C1 detections\footnote{XMM
  OBSIDs: 0037980301, 0037980701, 0037981001, 0037981101, 0037981201,
  0037981501, 0037981601, 0037981801, 0037982501, 0037982601,
  0109520201, 0109520301, 0109520601, 0111110301, 0111110401,
  0112680101, 0112680201, 0112680301, 0112680401, 0112680501,
  0147110101, 0147110201.} and fully reanalyzed them with the XMM-BCS
pipeline. We confidently detected all the C1 clusters and they are
among our highest ranked extended source detections.

\begin{figure*}[t!]
\begin{center}
\includegraphics[width=0.49\textwidth]{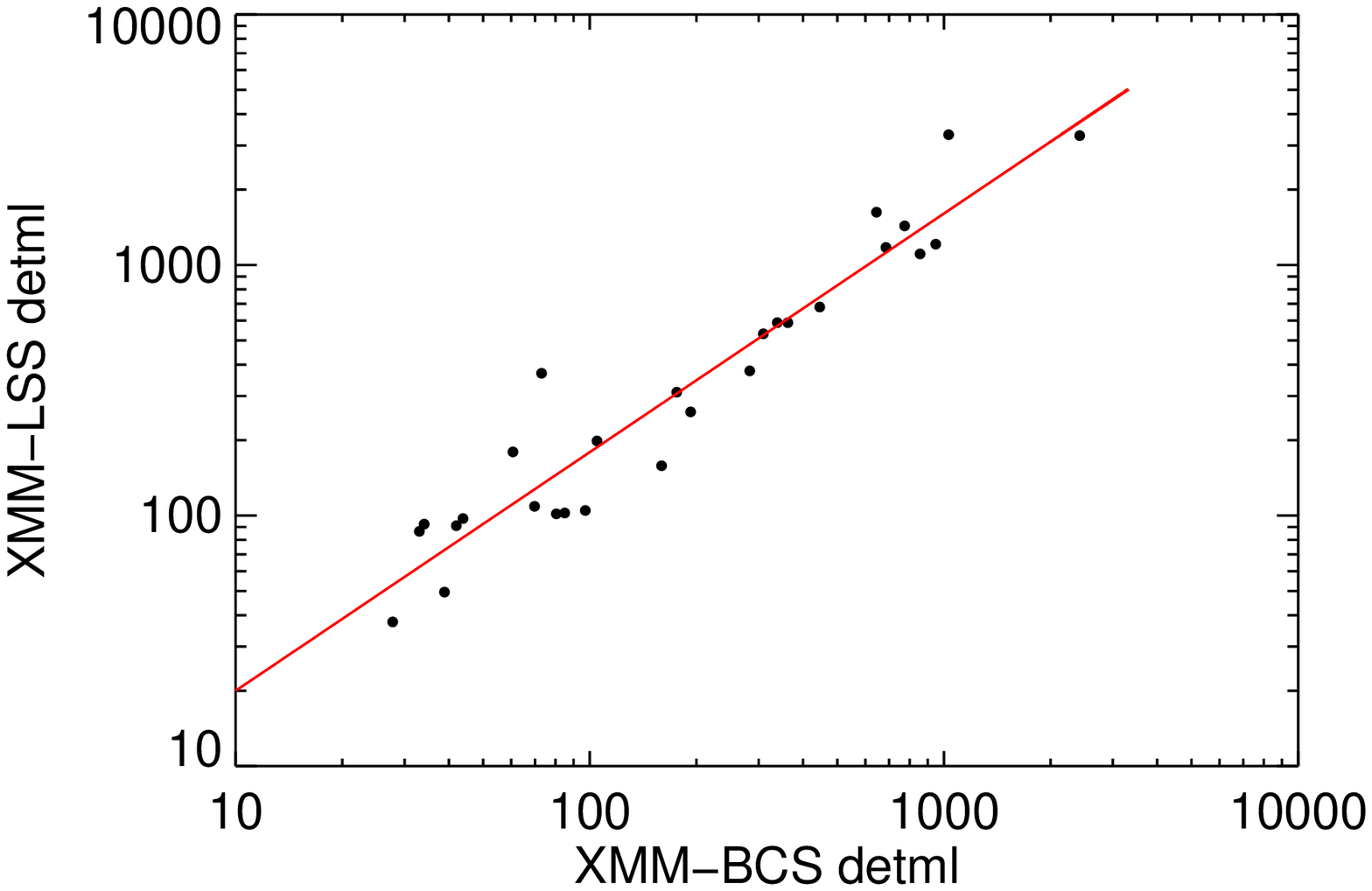}
\includegraphics[width=0.49\textwidth]{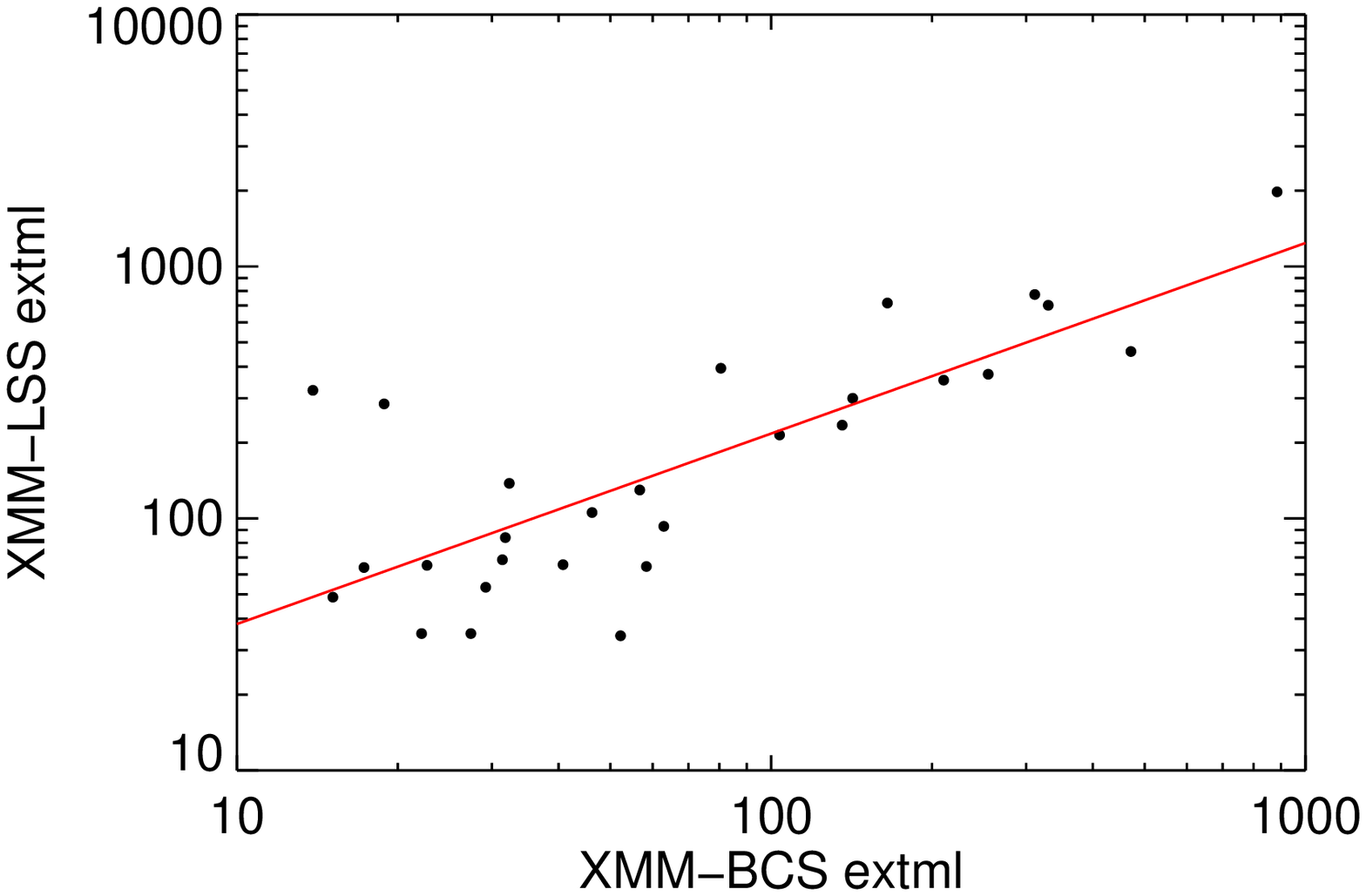}
\end{center}
\caption[Comparison of detection parameters with the XMM-LSS
  survey]{Comparison of detection (left panel) and extent likelihoods
  (right panel) between our pipeline (x-axis) and the XMM-LSS pipeline
  \citep[y-axis,][]{pacaud07}. The derived likelihoods are well
  correlated and the red line shows the best fit relations.}
\label{fig:lss-detml}
\end{figure*}

In Fig.~\ref{fig:lss-detml} we compare their detection and extent
likelihoods with their respective XMM-LSS variants
(\texttt{SB\_Detect\_Likelihood} and \texttt{SB\_Extent\_Likelihood}).
Both sets of parameters exhibit a strong correlation, showing a good
consistency between both detection approaches (XMM-LSS uses a single
band wavelet detection scheme). The scatter between the parameters is
caused by differences in the data reduction process, background
estimation and source detection algorithms.

The C1 sample is defined by \texttt{SB\_Detect\_Likelihood}$>32$,
\texttt{SB\_Extent\_Likelihood}$>33$. We fit a linear relation in the
two log-log planes and use these cuts to convert the XMM-LSS
thresholds to our parameters obtaining: \texttt{det\_ml}$>16.4$
(equivalent to $\sim5.4\sigma$ detection in our scheme) and
\texttt{ext\_ml}$>8.3$ (i.e. $\sim3.7\sigma$ extent significance).

\subsection{X-ray photometry comparison}
\label{sec:lsscomp_photometry}
\begin{figure}[ht]
\begin{center}
\includegraphics[width=0.5\textwidth]{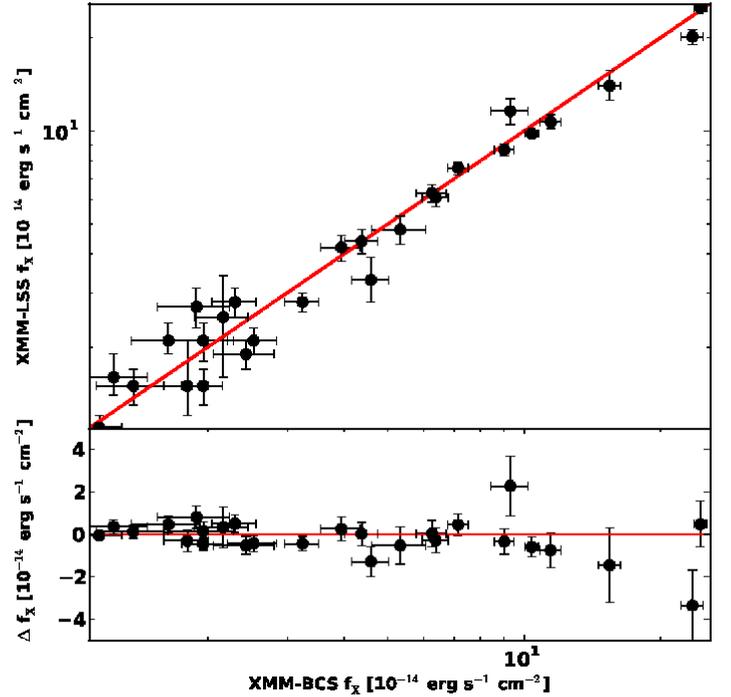}
\end{center}
\caption[Flux comparison with the XMM-LSS survey]{Comparison of
  measured X-ray fluxes for the C1 subsample of the XMM-LSS survey in
  the $0.5-2$~keV band and a $0.5$~Mpc aperture
  \citep[][y-axis]{pacaud07} and the fluxes measured by our pipeline
  (x-axis). The red line marks equality. The bottom panel shows the
  residuals $\Delta f_X = f_X^{XMM-LSS}- f_X^{XMM-BCS}$. See
  Sect.~\ref{sec:lsscomp_photometry} for details.}
\label{fig:xmmbcs-xmmlss-flux}
\end{figure}

In Fig.~\ref{fig:xmmbcs-xmmlss-flux} we compare the fluxes in the
$0.5-2$~keV band and $0.5$~Mpc aperture measured by the XMM-LSS and by
us using the growth curve method (Sect.~\ref{sec:gca}). Being
interested in the flux estimation we have fixed the cluster redshifts
to its spectroscopic value provided by XMM-LSS. We choose not to use
the information on the spectroscopic temperature, but rather we
estimate it from the scaling relation as we do for the XMM-BCS sample
(see Sect~\ref{sec:physpar}).

\begin{figure}[ht]
\begin{center}
\includegraphics[width=0.5\textwidth]{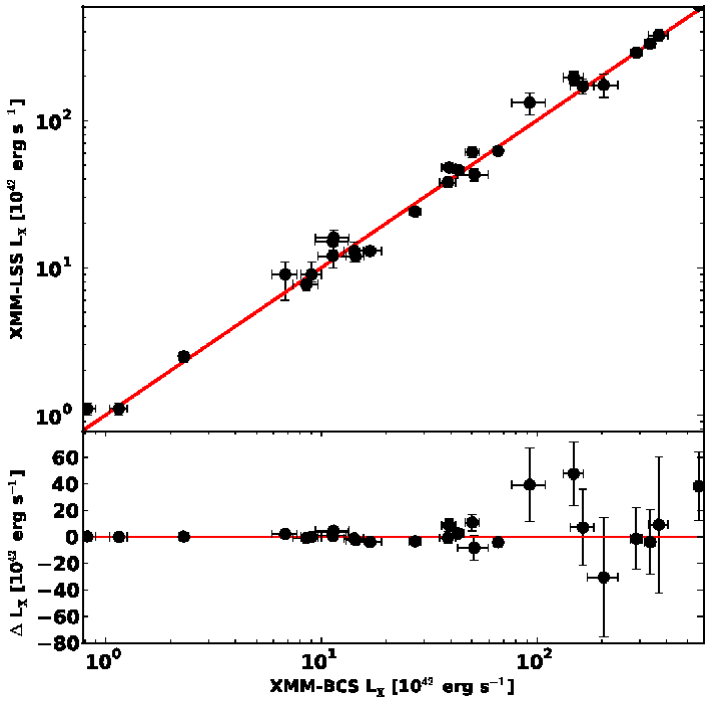}
\end{center}
\caption[Luminosity comparison with the XMM-LSS survey]{Comparison of
  bolometric X-ray luminosities for the C1 subsample of the XMM-LSS
  survey in the $r_{500}$ aperture \citep[][y-axis]{pacaud07} and the
  luminosities measured by our pipeline (x-axis). The red line marks
  equality. The bottom panel shows the residuals $\Delta L_X =
  L_X^{XMM-LSS}- L_X^{XMM-BCS}$. See
  Sect.~\ref{sec:lsscomp_photometry} for details.}
\label{fig:xmmbcs-xmmlss-lx}
\end{figure}

The $0.5-2$~keV fluxes in a 0.5~Mpc aperture are compared in
Fig.~\ref{fig:xmmbcs-xmmlss-flux}, where the residuals in the bottom
panel are defined as $\Delta f_X = f_X^{XMM-LSS}- f_X^{XMM-BCS}$. We
find the fluxes in agreement with the mean relative difference of
$\sim0.6\%$ and a scatter of $\sim17\%$ (averaged over the whole
sample). If we split the sample by the median flux ($fx\approx3.9
\times 10^{-14}$~erg cm$^{-2}$ s$^{-1}$) the XMM-LSS fluxes are on
average by $\sim4\%$ higher on the faint end while being slightly
lower by roughly the same amount on the brighter end. A systematic
difference of this magnitude would however cause only a $\sim1\%$
shift in the temperature estimates and $\sim2\%$ shift on mass.  An
analogous comparison for the bolometric luminosity in the $r_{500}$
aperture is displayed in Fig.~\ref{fig:xmmbcs-xmmlss-lx}. Here the
mean relative difference is $\sim6.6\%$ and a scatter of $\sim19\%$
(the corresponding temperature/mass difference are $2\%/3\%$).

The overall agreement is encouraging, if we take into account that the
two pipelines utilize principally different approaches to the flux
measurement. XMM-LSS utilizes a beta model fit to the cluster's
surface brightness integrated out to a fiducial radius, while our
method is completely non-parametric (except for a typically small
extrapolation factor if the required aperture is larger than the range
where the cluster emission is detected directly). Background
estimation in both approaches is also markedly different.

Since we did not use the information on the spectroscopic X-ray
temperature, it is interesting to note, that the mean temperature
residuals (spectroscopic compared to our estimates from the scaling
relations) are $<2\%$ with a standard deviation of $\sim23\%$
(i.e. comparable to measurement errors). Although the error bars are
large, this agreement indicates that the $L-T$ scaling relation and
its evolution adopted in this work \citep{pratt09} are suitable for
cluster samples drawn from surveys of this type.

The cluster mass is not a direct observable in either of the two
surveys. XMM-LSS gives rough estimates based on their spectroscopic
measurement and beta model fit using the relation from
\citet{ettori00}. Our estimates, using the $L-M$ relation of
\citet{pratt09}, give on average almost $40\%$ higher masses. We note
however, that the \citet{ettori00} relation was derived before the
advent of XMM-\emph{Newton} and \emph{Chandra}. It is thus not well
representing our current understanding of cluster mass estimation.

\begin{figure}[t!]
\begin{center}
\includegraphics[width=0.5\textwidth]{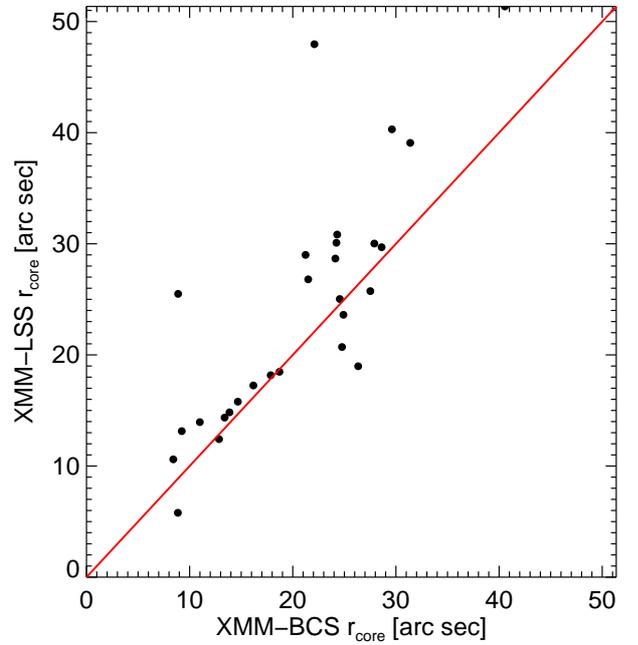}
\end{center}
\caption[Core radii comparison with the XMM-LSS survey]{Beta model
  core radii for the XMM-LSS C1 sample as estimated by our pipeline
  (x-axis) and by the XMM-LSS. The red line marks equality. The core
  radii are typically highly uncertain given the relatively low photon
  statistics. Despite this, the agreement between the two estimates is
  good. Note that the XMM-LSS values are fitted with the beta value as
  a free parameter, while we fix its value to 2/3.}
\label{fig:rcore}
\end{figure}

Finally, we also check the consistency of the beta model fits between
the two pipelines. Since the core radius $r_{\mathrm{core}}$ and the
$\beta$ exponent of the beta model are strongly degenerate, especially
for the case of low counts profiles, our fitting procedure keeps
$\beta$ fixed to the canonical value of 2/3. The XMM-LSS pipeline
carries out fits with both $r_{\mathrm{core}}$ and $\beta$ as free
parameters.  Despite this difference, we find good agreement between
the estimated core radii (Fig.~\ref{fig:rcore}).

\end{appendix}

\end{document}